\documentclass[preprint,12pt]{elsarticle}

\usepackage{amssymb}
\usepackage{graphicx}
\usepackage{amssymb,amsmath,bm}
\usepackage{textcomp}
\usepackage{multirow}
\usepackage{caption}
\usepackage{subfig}
\usepackage{adjustbox}
\usepackage{hyperref}
\usepackage{url}
\usepackage{amsmath}
\usepackage{booktabs}

\hypersetup{colorlinks=false, urlcolor=blue}

\begin{document}

\begin{frontmatter}



\title{Melody Extraction from Polyphonic Music by Deep Learning Approaches: A Review}


\author{Gurunath Reddy M, K. Sreenivasa Rao, Partha Pratim Das}

\address{Indian Institute of Technology Kharagpur, India \\
  {\small \tt \{mgurunathreddy, ksrao\}@sit.iitkgp.ernet.in, ppd@cse.iitkgp.ac.in }}

\begin{abstract}

Melody extraction is a vital music information retrieval task among music researchers for its potential applications in education pedagogy and the music industry. Melody extraction is a notoriously challenging task due to the presence of background instruments. Also, often melodic source exhibits similar characteristics to that of the other instruments. The interfering background accompaniment with the vocals makes extracting the melody from the mixture signal much more challenging. Until recently, classical signal processing-based melody extraction methods were quite popular among melody extraction researchers. The ability of the deep learning models to model large-scale data and the ability of the models to learn automatic features by exploiting spatial and temporal dependencies inspired many researchers to adopt deep learning models for melody extraction. In this paper, an attempt has been made to review the up-to-date data-driven deep learning approaches for melody extraction from polyphonic music. The available deep models have been categorized based on the type of neural network used and the output representation they use for predicting melody. Further, the architectures of the 25 melody extraction models are briefly presented. The loss functions used to optimize the model parameters of the melody extraction models are broadly categorized into four categories and briefly describe the loss functions used by various melody extraction models. Also, the various input representations adopted by the melody extraction models and the parameter settings are deeply described. A section describing the explainability of the block-box melody extraction deep neural networks is included. The performance of 25 melody extraction methods is compared. The possible future directions to explore/improve the melody extraction methods are also presented in the paper.     

\end{abstract}


\end{frontmatter}




\section{Introduction}
\label{intro}

Melody extraction is the task of automatically extracting the dominant melodic line in a polyphonic music signal. Here, polyphony refers to the music signal in which more than one instrument may sound concurrently (e.g., Vocals, Tanpura, and Tabla), or it can be a single instrument capable of producing multiple notes at a given time (e.g., violin). The word melody is a musicological term that is purely subjective. Hence, we can find many definitions of melody in various contexts. The melody representation adopted in melody extraction methods is the one proposed by Goto~\cite{goto2004real}, such as melody is the sequence of F0 (fundamental frequency or pitch) values that correspond to the dominant instrument's perceived pitch. The dominant instrument can be the human singing voice or any lead instrument in the polyphonic music signal. The accurate melody extraction remained a challenging and unsolved task in the research community because of its two-fold complexity~\cite{salamon2014melody}. Firstly, the polyphonic music signal is the superposition of many instruments which play simultaneously. Hence, it is hard to attribute specific frequency bands and energy levels to a specific instrument. Secondly, the task of determining the sequence of pitch values that constitutes the main melody is also tricky. This, in turn, poses mainly three challenges~\cite{salamon2014melody}: (\romannumeral 1) determining the melody regions in the music signal, (\romannumeral 2) ensuring the estimated F0 is in the correct octave range, and (\romannumeral 3) selecting the right melody pitch when there is more than one note present at the same time.

The accurately extracted melody can be used in many potential music applications such as automatic music transcription, query by humming, music de-soloing, singer identification, and many other music information retrieval tasks. Since the human singing voice is famous in popular music, vocal melody extraction is quite popular among melody extraction researchers. 

Until recently, classical signal processing-based melody extraction methods were quite popular among melody extraction researchers. We can find two major approaches to derive the vocal melody from polyphonic music signal by classical signal processing-based approaches viz. \textit{Source separation} and \textit{Salience} based methods. Source separation-based methods extract the F0 contour of the melody source by separating it from the rest of the music signal~\cite{durrieu2010source, tachibana2010melody, huang2012singing, rafii2013repeating}. On the other hand, a substantial number of saliency-based melody extraction methods are proposed based on time-frequency representation of music signal~\cite{salamon2014melody}. The salience-based melody extraction methods work on the principle of "understanding-without-separation" as described by Scheirer~\cite{scheirer2000machine}. Salience-based approaches follow generalized architecture to extract the melody from the mixture signal~\cite{salamon2013melody}. They involve mainly the following steps: (\romannumeral 1) preprocessing, (\romannumeral 2) spectral representation and filtering, (\romannumeral 3) designing pitch salience function, (\romannumeral 4) pitch tracking of the candidate salience peaks in the estimated salience function and (\romannumeral 5) making the vocal and non-vocal decision. Salience-based methods mostly differ by the way the salience function is computed, salience peaks are estimated, and the melody contours of the dominant source are created by pitch tracking methods~\cite{rao2010vocal,salamon2012melody}. A detailed review of signal processing-based melody extraction methods can be found in~\citep{salamon2014melody}.

Recently, the ability of the deep learning models, such as fully connected neural networks, convolutional neural networks (CNNs), and Recurrent Neural Networks (RNNs), to model the large scale data, and the ability to learn automatic features by exploiting spatial and temporal dependencies inspired many researchers to adopt deep learning models for melody extraction. In this paper, we review the popular data-driven supervised neural network models trained to minimize the discrepancy between the predicted and the manually annotated vocal F0 and the neural networks trained to separate the vocal source from the mixture to extract the melody from the polyphonic music.

\section{Melody Extraction Methods} \label{sec:melody_extraction_approaches}

We can broadly classify the data-driven melody extraction methods into classification~\cite{du2021singing, chou2018hybrid, gao2019multi, gao2020multi, hsieh2019streamlined, gao2021hrnet, kum2016melody, chen2019cnn, bittner2017deep, yu2021frequency, yu2021hanme, basaran2018main, park2017melody, kum2016melody, bittner2017deep, kum2020semi, rigaud2016singing, su2018vocal, gao2019vocal, lu2018vocal} and source separation~\cite{gao2021vocal, nakano2019joint, jansson2019joint, fan2016singing} based melody extraction methods. The possible vocal pitch range is quantized into a fixed number of classes (levels) in the classification-based approaches, mainly on the cent scale. In the source separation-based approaches, the melodic source signal (primarily vocals) is separated from input polyphonic music to extract the melody from the separated signal from the pitch detection algorithms or monophonic pitch detection neural network models. Based on the type of neural networks used, we can further classify the classification-based melody extractions methods into Feed Forward Neural Network (FFNN) models, CNN, RNN, Convolutional Recurrent Neural Network (CRNN), and semantic segmentation (U-net) based models. We can also group the melody extraction models based on vocal and general melody extraction models. Vocal melody extraction~\cite{du2021singing, chou2018hybrid, gao2020multi, gao2019multi, hsieh2019streamlined, gao2021hrnet, kum2019joint, kum2017classification, chen2019cnn, gao2021vocal, yu2021frequency, yu2021hanme, nakano2019joint, jansson2019joint, kum2016melody, bittner2018multitask, kum2020semi, rigaud2016singing, su2018vocal, gao2019vocal, lu2018vocal, fan2016singing} models extract the melody or F0 of the vocal source of the polyphonic vocal music. Whereas general~\cite{hsieh2019streamlined, bittner2018multitask, bittner2017deep, basaran2018main, park2017melody} melody extraction models are capable of extracting melody from both vocal and melodic instruments of the polyphonic music.

\subsection{Classification based Melody Extraction Models} \label{ss:classification}

\begin{figure}[!tbp]
        \centering
		\resizebox{8cm}{8cm}{\includegraphics{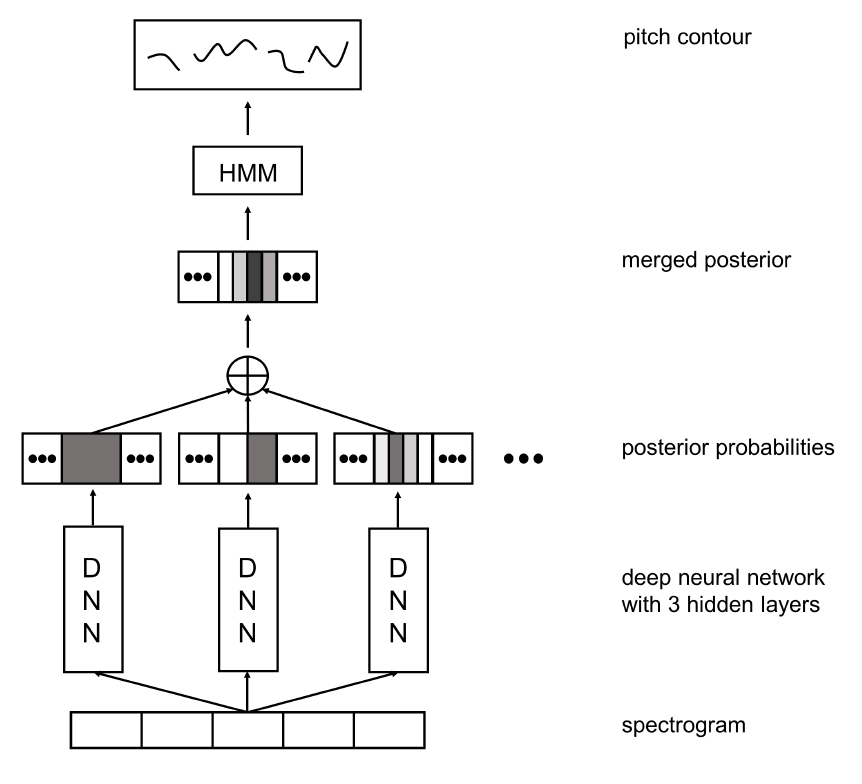}}  
        \caption{Block diagram of the multicolumn deep neural networks for melody extraction~\cite{kum2016melody}.}
        \label{fig:arch_kum2016melody}
\end{figure} 

Multi-column deep neural networks are proposed for vocal melody extraction from polyphonic music in~\cite{kum2016melody}. The proposed approach trains each neural network to predict the melody label with different pitch resolutions. The final melody contour is obtained by combining the outputs of all the neural networks and post-processing with a hidden Markov model. Also, the training data is augmented with pitch shifting to overcome the model overfitting. The proposed multi-column deep neural network (MCDNN) is shown in Fig.~\ref{fig:arch_kum2016melody}. The input to the network is the odd-numbered contextual spectrogram frames. The DNN is trained to predict the pitch for the middle frame. As shown in the figure, the network consists of identical DNNs with the output layers configured to predict the pitch at different resolutions. For example, the leftmost DNN is configured to predict pitch at semitone, other DNNs with half a semitone, quarter semitone, and so on. Finally, the posterior probabilities of all columns are merged in the maximum likelihood sense. After that, temporal smoothing is performed by HMM on the combined output from all the columns. Since the model is trained with only voiced frames, a simple energy-based singing voice detector is employed for vocal/non-vocal classification.

\begin{figure}[!tbp]
        \centering
		\resizebox{8cm}{8cm}{\includegraphics{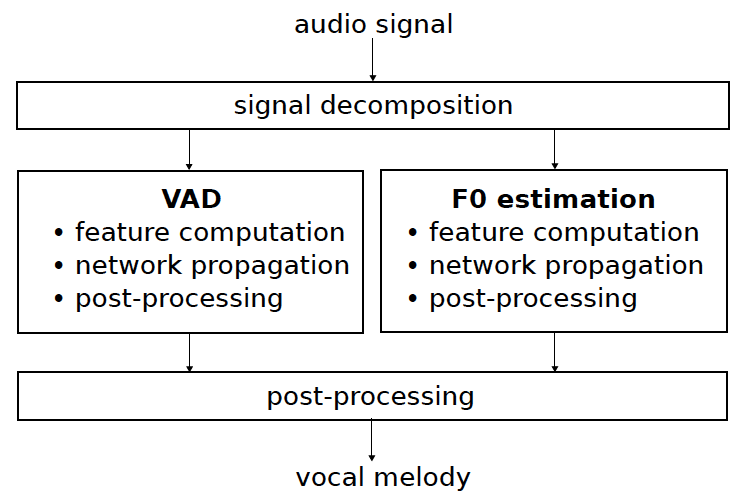}}  
        \caption{Block diagram of the singing voice detection and melody extraction model~\cite{rigaud2016singing}.}
        \label{fig:arch_rigaud2016singing}
\end{figure} 

In this~\cite{rigaud2016singing} paper, two separate DNNs are proposed for vocal melody extraction and singing voice detection, respectively. The overview of the proposed method is shown in Fig.~\ref{fig:arch_rigaud2016singing}. A three-layer BiLSTM network of 50 units each with the last single logistic unit is used for voice activity detection (VAD). The input to the VAD network is the Mel-frequency spectrogram computed from the output of Harmonic Percussion Source Separation (HPSS). The output of the VAD network is post-processed with a threshold of 0.5 for vocal and non-vocal detection. The input to the melody extraction network is the harmonic enhanced HPSS signal. The melody extraction network consists of two hidden layers of 500 sigmoid units and a final softmax layer. Finally, a Viterbi pitch tracking is applied on the softmax matrix output to smoothen the pitch contour. Audio degradation toolbox~\cite{mauch2013audio} is used to augment the dataset to avoid overfitting by degrading the audio with random degradation function in the toolbox. 	

\begin{figure}[!tbp]
        \centering
		\resizebox{12cm}{6cm}{\includegraphics{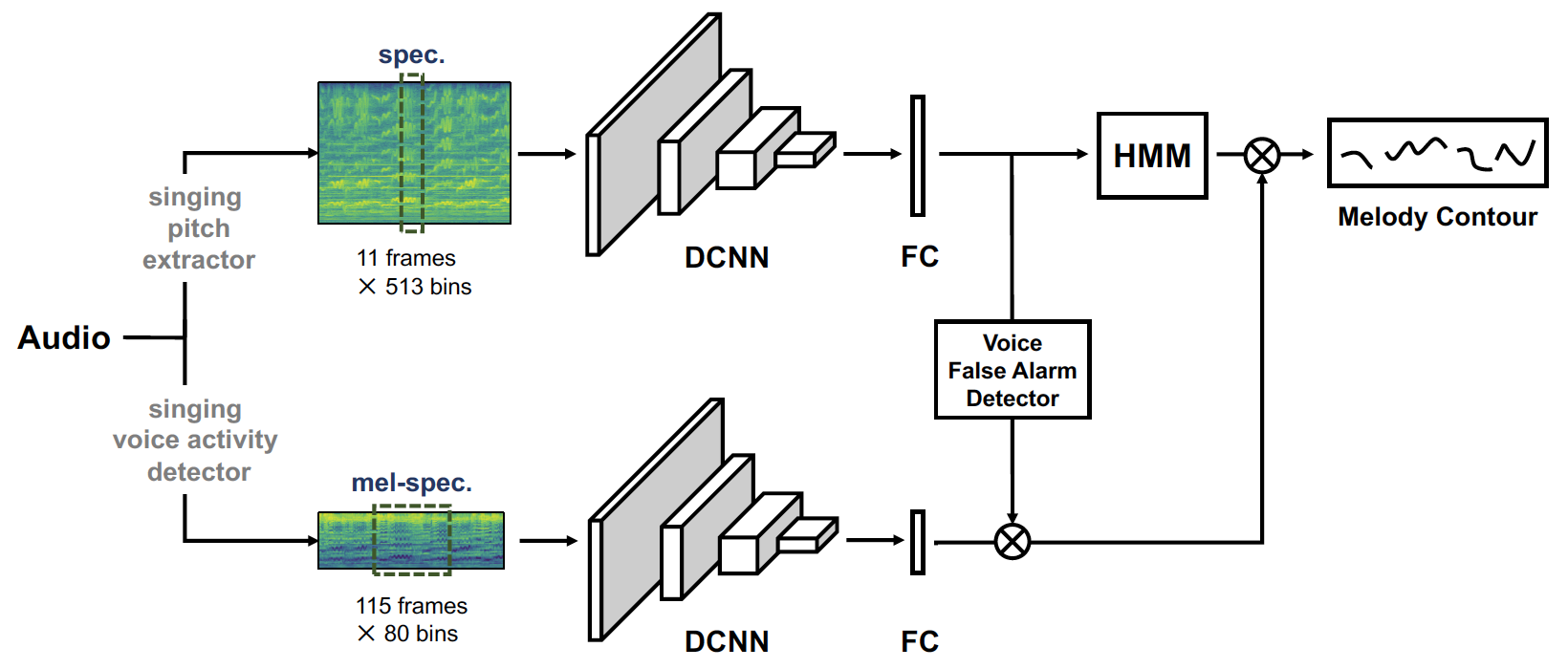}}  
        \caption{The joint singing voice detection and melody extraction model~\cite{kum2017classification}.}
        \label{fig:arch_kum2017classification}
\end{figure} 

Authors of~\cite{kum2017classification} proposed the deep CNN-based parallel networks for singing pitch extraction and singing voice detection. The output of the singing pitch extractor is smoothed by a post-processing approach. Further, the false alarm rate of the voicing detection at the boundaries is reduced by proposing a voicing false alarm detector with the help of the predicted pitch. As shown in Fig.~\ref{fig:arch_kum2017classification}, the proposed model consists of singing pitch extractor (SPE) and the singing voice activity detection (SVAD) networks. The SPE network consists of four convolution blocks followed by fully connected layer pitch label prediction. Interestingly, they apply average pooling in the time axis and max-pooling in the frequency axis in each convolution block. The intuition behind applying average pooling along the time axis is justified, with the pitch mainly being smooth and continuous along the time axis. The SPE network is trained with the log-spectrogram input and quantized pitch labels as the target. Similar to the SPE network, the SVAD network consists of four blocks of convolutions. The final block contains 1x1 convolutions followed by global average pooling for voice/non-voice classification. Further, the output of the SVAD is smoothed with median filtering. The input to the SVAD is the 115 frames length Mel-spectrogram. Further, the voicing false alarm rate due to long segments of the Mel-spectrogram at the boundaries is reduced by thresholding the probability of the pitch class with the highest class probability of the output of SPE. That is, if the melody pitch exists for a frame, then the probability for that pitch class will be very high compared to a non-melodic frame where the probabilities tend to be low for all classes. The final voicing detector is obtained by multiplying the output of the SVAD and the voice false alarm detector. Further, the output of the SPE is smoothed by HMM-based Viterbi decoding. 

\begin{figure}[!tbp]
        \centering
		\resizebox{12cm}{4cm}{\includegraphics{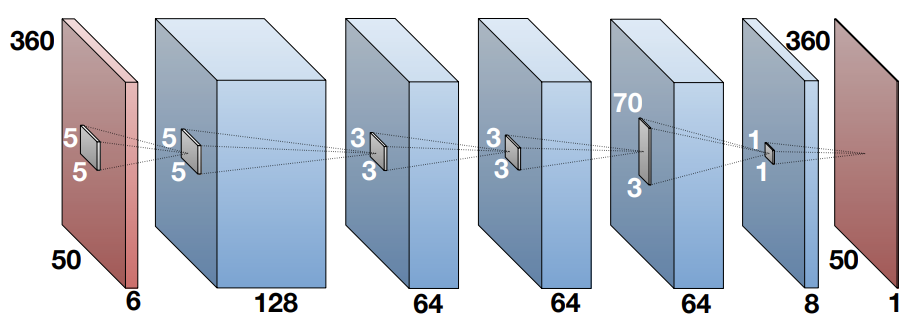}}  
        \caption{CNN architecture to map the input HCQT to F0 saliency~\cite{bittner2017deep}.}
        \label{fig:arch_bittner2017deep}
\end{figure} 

A fully convolutional neural network is proposed to learn the saliency representation of the polyphonic music to extract the melody is proposed by~\cite{bittner2017deep}. The Harmonic Constant-Q Transform (HCQT) is used as the input representation to capture the harmonic relationship in polyphonic music. The model architecture to map the three-dimensional HCQT to saliency representation whose bin with the highest magnitude represents the fundamental frequency is shown in Fig.~\ref{fig:arch_bittner2017deep}. The model consists of five CNN layers, where the first two-layer contains 5x5 kernels. The following two layers have 3x3 kernels, and the final layer has 70x3 kernels to capture the relationship between frequency components within the octave. The feature maps are zero-padded at each layer to preserve the input dimension. The output of the last layer is fed to logistic activation to predict the silence of the frequency bin. Also, pooling layers are not used in the network to preserve the slight shifts in the time frequency. The melody is estimated from the learned saliency representation by choosing a frequency bin with maximum magnitude for each frame. The voicing decision is performed by a fixed threshold estimated by the validation set.

\begin{figure}[!tbp]
        \centering
		\resizebox{10cm}{6cm}{\includegraphics{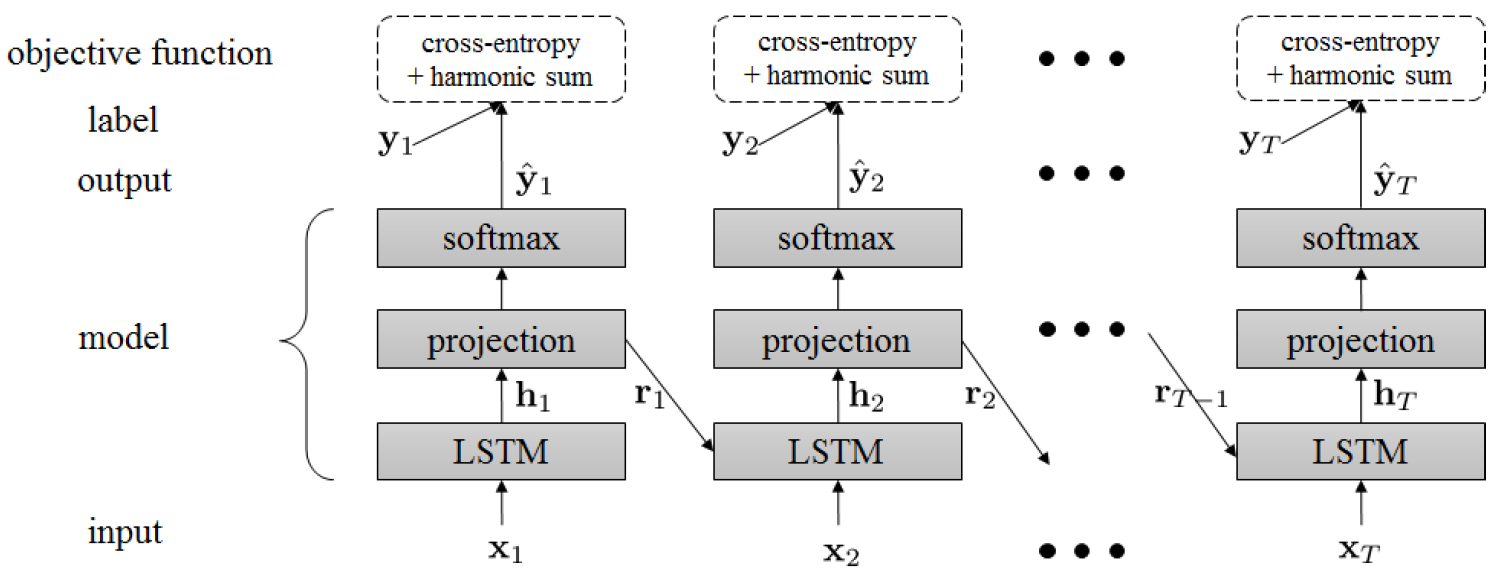}}  
        \caption{Illustration of the LSTM-RNN based melody extraction model~\cite{park2017melody}.}
        \label{fig:arch_park2017melody}
\end{figure} 

An LSTM-RNN based vocal melody extraction and detection from polyphonic music are proposed in~\cite{park2017melody}. A novel harmonic sum loss function is proposed to capture the harmonic structure of the vocals to minimize the octave errors and interference from background music. The proposed LSTM-RNN architecture for the melody extraction and detection is shown in Fig.~\ref{fig:arch_park2017melody}. The sequential network takes magnitude STFT as input and predicts the quantized pitch classes through the softmax function. The model is optimized through cross-entropy as the primary loss function and harmonic sum loss as a regularizer. 

\begin{figure}[!tbp]
        \centering
		\resizebox{7cm}{8cm}{\includegraphics{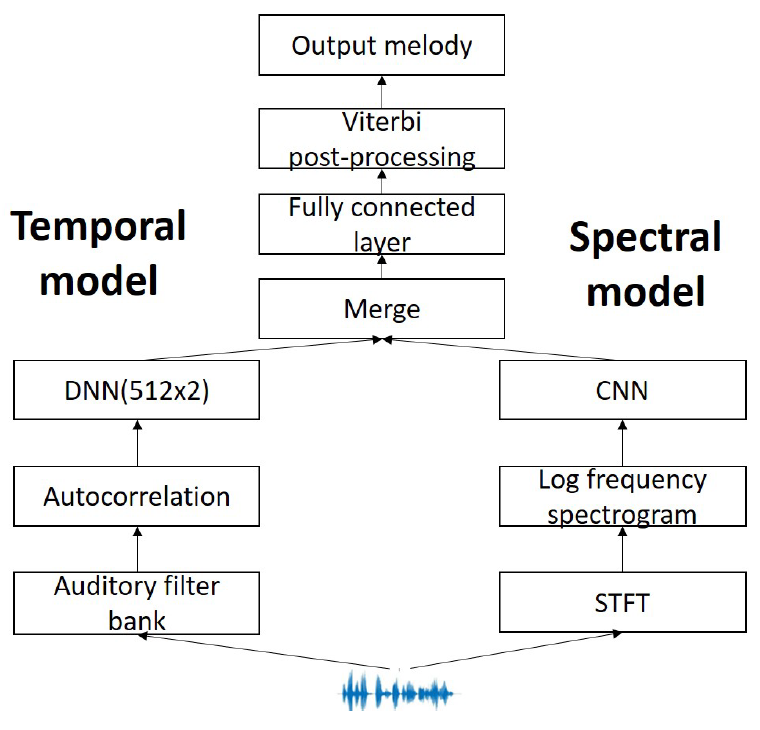}}  
        \caption{Hybrid neural network model based on the human perception of pitch for singing melody extraction from polyphonic music~\cite{chou2018hybrid} which consists of spectral and temporal branches.}
        \label{fig:arch_chou2018hybrid}
 \end{figure} 

A hybrid neural network is proposed in~\cite{chou2018hybrid} based on the human perception of the pitch for singing melody extraction from polyphonic music. Based on the spectral and temporal pitch perception theory for resolved and unresolved harmonics, authors designed a duplex neural network in which a subnetwork consists of CNNs which mimics the spectral model of human pitch perception through learned kernels which behave like templates to shifting harmonic components as shown in Fig.~\ref{fig:arch_chou2018hybrid}. The other subnetwork consists of FFNs to simulate the temporal model using the integrated autocorrelation as input. The temporal and spectral subnetworks are combined to form a hybrid network to integrate the complementary information in both models to predict the melody. The output of the duplex model is post-processed by Veterbi post-processing for smoothing the discontinuous pitch contour.  

\begin{figure}[!tbp]
        \centering
		\resizebox{7cm}{5cm}{\includegraphics{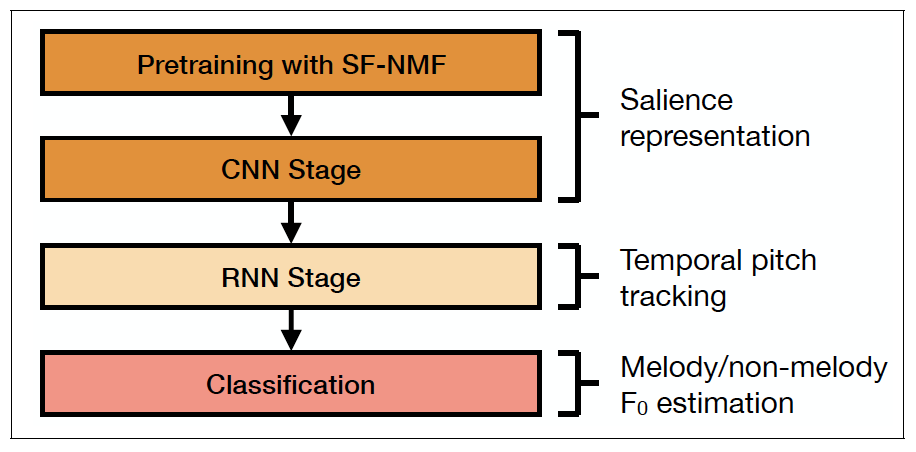}}  
        \caption{Illustration of the SF-NMF CRNN approach for melody extraction~\cite{basaran2018main}.}
        \label{fig:arch_basaran2018main}
\end{figure}

In this~\cite{basaran2018main} paper, a CRNN based main melody extraction approach is proposed. Unlike other approaches, in this work, musically motivated source-filter non-negative matrix factorization (NMF) decomposition source representation is chosen as input to capture the lead instrument's pitch and timbre content compared to classical time-frequency representations. The block diagram of the proposed method is shown in Fig.~\ref{fig:arch_basaran2018main}. Initially, the saliency representation of the input music is obtained by decomposing with source-filter NMF, where the mixture signal is modeled as the source, filter, and accompaniment parts. The source part, which consists of the salience representation of the melody, is used as input to the model. Later, the salience representation is passed through the convolution layers to improve melody saliency further by learning local spatial dependencies. The feature maps of the CNN output layer are fed to the BiGRU for temporal modeling of the melody. The output of the BiGRU is fed to the classification layer for melody F0 estimation and voiced/unvoiced frame classification. Note that the CNN and RNN networks are trained independently in this work.

\begin{figure}[!tbp]
        \centering
		\resizebox{14cm}{4cm}{\includegraphics{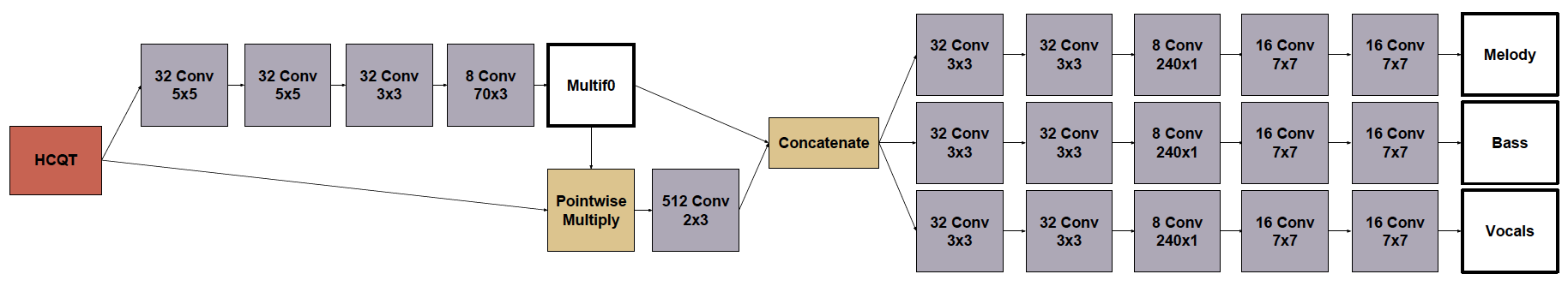}}  
        \caption{Multitask saliency extraction model for various F0 extraction tasks~\cite{bittner2018multitask}.}
        \label{fig:arch_bittner2018multitask}
\end{figure} 

A multitask approach that jointly estimates multi f0, melody, vocals, and the bass line is presented in~\cite{bittner2018multitask}. The paper's significant contribution is the proposed method to semi-automatically create F0 annotations for the proposed task learning using multitrack datasets. The multitask architecture that uses multiple F0 estimation outputs to improve the other auxiliary tasks is shown in Fig.~\ref{fig:arch_bittner2018multitask}. Initially, the model predicts the multiple-F0 saliency map from the HCQT. The multi-F0 saliency map is used to mask each HCQT channel. The masked HCQT features are passed through the convolution layer to capture the timbral information. This output is concatenated with the multiple-F0 saliency maps. The concatenated feature maps are passed through identical convolutional subnetworks to predict the melody, bass, and vocal saliency.

\begin{figure}[!tbp]
        \centering
		\resizebox{8cm}{5cm}{\includegraphics{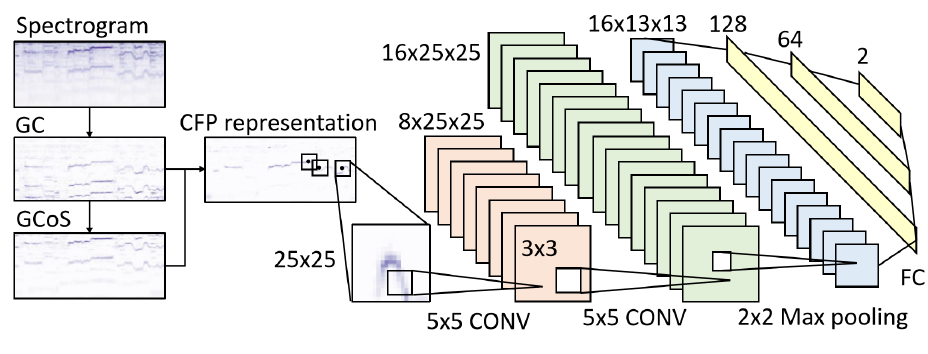}}  
        \caption{Block diagram of the patch based melody extraction~\cite{su2018vocal}.}
        \label{fig:arch_su2018vocal}
\end{figure} 

Inspired by the object detection in the image from computer vision, a patch-based CNN model is proposed in~\cite{su2018vocal} for vocal melody extraction from polyphonic music. A strategy similar to Region-based Convolutional Neural Networks (R-CNN) is used for vocal melody extraction. Specifically, a CNN is trained with the candidate patches extracted from the CFP representation of the music signal to determine if a patch contains a singing voice or not. Then the vocal melody is localized in time and frequency from the predicted vocal patch. The patch-based CNN consists of data representation, patch selection, and CNN modeling as shown in Fig.~\ref{fig:arch_su2018vocal}. Assuming that every peak in the frame of a CFP representation is candidate vocal melody, a simple strategy such as a $25\times25$ patch size having a peak at the center is selected as a patch. Since the patches with melody peaks will be less than non-melody patches, 10\% non-melody patches are selected to train the CNN model. The CNN model consists of two convolution layers followed by three fully connected layers for classifying if the input patch contains a singing voice or not. The predicted patches are arranged according to time and frequency based on the patch center. The following strategies are employed for obtaining melody from the rearranged patches: the frequency index for which the CFP is maximal from the patches having output probability greater than 0.5 and taking the frequency index with the highest output probability.

\begin{figure}[!tbp]
        \centering
		\resizebox{8cm}{5cm}{\includegraphics{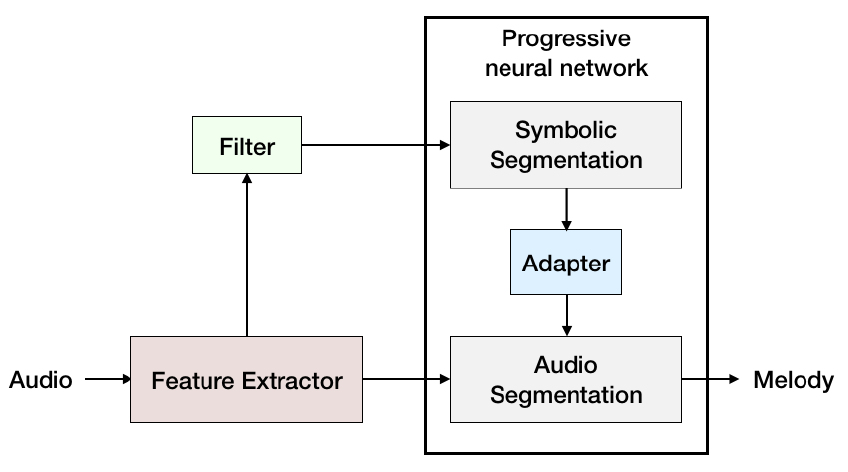}}  
        \caption{The semantic segmentation and audio-symbolic domain transfer learning framework for melody extraction~\cite{lu2018vocal}.}
        \label{fig:arch_lu2018vocal}
\end{figure} 

A dilated convolution-based semantic segmentation model and an adaptive progressive neural network are proposed to transfer the symbolic domain semantic segmentation network to the audio domain in~\cite{lu2018vocal}. As shown in Fig.~\ref{fig:arch_lu2018vocal}, the model consists of a CFP feature extractor. A progressive neural network consists of two semantic segmentation models. One model is trained on the symbolic data and the other on audio data, and a filter is used to match the dimensions of the audio representation with the symbolic segmentation model. The semantic segmentation model is the improvised version of DeepLab V3~\cite{chen2017rethinking, chen2018encoder}. DeepLab V3 is a fully convolution neural network with encoder-decoder architecture. The encoder is implemented with ResNet followed by atrous spatial pyramid pooling layers, and the decoder is implemented with the transpose convolution layers with skip connections. First, the segmentation model is trained on symbolic data. The model trained on symbolic data is connected to another segmentation model trained progressively with audio data by freezing the parameters of the symbolic segmentation model.

 \begin{figure}[!tbp]
        \centering
		\resizebox{8cm}{6cm}{\includegraphics{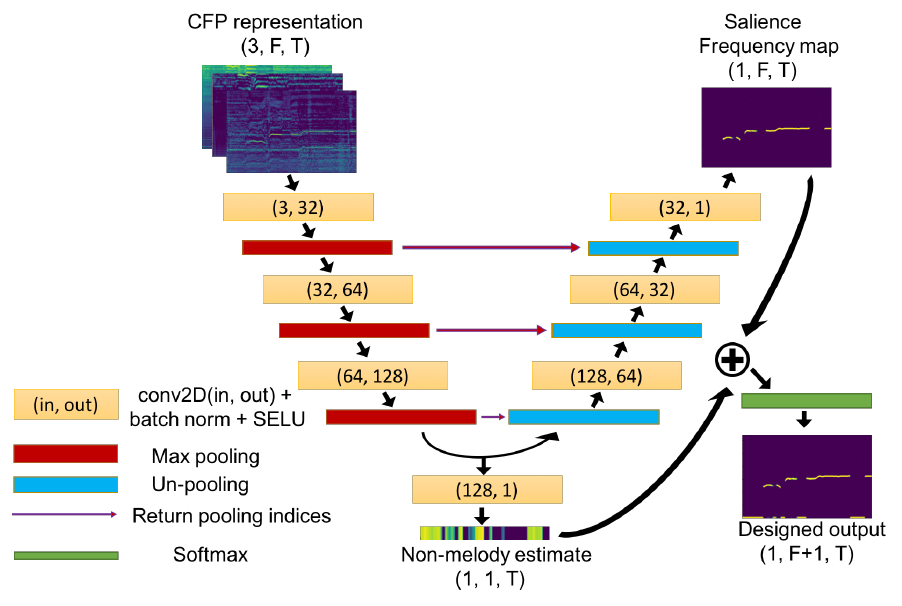}}  
        \caption{Streamlined encoder-decoder for vocal melody extraction~\cite{hsieh2019streamlined}.}
        \label{fig:arch_hsieh2019streamlined}
 \end{figure}

\begin{figure}[!tbp]
        \centering
		\resizebox{8cm}{5cm}{\includegraphics{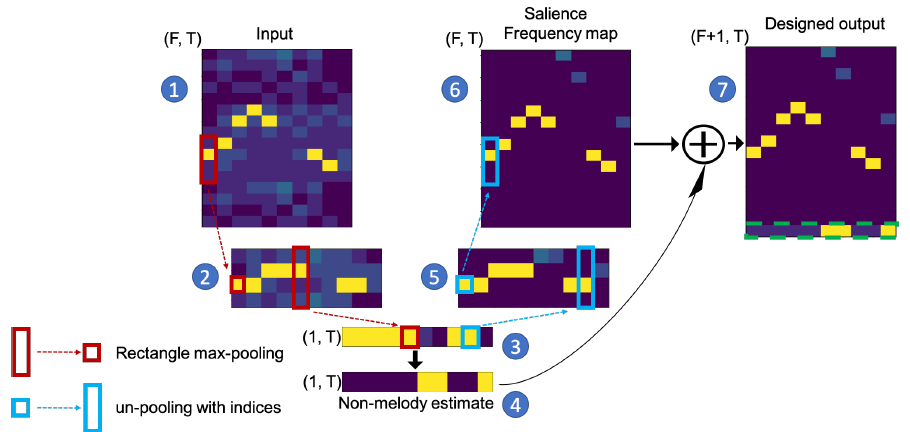}}  
        \caption{Streamlined encoder-decoder for vocal melody extraction model in action~\cite{hsieh2019streamlined}.}
        \label{fig:arch_model_in_action_hsieh2019streamlined}
 \end{figure} 

A streamlined encoder-decoder semantic segmentation melody extraction model inspired by computer vision semantic pixel-wise segmentation neural networks is proposed in~\cite{hsieh2019streamlined}. Also, the proposed model uses the bottleneck layer to estimate the presence of melody for each frame. Further, unlike predecessor methods, this method uses a simple argmax to obtain the binary melody matrix instead of using thresholding. The simple encoder-decoder melody extraction model is shown in Fig.~\ref{fig:arch_hsieh2019streamlined}. The encoder consists of three convolution and three max-pooling layers. The output of the encoder is branched to the decoder, which consists of three up-convolutional layers and three un-pooling layers to estimate the frequency salience map. Inspired by SegNet~\cite{badrinarayanan2017segnet}, the exact pooling indices of the encoder are passed to the decoder un-pooling layers to localize the melody in frequency. The output of the encoder is also fed to a convolution layer to estimate the existence of melody per frame, i.e., a non-melody estimator. The frequency salience map and the non-melody detector outputs are concatenated along the frequency axis. An argmax function is applied along the frequency to estimate the binary-valued melody line. The model in action through the visual representation of the feature maps is shown in Fig.~\ref{fig:arch_model_in_action_hsieh2019streamlined}. The output of the last encoding layer contains values close to 1 for the frames with melody notes (3). This feature map is bit reversed (4) and concatenated with the saliency feature map (6) so that an argmax can be applied along the frequency axis to obtain the melody line.

\begin{figure}[!tbp]
        \centering
		\resizebox{12cm}{8cm}{\includegraphics{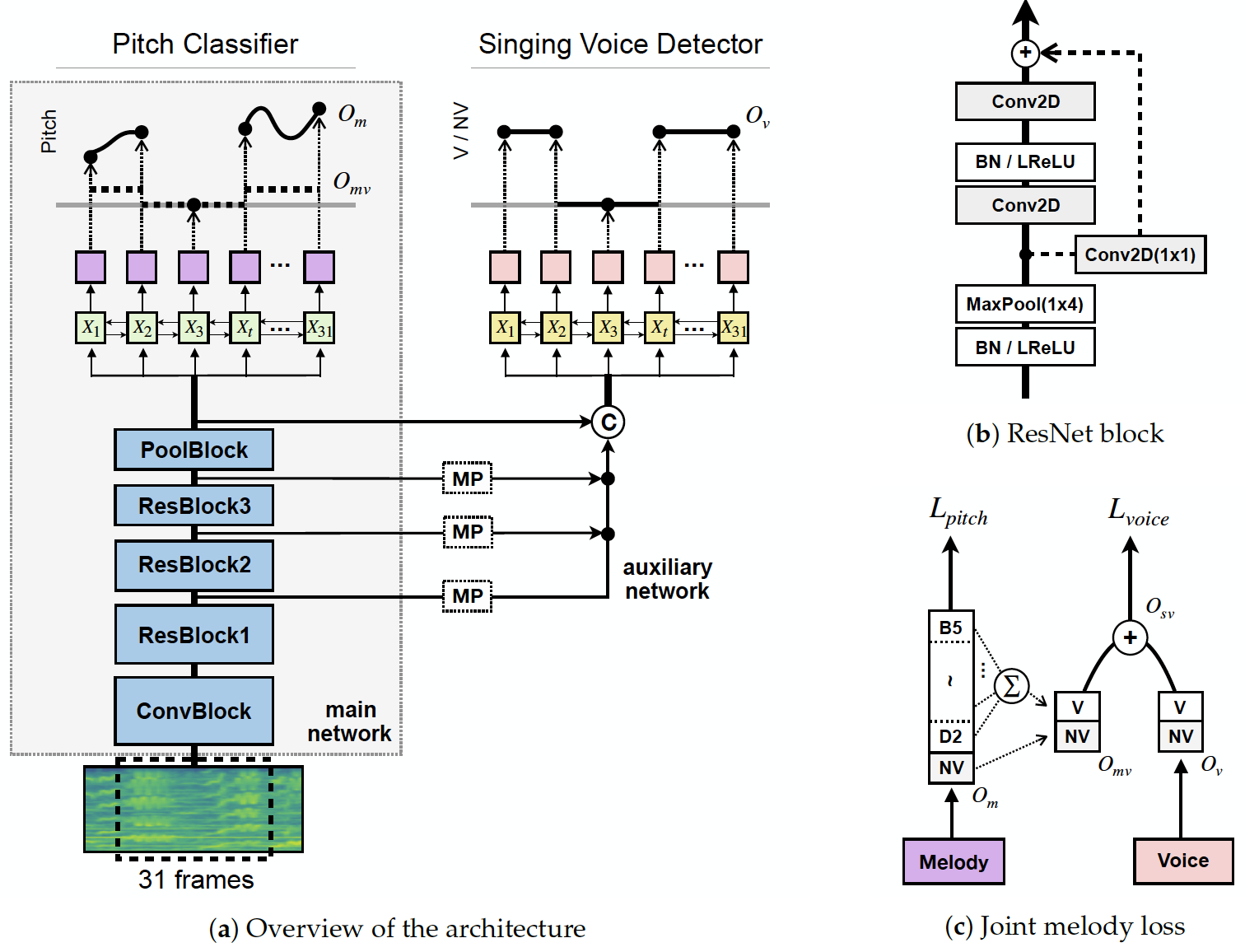}}  
        \caption{Illustration of the overall architecture, the ResNet block and the joint loss estimation of the joint detection and classification approach~\cite{kum2019joint}.}
        \label{fig:arch_kum2019joint}
\end{figure}

The singing melody extraction and singing detection are related tasks. Hence,~\cite{kum2019joint} proposes a multi-task joint detection and classification supervised approach for vocal melody extraction from polyphonic music. Specifically, they propose convolutional recurrent neural networks (CRNN) with residual connections and BiLSTM layers as the primary network that performs high-resolution melody classification and unvoiced frame detection and an auxiliary network that performs singing voice detection. The joint model is trained to minimize the melody and singing voice total loss function. The joint detection and classification model architecture is shown in Fig~\ref{fig:arch_kum2019joint}. The main network consists of initial convolution layers followed by residual blocks~\cite{he2016identity}. The output of the final residual block is connected to BiLSTM for melody and unvoiced classification. The auxiliary network uses the features from various layers of the main network for sequential detection of singing voice frames through the BiLSTM layer. The resnet block and the computation of joint melody loss are also shown in Fig.~\ref{fig:arch_kum2019joint}.

\begin{figure}[!tbp]
        \centering
		\resizebox{14cm}{5cm}{\includegraphics{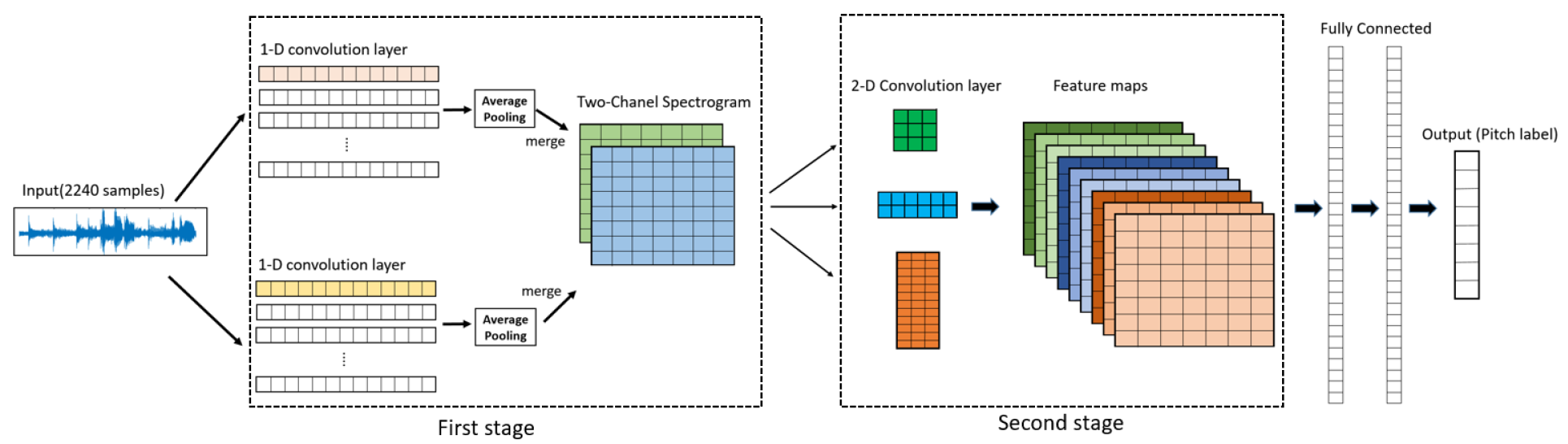}}  
        \caption{Architecture of the two stage multi-resolution auditory perception based model~\cite{chen2019cnn}.}
        \label{fig:arch_chen2019cnn}
\end{figure} 

Inspired by human hearing perception, a two-stage multi-resolution end-to-end neural network model is proposed for singing melody extraction in~\cite{chen2019cnn}. According to the two-stage auditory model~\cite{chi2005multiresolution}, the first stage performs spectrum estimation using a bank of constant-Q filters by mimicking the frequency selectivity of the cochlea. The second stage mimics the function of the auditory cortex, which analyzes the spectro-temporal envelope of the sound using a bank of 2D spectro-temporal modulation filters. Inspired by these perceptual properties, a two-stage end-to-end CNN model is proposed. In the first stage, 1D CNNs are used to mimic a spectrum estimator, and in the second stage, 2D CNNs are used to analyze the spectro-temporal analysis of the sound. The architecture of the two-stage melody extraction is shown in Fig.~\ref{fig:arch_chen2019cnn}. The first stage uses 1D CNNs to extract the spectrogram-like matrix instead of the conventional Fourier spectrogram from the input audio samples. Specifically, they use parallel 1D CNNs with kernel sizes of 960 and 64 with 16 sample strides with 200 kernels in each branch to estimate the spectrogram-like multi-resolution feature maps. Further, an average pooling is used to downsample the output of the 1D CNNs. The 1D kernels of the first stage are initialized from the pretrained 1D CNN melody extraction models trained with kernel sizes of 960 and 64. The sorted kernels with maximum magnitude frequency responses for 960 and 64 kernel sizes are used as pretrained weights for 1D kernels in the first stage for end-to-end training. The second stage consists of parallel three branches of 2D CNN kernels with different sizes with eight filters in each branch to filter the spectrogram-like feature maps of the first stage to extract the spectro-temporal melody features. The feature maps from the second stage are cascaded and fed to two fully connected neurons to predict the melody. 

\begin{figure}[!tbp]
        \centering
		\resizebox{10cm}{2cm}{\includegraphics{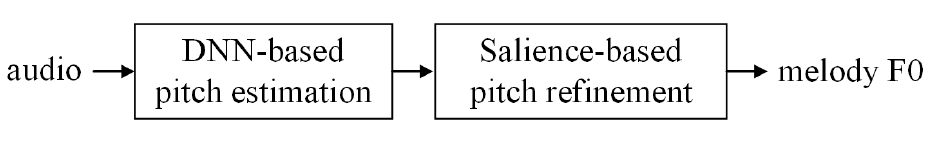}}  
        \caption{Framework of the DNN based pitch detection and the salience based pitch refinement~\cite{gao2019vocal}.}
        \label{fig:arch_gao2019vocal}
\end{figure} 

In this~\cite{gao2019vocal} paper, to address the limited availability of the labeled melody data, the authors proposed to use abundantly available MIDI melody files as a source of data to train the DNN model for melody extraction. Since the output of the DNN model trained on MIDI files will have pitch at low resolution at semitone level, a saliency-based method is proposed to refine the low-resolution pitch to high resolution at 10 cents. The proposed framework is shown in Fig.~\ref{fig:arch_gao2019vocal} it contains two steps: training the DNN model on melody MIDI files to extract the melody sequence at semitone resolution and later, refining the semitone level pitch to high 10 cents resolution with a saliency-based approach. The DNN model consists of an input layer, three fully connected hidden layers, and a classification layer. The input to the DNN model is the 41 frames contextual CQT matrix. The DNN model predicts the note number as output which is further smoothed by median filtering. The output of the DNN model is further refined by computing the saliency function with 600 bins as in~\cite{salamon2012melody} at 10 cents resolution. Specifically, the predicted notes are first converted into frequency and then to bins on the saliency function. Then, the bin corresponding to the maximum saliency value around the 2K+1 bins of the saliency function around the bin value calculated for each frame from the predicted pitch is flagged as refined melody pitch value, and then the refined bin value is converted back to frequency.

\begin{figure}[!tbp]
        \centering
		\resizebox{5cm}{10cm}{\includegraphics{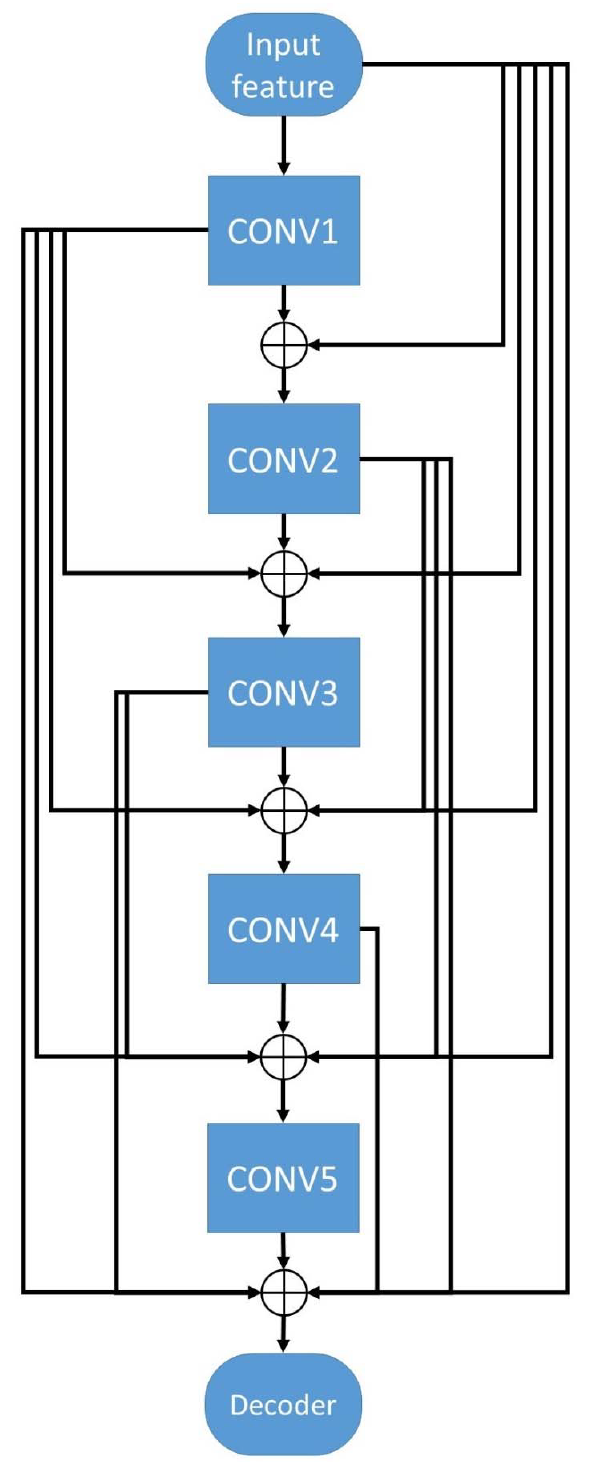}}  
        \caption{Illustration of the CNN encoder with the CNN layers connected densely with atrous spatial pyramid pooling.}
        \label{fig:arch_gao2019multi}
 \end{figure} 

 In~\cite{gao2019multi}, an encoder-decoder model for melody extraction inspired by the semantic segmentation, DenseNet, and atrous spatial pyramid pooling from computer vision to extract melody from polyphonic music is proposed. 
 The encoder consists of five dilated convolution layers which are connected densely with atrous spatial pyramid pooling as shown in Fig.~\ref{fig:arch_gao2019multi}. The atrous spatial pyramid pooling is used to increase the effective size of receptive fields with various dilation rates to learn the correlation between far-away spectral components. Further, the multi-dilation results in multi-scale features to improve the melody extraction performance. Also, the input of the encoder at each layer is sent to all successive layers for feature reuse inspired by~\cite{huang2017densely}. To reduce the feature map redundancy, each layer is designed to be “thin” to learn fewer feature maps. At each layer, they use SELU as an activation function. The decoder is similar to the one used in~\cite{gao2020multi} where a softmax function is used to obtain both voicing decision and melody class labels. Also, the authors claim that the proposed model uses significantly fewer parameters than CNN-based models compared in their paper.         
 
 \begin{figure}[!tbp]
        \centering
		\resizebox{14cm}{6cm}{\includegraphics{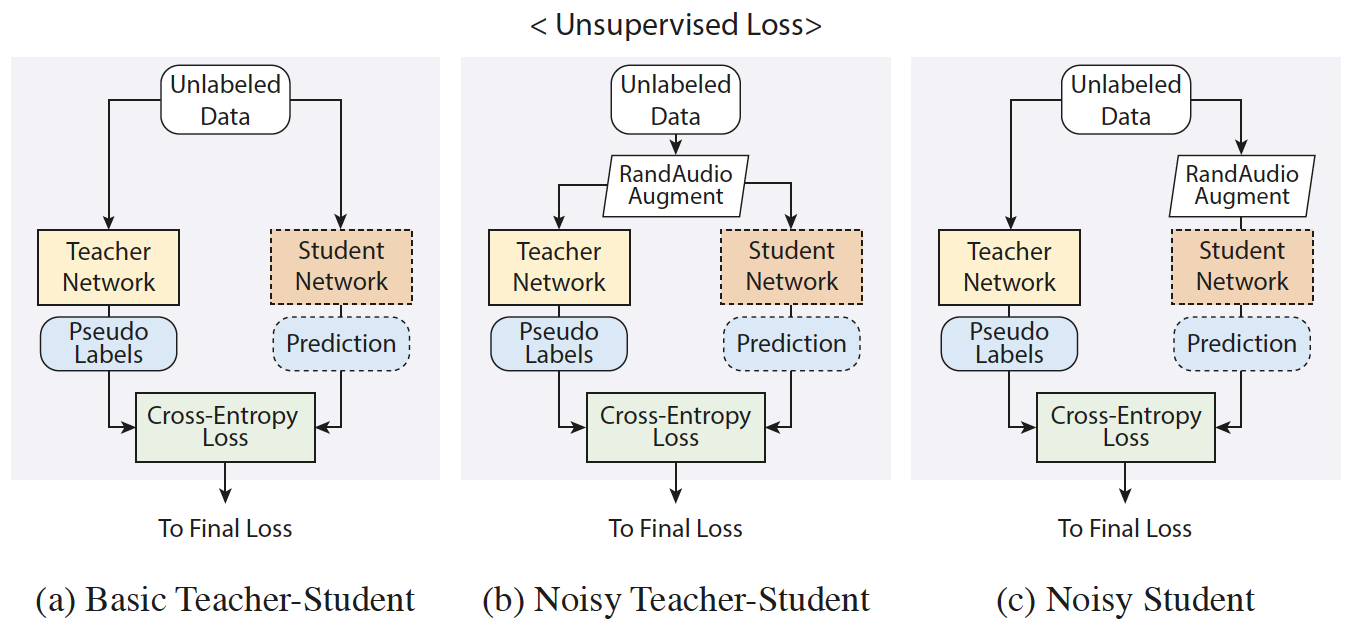}}  
        \caption{Block diagrams of the proposed teacher-student models for melody extraction~\cite{kum2020semi}.}
        \label{fig:arch_kum2020semi}
\end{figure} 

A semi-supervised learning (SSL) approach using a teacher-student model is proposed for vocal melody extraction to handle the limited amount of melody annotated data available for training deep models in~\cite{kum2020semi}. The teacher model is pre-trained with labeled data. The pre-trained teacher guides the student model to predict the labels on the unlabeled data in a self-training setting. Three different configurations of the teacher-student models are trained with different data augmentations and loss functions for melody extraction in an SSL setting. The proposed SSL model is trained using a self-training approach where the teacher network is trained with the labeled data. Then the student model is trained with the artificial labels generated from the teacher using unlabeled data. The CRNN~\cite{kum2019joint} model, which consists of four ResNet blocks and a BiLSTM layer, is used as a backbone architecture for training SSL models. They propose three configurations of teacher-student models as shown in Fig.~\ref{fig:arch_kum2020semi} (i) Basic teacher-student: It uses unlabeled data to train the model. The pseudo labels from the pre-trained teacher model are generated from unlabeled data to train the student model. (ii) Noisy teacher-student: This model utilizes noisy unlabeled data for teacher and student models. (iii) Noisy student: In this configuration, the teacher model generates the pseudo labels from the clean unlabeled data, but the student network takes the noisy unlabeled data for training. Data augmentation by pitch-shifting is performed on the labeled data. For unlabeled data, the data augmentation is performed by RandAudioAugment (RAA)~\cite{cubuk2019randaugment}.

\begin{figure}[!tbp]
        \centering
		\resizebox{14cm}{4cm}{\includegraphics{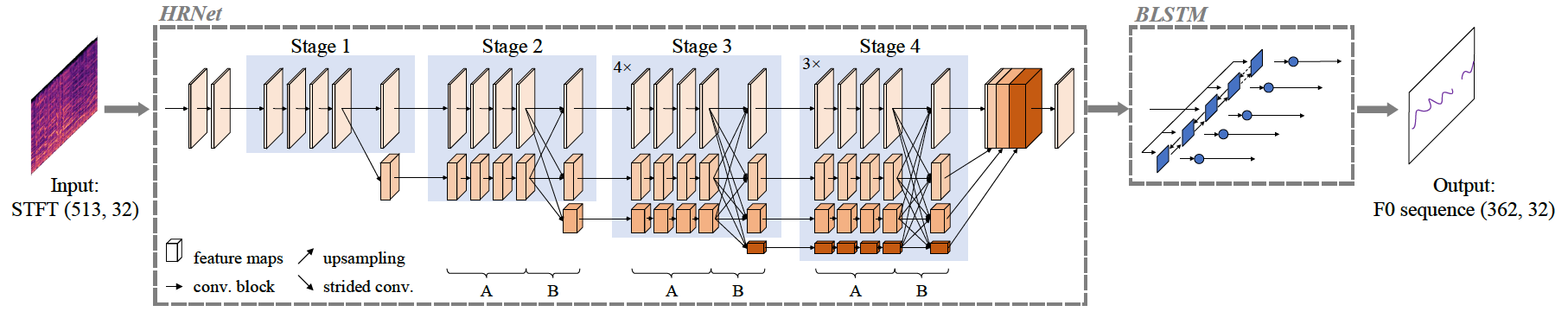}}  
        \caption{Illustration of the HRNet-BLSTM architecture melody extraction~\cite{gao2021hrnet}.}
        \label{fig:arch_gao2021hrnet}
\end{figure} 

Melody MIDI is used as a source of data to alleviate the scarcity of labeled melody data for training models similar to~\cite{lu2018vocal} and~\cite{gao2019vocal} in~\cite{gao2021hrnet}. Different from~\cite{lu2018vocal} and~\cite{gao2019vocal} where they directly use melody MIDI as the target to the model, the proposed model uses the pitch refinement method to refine the semitone scale pitch sequence decoded from MIDI to cent scale. The refined pitch sequence and the corresponding audio are used to train the deep learning model for melody extraction. Specifically, the authors claim three major contributions in this paper: (i) use of high-resolution feature maps~\cite{wang2020deep} for HRNet-BLSTM model for melody extraction, (ii) a new loss function to solve class imbalance problem, and (iii) a two-stage training strategy to train the model that is, first using MIDI refined data and then using manually labeled data. The high resolution HRNet-BLSTM melody extraction model is shown in Fig~\ref{fig:arch_gao2021hrnet}. It takes 32 frame size STFT as input to the model. The highlight of the HRNet model is the parallel multi resolution convolutional layers (shown as A in Fig.~\ref{fig:arch_gao2021hrnet}) and multi-resolution fusion layers to learn high-resolution feature maps~\cite{sun2019deep,wang2020deep}. The output of the HRNet is connected to the BLSTM layers with 512 hidden units to take care of the temporal dependency of the melody. Further, the output of the BLSTM layers is connected to the fully connected layer with softmax activation to obtain the per frame posterior distribution for each pitch class. The model is trained in two steps. In the first step, the refined pitch labels are obtained automatically by matching the semitone level MIDI note with the corresponding peak in the saliency function~\cite{salamon2012melody} of the audio to obtain the cent level pitch. The refined pitch thus obtained is used along with the corresponding audio to train the HRNet-BLSTM model in the first pass. In the second pass, a subset of manually labeled data from the MedlyDB is used to fine tune the model trained in the first stage.

\begin{figure}[!tbp]
        \centering
		\resizebox{8cm}{4cm}{\includegraphics{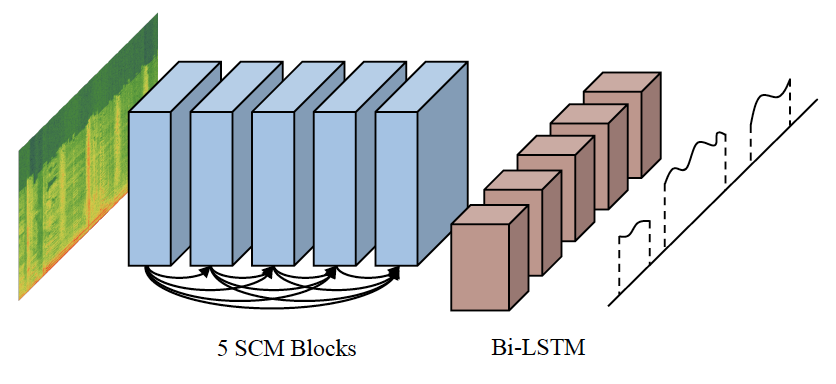}}  
        \caption{Spectrum correlation module~\cite{du2021singing}.}
        \label{fig:arch_scm_block_du2021singing}
 \end{figure} 

\begin{figure}[!tbp]
        \centering
		\resizebox{10cm}{4.5cm}{\includegraphics{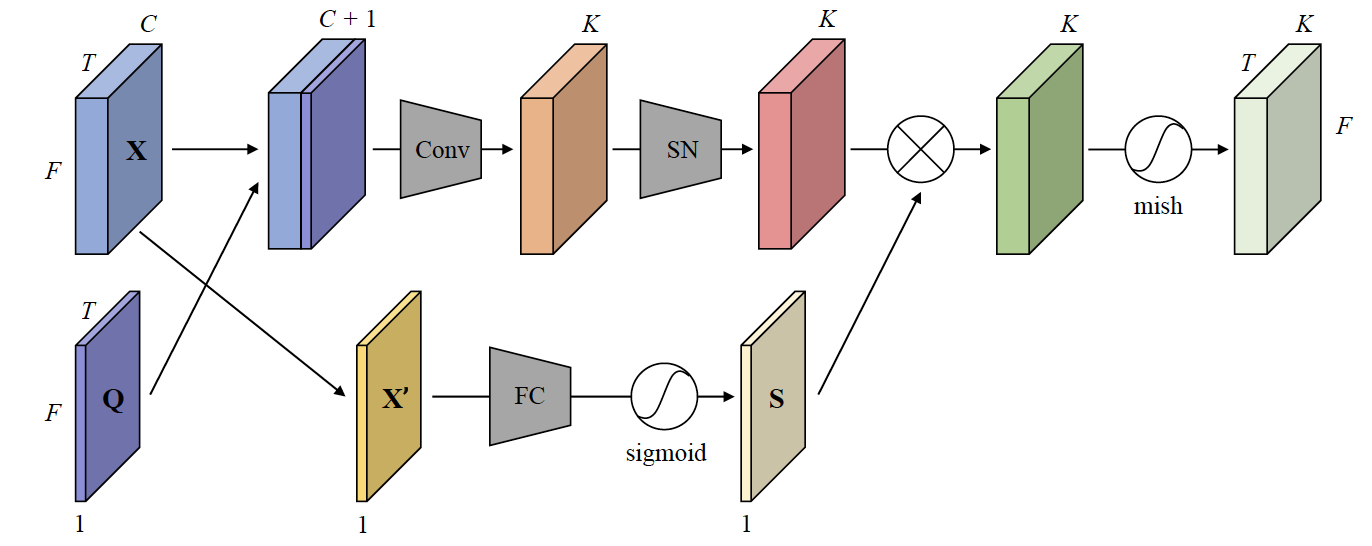}}  
        \caption{The architecture of the SCM block which consists of a convolutional branch and an attention branch~\cite{du2021singing}.}
        \label{fig:arch_scm_du2021singing}
 \end{figure} 

Singing voice has a wide frequency spectrum that covers thousands of hertz. Further, the frequency components are related to each other in the form of harmonics. In~\cite{du2021singing}, a crucial observation is made that is the existing CNN-based melody extraction methods use small kernels of size 5x5 or 5x1 to learn the filters. However, the receptive fields of such kernels are confined to small regions or local information within the kernel, neglecting the spectral correlation at different spectral components of the singing voice, especially far apart frequencies. That is, correlations at higher frequencies and other spectral components of the singing voice are difficult to learn with small kernel sizes. Hence, they propose the spectrum correlation (SCM) module to learn the correlation among all frequency bands to encode the global spectral information into the learned feature representation. The proposed model consists of 5 SCM blocks which are densely connected like DenseNet~\cite{huang2017densely}, a bi-directional long short term (Bi-LSTM) layer, and a fully connected layer as shown in Fig.~\ref{fig:arch_scm_block_du2021singing}. The architecture of the SCM block is shown in Fig.~\ref{fig:arch_scm_du2021singing}. It consists of two branches: a convolutional branch and an attention branch. The input feature map X is concatenated with center frequency encoded feature map Q. The concatenated feature maps are passed to the convolutional layer and normalized by the switchable normalization layer. On the other hand, the feature maps X are compressed to a single-channel feature map X’ by averaging along the channel axis. Further, the feature map X’ is passed through a fully connected layer followed by a sigmoid activation function to learn the correlation among different frequency bands. The output of the convolution and the attention branches are combined by the Hadamard product and passed to Mish~\cite{misra2019mish} activation function to get the final feature maps.

\begin{figure}[!tbp]
        \centering
		\resizebox{12cm}{5cm}{\includegraphics{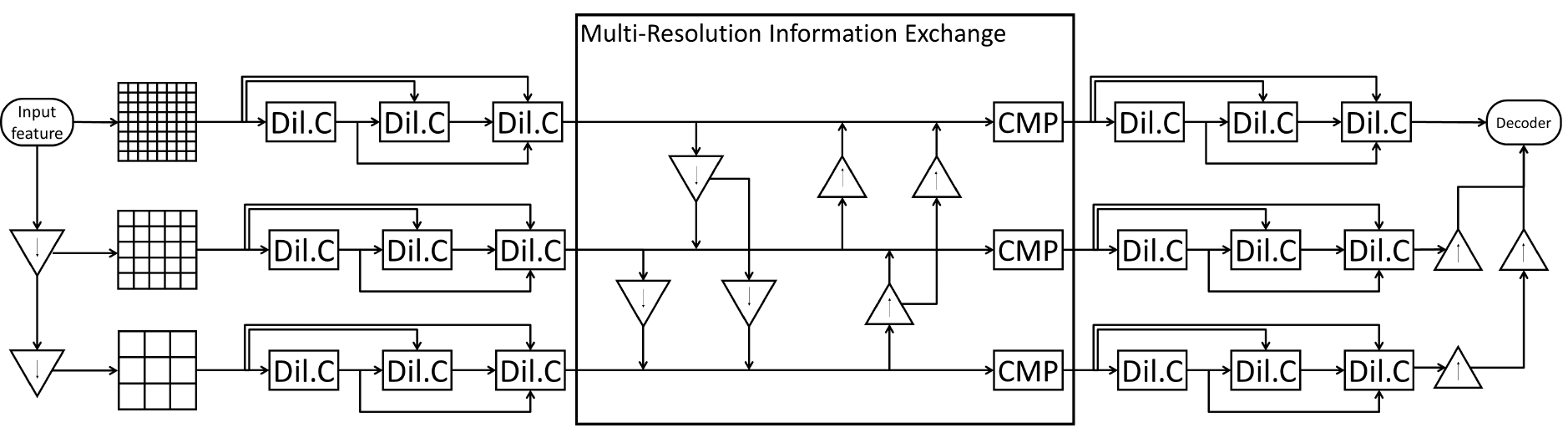}}  
        \caption{Block diagram of the encoder of the multi-dilation plus multi-resolution model. Inverted triangles with downward arrow indicates high resolution to low resolution transition while the upward arrows indicates the vice-versa. Dil.c represents the dilated convolutions and CFP: fusion of feature maps through concatination~\cite{gao2020multi}.}
        \label{fig:arch_enc_gao2020multi}
 \end{figure}

\begin{figure}[!tbp]
        \centering
		\resizebox{6cm}{10cm}{\includegraphics{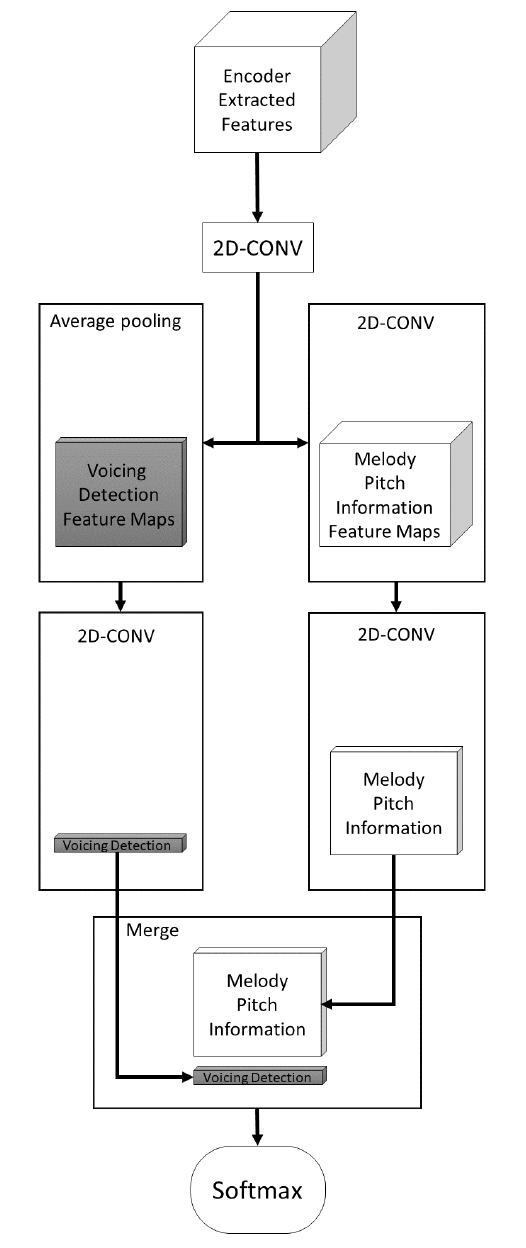}}  
        \caption{The two branch melody extractor and voicing detection decoder model~\cite{gao2020multi}.}
        \label{fig:arch_dec_gao2020multi}
 \end{figure}

In~\cite{gao2020multi}, an auditory model inspired spectro-temporal multi-resolution encoder and a semantic segmentation decoder to address the bottom-up process, and a top-down model of hearing of singing melody extraction is proposed. Specifically, the bottom-up process of multi-resolution decomposition of time-frequency representation of the audio by cortical function~\cite{chi2005multiresolution} is achieved by multi-dilation plus multi-resolution encoder as shown in Fig.~\ref{fig:arch_enc_gao2020multi}. The log-frequency magnitude spectrogram is down-sampled with the 3x3 convolution with 2x2 strides to achieve the low-resolution feature maps with input spectrogram. Further, the feature maps are passed to dilated convolutions. The multi-resolution feature maps are exchanged similar to the multi-resolution blocks of~\cite{sun2019deep, sun2019high}. That is, 2D CNN processes high-resolution feature maps with 2x2 stride making low-resolution feature maps.  While transpose convolutions process low-resolution feature maps to become high-resolution feature maps. These multi-resolution feature maps are concatenated. Further, the number of channels of features is increased by three times. A 2D convolution with $1\times1$ kernel is used to integrate the multi-resolution feature maps from the previous stage to restore the feature to the original input size. The decoder consists of parallel voicing detection and the melody extraction subnetworks as shown in Fig.~\ref{fig:arch_dec_gao2020multi}. The input original feature map and the output of all encoder feature maps are used as input to the decoder. The voicing subnetwork consists of an average pooling layer and a 2D Conv layer to integrate the cross-channel information to collapse into a single channel. On the other hand, the melody extraction subnetwork contains three 2D Conv layers to refine the feature maps and to reduce them to a single channel. The output of both the subnetworks are merged and fed to the softmax layer for final estimation.

\begin{figure}[!tbp]
        \centering
		\resizebox{10cm}{5cm}{\includegraphics{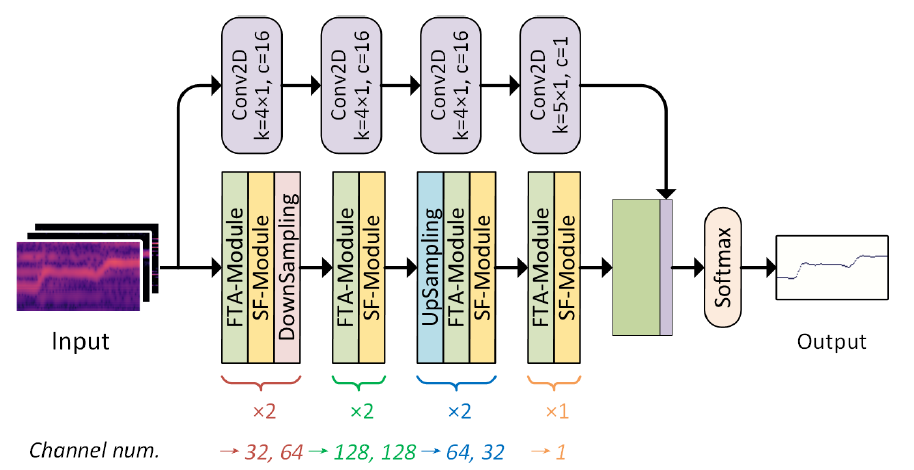}}  
        \caption{Overall melody extraction architecture. Top branch predicts the presence of melody. Bottom branch represents the frequency-temporal attention and selective fusion subnetwork~\cite{yu2021frequency}.}
        \label{fig:arch_overall_yu2021frequency}
\end{figure} 

\begin{figure}[!tbp]
        \centering
		\resizebox{8cm}{5cm}{\includegraphics{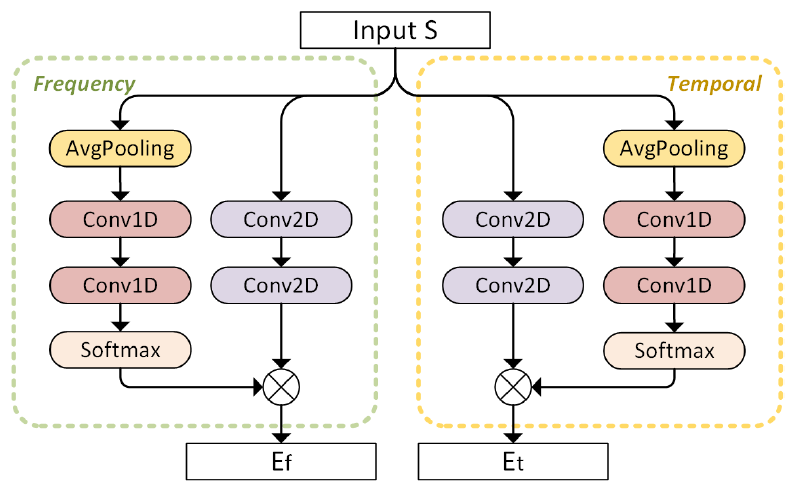}}  
        \caption{Detailed frequency-temporal attention network. $E_{f}$ and $E_{t}$ stands respectively for frequency and temporal attention~\cite{yu2021frequency}.}
        \label{fig:arch_f_t_att_yu2021frequency}
\end{figure} 

\begin{figure}[!tbp]
        \centering
		\resizebox{8cm}{5cm}{\includegraphics{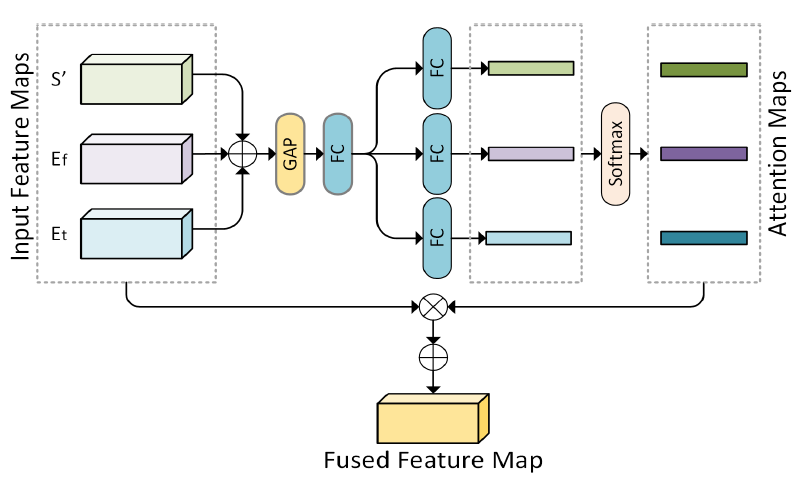}}  
        \caption{The architecture of the selective fusion network~\cite{yu2021frequency}.}
        \label{fig:arch_fusion_yu2021frequency}
\end{figure} 

A frequency-temporal attention method is proposed to mimic the auditory perception of sound in~\cite{yu2021frequency}. Specifically, the authors model the human auditory periphery mechanism of encoding frequency, time and magnitude with neural codes~\cite{yost2009pitch}. That is, different regions of the cochlea of the human ear are stimulated differently by different frequency bands. Then, the human picks the pitch from the temporal patterns generated by the unresolved harmonics in the auditory periphery~\cite{schouten1962pitch}. Inspired by selective kernel networks~\cite{li2019selective}, the authors propose a “selective fusion model to dynamically select the features from temporal and frequency attention networks” for singing melody extraction. The overall architecture of the frequency-temporal attention neural network is shown in Fig~\ref{fig:arch_overall_yu2021frequency}. It consists of a voicing detector, frequency-temporal attention (FTA), and selective fusion (SF) modules. 1-D convolutions are used to obtain the frequency and temporal attention maps instead of conventional attention layers~\cite{bahdanau2014neural, vaswani2017attention}. The process of obtaining frequency and temporal attention maps from the input through 1-D and 2-D convolution kernels is shown in Fig~\ref{fig:arch_f_t_att_yu2021frequency}. The detailed architecture of selective fusion is shown in Fig~\ref{fig:arch_fusion_yu2021frequency}. The SF module dynamically selects spectral and temporal features and fuses them later for melody extraction. Specifically, the input feature maps (temporal and frequency attention and input representations) are multiplied with the attention maps obtained from the same inputs to obtain the fused feature maps. 

\begin{figure}[!tbp]
        \centering
		\resizebox{10cm}{4cm}{\includegraphics{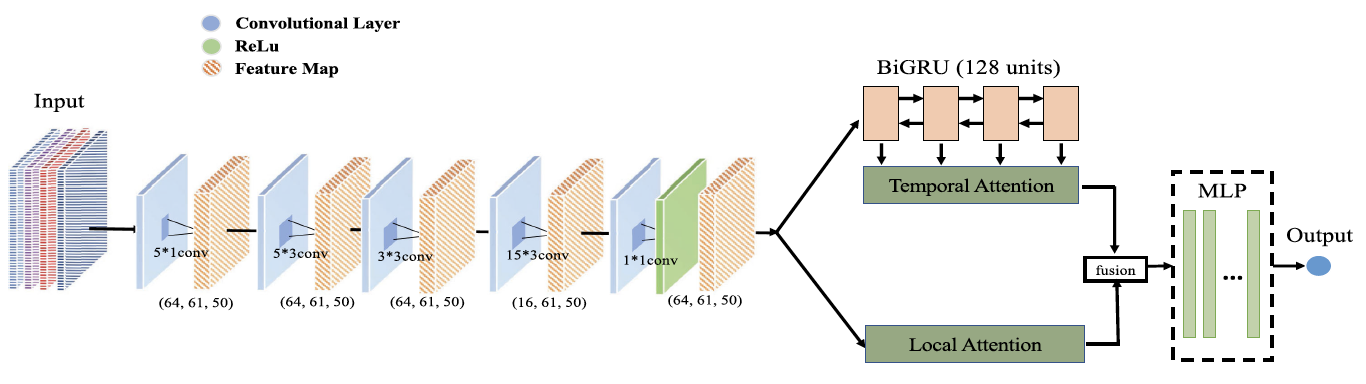}}  
        \caption{Overall architecture of the hierarchical attention network~\cite{yu2021hanme}.}
        \label{fig:arch_overall_yu2021hanme}
\end{figure} 

\begin{figure}[!tbp]
        \centering
		\resizebox{6.5cm}{4.5cm}{\includegraphics{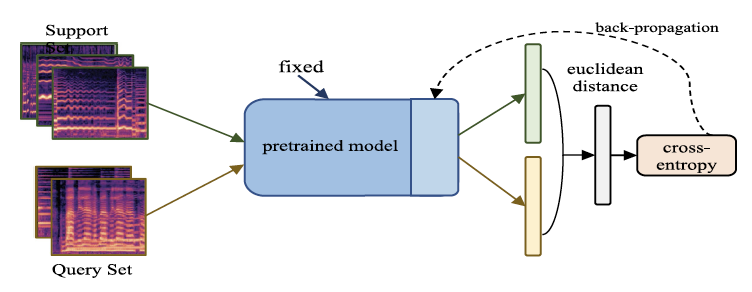}}  
        \caption{Illustration of partial parameter adaptation network~\cite{yu2021hanme}.}
        \label{fig:arch_part_parm_yu2021hanme}
\end{figure}

Observing the limitation of the existing methods that the models give equal importance to the contextual frames of the input representation, the authors of the paper~\cite{yu2021hanme} proposes Hierarchical Attention Network to learn the context-aware input features for singing Melody Extraction named as HANME. Specifically, they propose an attention layer that learns the context vector based on the spatial features extracted through residual CNNs. A second attention layer is proposed to learn temporal context vectors from the long-term features extracted by BiLSTM. Further, in order to alleviate the scare labels for each pitch class, the partial parameter adaptation~\cite{wang2020generalizing} method is proposed to fine-tune the melody classifier weights. The backbone of the HANME is the CRNN model. As shown in Fig.~\ref{fig:arch_overall_yu2021hanme}, it consists of CNN feature extraction layers and temporal and local attention subnetworks followed by multilayer perceptron for singing melody prediction and detection. The local attention network learns the importance of the high semantic features extracted by the CNN layer and encodes these features as a local context vector for the melody extraction. Similarly, the temporal attention contextual vector is computed with BiLSTM. For melody prediction, the local spatial and temporal context vectors are concatenated and fed to MPL. The partial parameter adaptation based on the few-shot learning~\cite{wang2020generalizing} to improve the performance of the underrepresented class by selectively updating the weights associated with these classes is shown in Fig.~\ref{fig:arch_part_parm_yu2021hanme}. Specifically, in this approach, the samples of rare classes are divided into support sets and query sets. Then, from the pre-trained model, the embeddings are obtained to compute the prototypical vector and Euclidean distance~\cite{snell2017prototypical}. Further, the classifier weights of the rare classes are updated with the cross-entropy loss function.   


\subsection{Separation based Melody Extraction Models} \label{ss:seapration}

\begin{figure}[!tbp]
        \centering
		\resizebox{10cm}{6cm}{\includegraphics{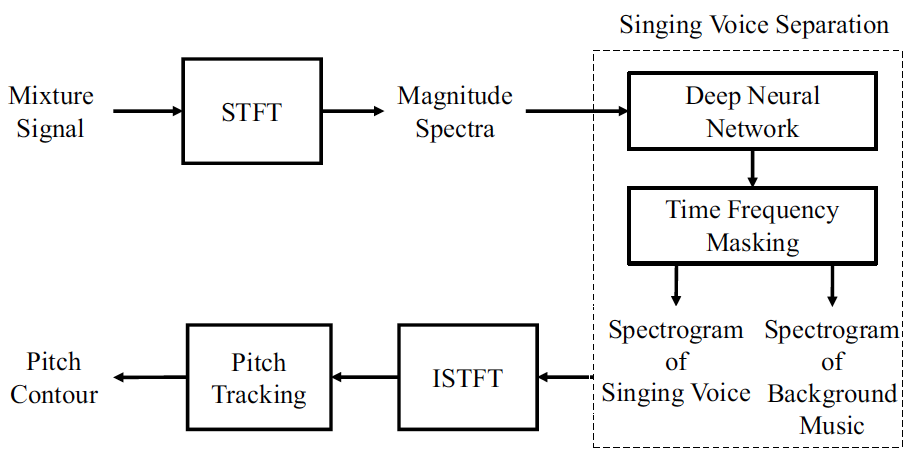}}  
        \caption{A two stage singing voice separation and dynamic programming based melody extraction from the separated vocals~\cite{fan2016singing}.}
        \label{fig:arch_fan2016singing}
\end{figure} 

A two-stage singing voice separation and melody extraction from the separated singing voice is proposed in~\cite{fan2016singing}. The framework of the two-stage melody extraction approach is shown in Fig.~\ref{fig:arch_fan2016singing}. As shown in the figure, the input magnitude spectrogram is passed through the fully connected layers to predict the vocal and instrument spectrograms. The predicted spectrograms are then used to obtain the soft masks for vocal and instrument spectrograms. The soft masks are multiplied with the input spectrograms to recover the magnitude and phase of the vocal and instrument components. The separated sources are then inverse Fourier transformed to obtain the time domain vocal and instrument audio signals. Finally, the pitch from the separated time-domain signal is obtained by dynamic programming~\cite{chen2008trues}.  

\begin{figure}[!tbp]
        \centering
		\resizebox{8cm}{3cm}{\includegraphics{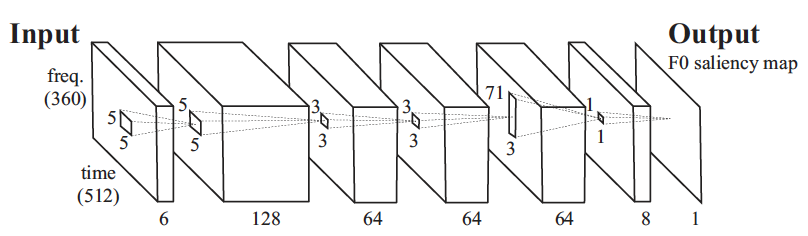}}  
        \caption{F0 saliency estimation network~\cite{nakano2019joint}.}
        \label{fig:arch_f0_saliency_nakano2019joint}
\end{figure} 

\begin{figure}[!tbp]
        \centering
		\resizebox{6cm}{6cm}{\includegraphics{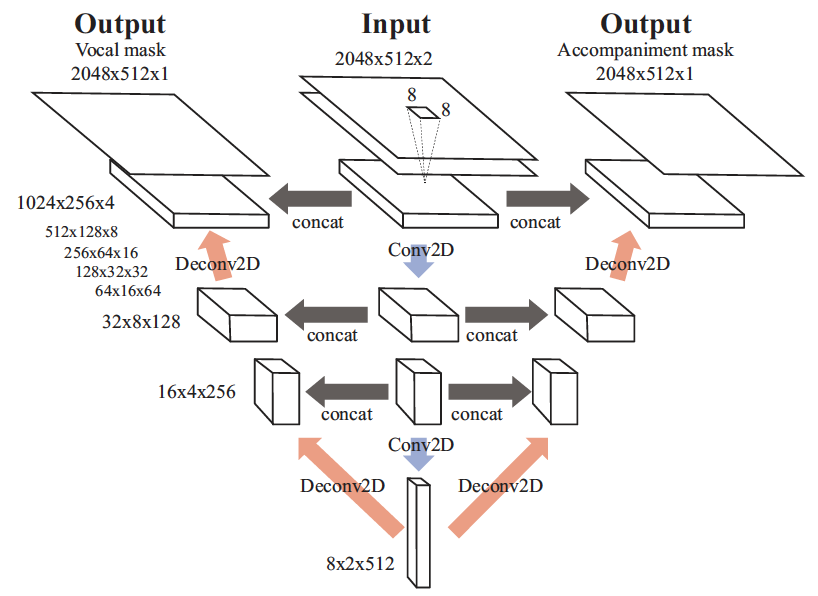}}  
        \caption{U-net based vocal and accompaniment separation network~\cite{nakano2019joint}.}
        \label{fig:arch_sepa_nakano2019joint}
\end{figure} 

A multi-task approach to joint singing voice separation and singing melody extraction is proposed in~\cite{nakano2019joint}. A unified CNN and U-net network is proposed for vocal F0 saliency extraction, and a shared encoder and two decoders for separating vocals and accompaniment are proposed. A differentiable layer that converts the F0 saliency into a harmonic mask is introduced between the two networks to integrate the information of harmonic structure in the networks. As shown in Fig.~\ref{fig:arch_f0_saliency_nakano2019joint}, the F0 saliency map extractor consists of five CNN layers based on~\cite{bittner2017deep}, that maps the input HCQT feature maps to the F0 saliency map. In contrast, the separation network takes magnitude spectrogram as input and outputs the vocal and accompaniment soft masks as output using harmonic structure information. The vocal and accompaniment separation network is shown in Fig.~\ref{fig:arch_sepa_nakano2019joint}. A multi-task U-net architecture takes the input as a stacked magnitude spectrogram and the harmonic spectrogram to predict the vocal and accompaniment masks. The harmonic spectrogram is obtained by multiplying the thresholded F0 saliency matrix and the weighted dictionary of harmonic spectra. The weighted harmonic dictionary spectra are obtained by placing the weighted sum of Gaussians placed on the integral multiples of F0 at each column. Both the models are trained end-to-end to minimize the sum of cross-entropy and L1 loss for F0 saliency prediction and vocal and accompaniment separation.    

\begin{figure}[!tbp]
        \centering
		\resizebox{12cm}{10cm}{\includegraphics{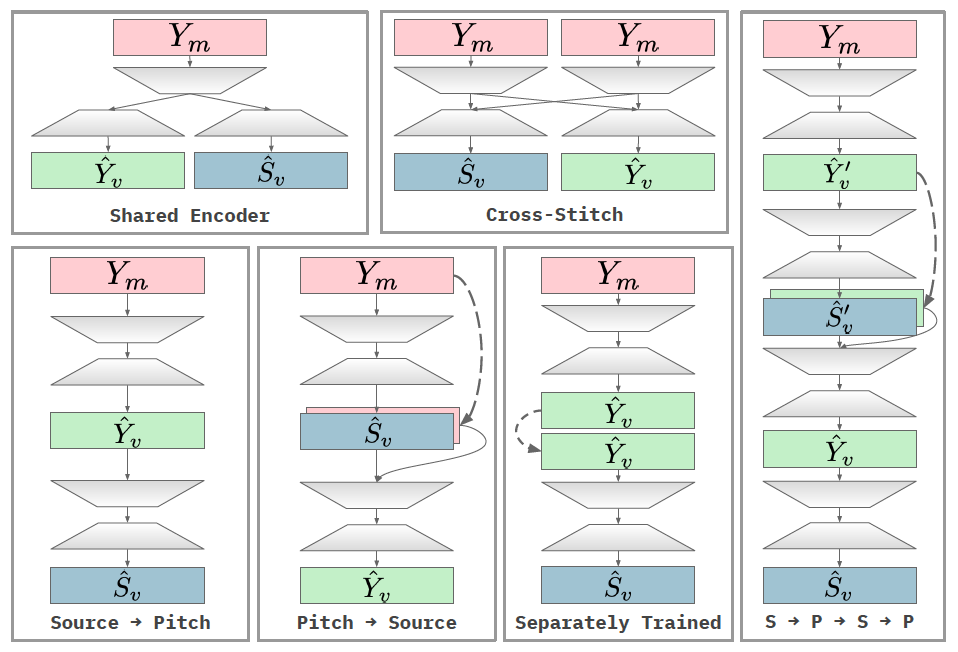}}  
        \caption{Various joint U-net models proposed in~\cite{jansson2019joint} for joint separation and vocal melody extraction.}
        \label{fig:arch_jansson2019joint}
\end{figure} 

The vocal separation and vocal F0 extraction are related tasks. Based on this fact, several experiments are proposed for joint separation and vocal melody extraction from polyphonic music in~\cite{jansson2019joint}. Further, they show that joint optimization is beneficial for these tasks, and the stacked network that performs vocal separation first is better than other models. The joint model architectures based on U-net architecture explored in the paper, which performs both separation, and F0 salience estimation, are shown in Fig.~\ref{fig:arch_jansson2019joint}. The explored architectures share the information through shared parameters in a standard multi-task setup, or the output of the one stage of the network is given as input to the other stage of the network as input: (i) conventional multi-task models which share weights (encoder) for both tasks (decoders). (ii) cross stitch model: each output consists of a separate encoder-decoder model, but the encoders are concatenated before passing the information to the individual decoders. (iii) stacked models: in this method, the tasks are learned in a cascaded manner, e.g., first, predict vocal spectrogram from the mixture and then predict the F0 saliency given the predicted vocals. (iii) stacked refinement~\cite{he2019deepotsu}: in this architecture, the output of a network is fed to the identical network with the different parameters. The vocal/non-vocal decision is taken based on thresholding the predicted saliency map with maximum likelihood at each frame.

\begin{figure}[!tbp]
        \centering
		\resizebox{12cm}{5cm}{\includegraphics{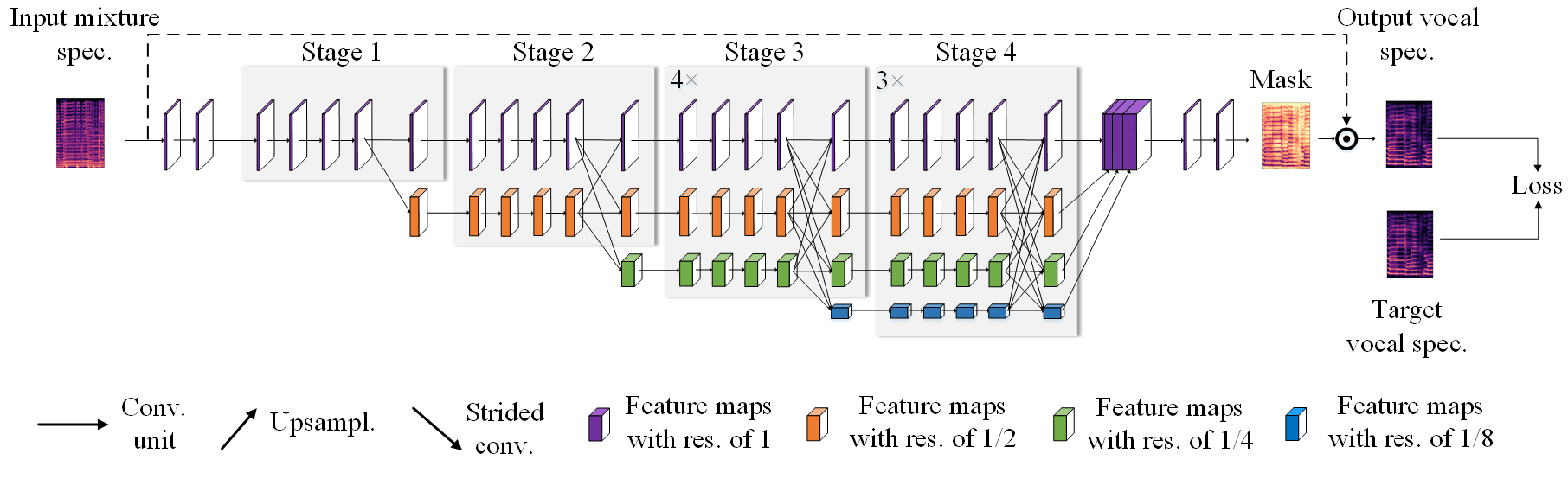}}  
        \caption{High resolution HRNeT for singing voice separation~\cite{gao2021vocal}.}
        \label{fig:arch_sepa_gao2021vocal}
\end{figure} 

\begin{figure}[!tbp]
        \centering
		\resizebox{12cm}{4cm}{\includegraphics{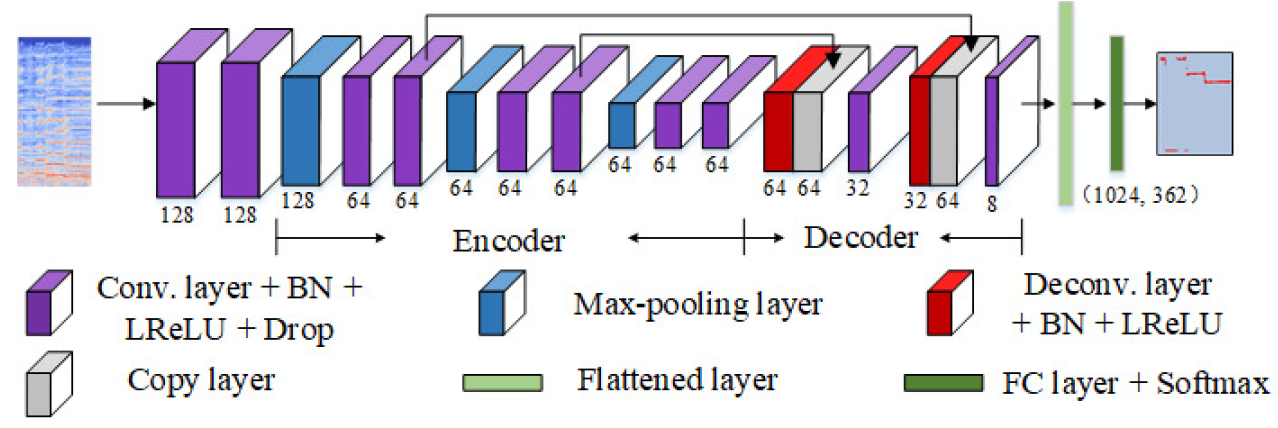}}  
        \caption{Encoder-decoder based melody extraction model~\cite{gao2021vocal}.}
        \label{fig:arch_ext_gao2021vocal}
\end{figure} 

An HRNeT based singing voice separation and a supervised encoder-decoder-based melody extraction model is proposed in~\cite{gao2021vocal} to extract the melody from the separated vocal source. The HRNet based singing voice separation (SVS) model is shown in Fig~\ref{fig:arch_sepa_gao2021vocal}. The HRNet consists of ResNet-50 based on multiple stages of multiple high-resolution paths to predict the soft vocal mask as shown in Fig~\ref{fig:arch_sepa_gao2021vocal}. The predicted vocal soft mask is multiplied with the input mixture spectrogram to obtain the vocal spectrogram. During inference, the predicted vocal spectrogram is combined with the phase of the input mixture spectrogram to obtain the vocal audio signal. The encoder-decoder network to map the input spectrogram to vocal melody F0 (VME) is shown in Fig~\ref{fig:arch_ext_gao2021vocal}. The input to the network is the magnitude spectrogram of the input audio signal. It consists of encoder convolution layers and the corresponding deconvolution layers in the decoder. For better feature reuse, the encoder-decoder layers are connected through skip-connections. The output of the last convolution layer at the decoder is flattened and connected to a fully connected layer to predict the per frame pitch class labels.

\section{Datasets} \label{sec:dataset}

We can find multiple openly available melody annotated datasets for training deep learning models. The openly available datasets consist of parallel data consisting of audio clips and the corresponding melody F0 ground truth labels for training supervised melody extraction models. The following openly available datasets, which consist of polyphonic music and the labeled melody F0 at per-frame level, are used to train and evaluate the melody extraction models. In some cases, the authors also use their own in-house data for training the models. In contrast, openly available data is used for evaluating the proposed models for comparison with other models. 

\noindent \textbf{ADC2004}: This dataset consists of 20 excerpts with an average length of each song is 20 s. The dataset consists of music in the genres of Pop, Opera, and Jazz. It includes real recordings, synthesized singing voices, and the songs generated from MIDI files. The total duration of the dataset is 369 s.

\noindent \textbf{MIREX05}: It consists of 25 excerpts with a total dataset duration of 686 s. The duration of each music clip spans between 10-40 s. The music in the dataset comes from Rock, Pop, Jazz, Classical Piano, and R\&B. It consists of both musics synthesized from MIDI and real recordings.

\noindent \textbf{iKala}: This dataset consists of 252 audio files with each audio clip in the duration of 30 s long. The audio clips are sampled from 206 iKala songs featuring professional musicians and singers. The audio clips include accompaniment and the singing voice recorded in separate channels.  

\noindent \textbf{MedlyDB}: It is a multi-track dataset, meaning that all instruments and vocals are available in isolation. The dataset consists of 196 multi-track songs, of which 108 multi-tracks have melody annotations. MedleyDB consists of songs with a variety of musical genres. It mainly consists of full-length songs, with the majority of the songs having a duration between 3-5 minutes. MedleyDB is considered the most challenging dataset for melody extraction for its variety of genres, full-length songs, and approximately three hours of duration of data.              

\noindent \textbf{RWC synth}: This dataset consists of 80 popular Japanese songs and 20 Western songs. These songs are recorded by 148 musicians sung by both male and female professional singers.   

\noindent \textbf{MIR\_1K}: This dataset consists of 1000 Chines Pop song clips with music accompaniment and the vocals recorded in separate channels. The duration of each audio clip ranges between 4-13 s long. The audio clips are extracted from 100 karaoke songs. The excepts consists of songs sung by both male and female amateur singers.  

Different datasets follow different timestamp resolutions for pitch annotation. For example, in MIR\_1K dataset, we can find a pitch label for every 10 ms. Whereas in MedleyDB we can find pitch labels at 5.8 ms time steps. Also, datasets come with various sampling rates. For example, MIR\_1K audio clips sampling rate is 16 kHz. Whereas iKala is sampled at 44.1 kHz. Some datasets contain actual pitch labels in Hz, e.g., ADC2004. In contrast, some datasets contain melody in MIDI pitches. Since the openly available datasets do not follow common protocols or guidelines, one must perform at least some pre-processing steps: sample rate conversion, MIDI to pitch conversion and vice versa, and frame rate conversion before using the data for training deep learning models for melody extraction. The multi-track datasets such as iKala, MIR\_1K and MedleyDB can also be used for training pre-processing models such as singing voice detection and vocal separation. Please note that not all datasets come with predominant melody instrument music clips, e.g., iKala and MIR\_1K contain only music with vocal melody annotations.              

\begin{figure}[!tbp]
        \centering
		\resizebox{6cm}{5cm}{\includegraphics{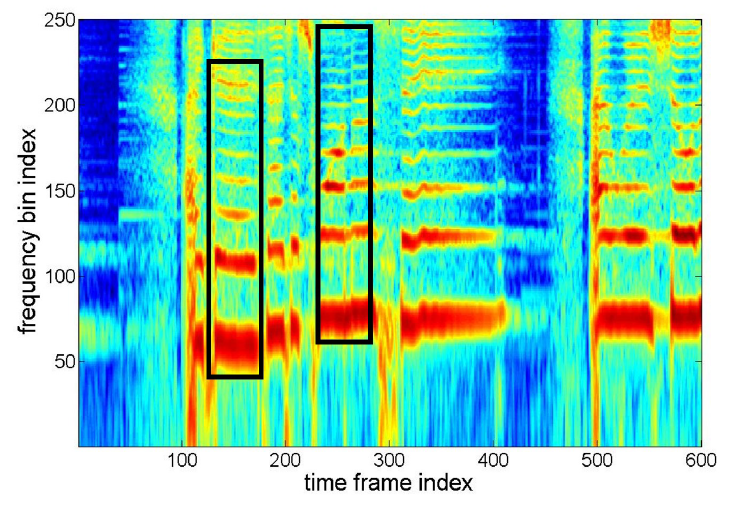}}  
        \caption{The log-frequency spectrogram which show invariant harmonic pattern to change in pitch~\cite{chou2018hybrid}.}
        \label{fig:log_freq_spect}
 \end{figure} 

\begin{figure}[!tbp]
        \centering
		\resizebox{8cm}{8cm}{\includegraphics{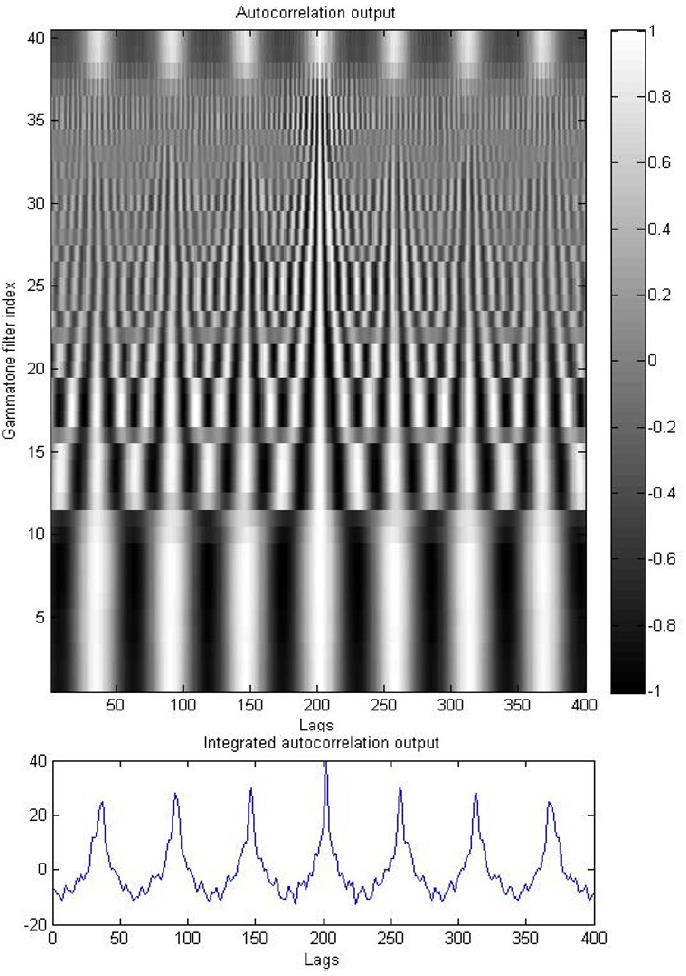}}  
        \caption{The gammatone filtered autocorrelation output (top figure) and the integrated autocorrelation (bottom figure)~\cite{chou2018hybrid}.}
        \label{fig:gammatone_integrated}
 \end{figure}

\begin{figure}[!tbp]
        \centering
		\resizebox{12cm}{6cm}{\includegraphics{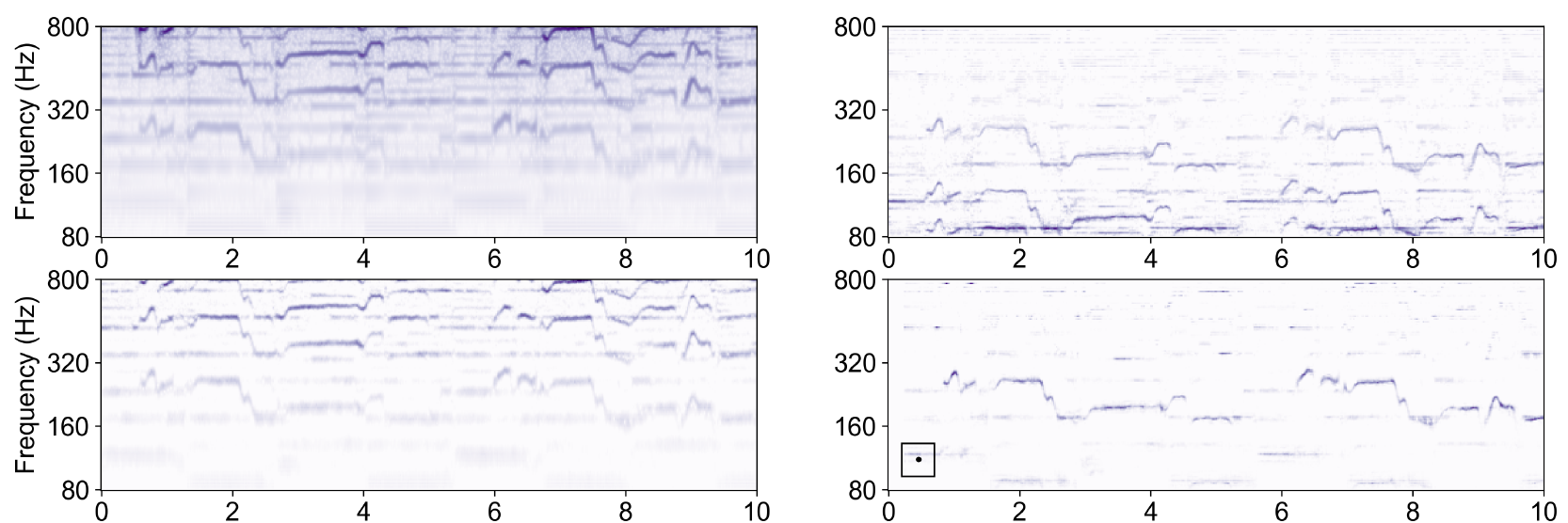}}  
        \caption{Illustration of the steps in computation of CFP representation~\cite{su2018vocal}. power-scaled spectrogram (top left), generalized cepstrum (GC) (top right: ), generalized cepstrum of spectrum (GCoS) (bottom left:), CFP representation (bottom right). As we can see in the CFP representation, the pitch content is more prominent compared to other components~\cite{lu2018vocal}.}
        \label{fig:cfp_rep}
 \end{figure} 

\section{Input representation} \label{sec:input_rep} 

The input representation to the melody extraction models is mostly the variations of the frequency spectrum that represents the harmonic structure of the melodic source. 

Magnitude spectrogram is the most commonly used representation for melody extraction computed by taking the magnitude of the short-time Fourier transform. In~\cite{du2021singing}, the magnitude spectrogram is used as input to the model. The log-frequency spectrogram, which is the pitch invariant harmonic pattern, is used as input to the spectral model, and temporal autocorrelation is used as input to the temporal model in~\cite{chou2018hybrid}. The temporal autocorrelation is computed by passing the audio signal first through 40 gammatone filters whose center frequencies are distributed on the ERB scale between 80 Hz and 6 kHz. Then, normalized autocorrelation is computed within each filter channel with a 48 ms window. Further, the information in individual channels is integrated and collapsed to a single channel to obtain the integrated autocorrelation signal. The integrated autocorrelation is fed as input to the temporal model. The log-frequency, gammatone filter autocorrelation and the integrated autocorrelation representation used by the model proposed in~\cite{chou2018hybrid} are shown in Fig.~\ref{fig:log_freq_spect} and Fig.~\ref{fig:gammatone_integrated}. The log-frequency spectrogram is also used in~\cite{gao2020multi} as input representation. The combined frequency and periodicity (CFP) representation obtained by simply multiplying the time-domain and frequency-domain representations that can suppress the harmonic and subharmonic peaks to yield a representation that localizes the pitch (F0) in both time and frequency is used as input in~\cite{gao2019multi, hsieh2019streamlined, yu2021frequency, yu2021hanme, su2018vocal, lu2018vocal}. The CFP representation is obtained by computing the power spectrum, generalized cepstrum (GC), generalized cepstrum of the spectrum (GCoS), and finally multiplying the GCoS by GC to obtain the CFP, which emphasizes the fundamental frequency of melodic source and instruments by suppressing harmonics and subharmonics. GC is a time-based representation. It exhibits strong subharmonics in the lower frequency range. Whereas GCoS is a modified spectrum in the frequency domain. The GCoS and GC are a compliment to each other. The GCoS represents the pitch by its $F0$ and the harmonics ($nF0$) whereas GC represents the pitch by its $F0$ and its sub-harmonics ($F0/n$). The multiplication of GC and GCoS effectively suppresses the harmonic and sub-harmonics and localizes the pitch. An illustration of the steps in CFP computation is shown in Fig.~\ref{fig:cfp_rep}. We can clearly observe that the CFP representation emphasizes pitch content compared to the other components in the spectrogram. Log compressed magnitude spectrogram is used in~\cite{kum2019joint}. Whereas log compressed magnitude spectrogram for melody extraction and Mel-spectrogram for singing voice detection is used in \cite{kum2017classification}. Unlike other approaches, raw audio samples are used as input to the model in \cite{chen2019cnn}. 

\begin{figure}[!tbp]
        \centering
		\resizebox{12cm}{5cm}{\includegraphics{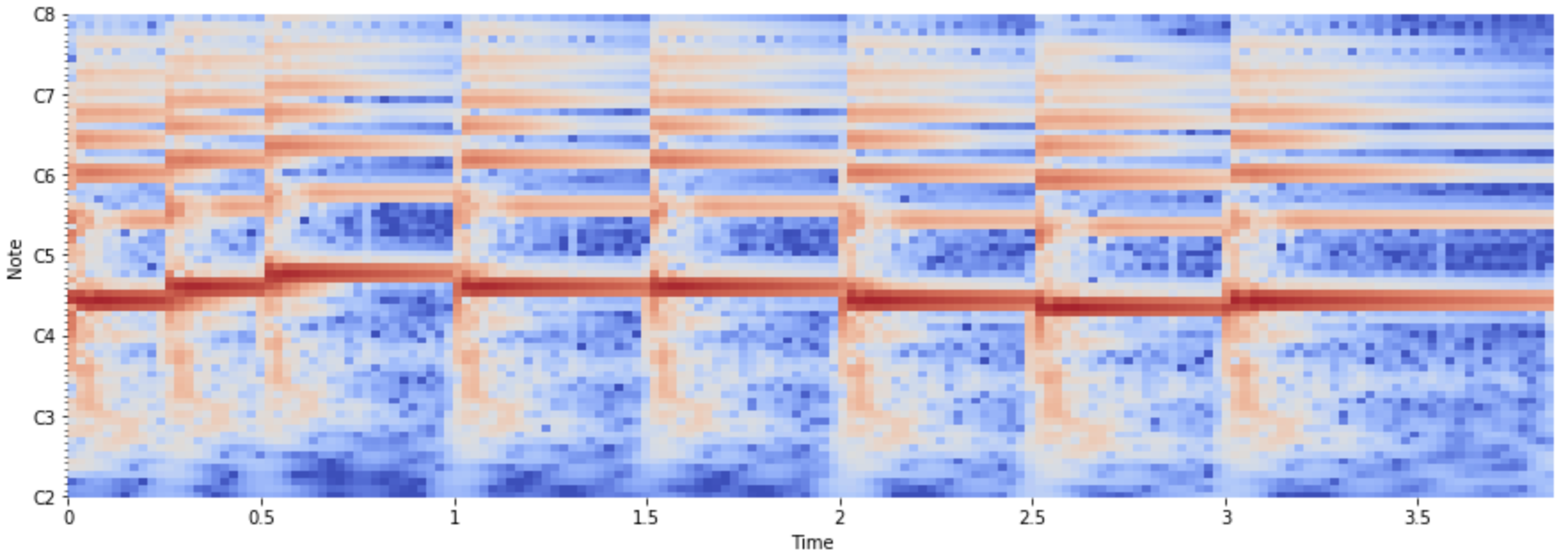}}  
        \caption{Illustration of the CQT representation of a piece of piano excerpt~\cite{mircourse}.}
        \label{fig:cqt_rep}
 \end{figure} 

Constant-Q transform (CQT)~\cite{schorkhuber2010constant} is a technique to transform the time-domain signal to the time-frequency domain such that the frequency bins of the center frequencies are geometrically spaced, and with the equal Q factors of the frequency bins are used as input representation in~\cite{gao2019vocal}. Here, the Q factors are the ratio of the center frequencies to the bandwidths. Similar to the mel scale, the CQT also uses the logarithmically spaced frequency bins. The primary advantage of the CQT is that it gives good low-frequency resolution and a better high-frequency time resolution. An illustration of the CQT representation of the piano music sample where each frequency bin corresponds to a MIDI pitch is shown in Fig.~\ref{fig:cqt_rep}.

\begin{figure}[!tbp]
        \centering
		\resizebox{10cm}{5cm}{\includegraphics{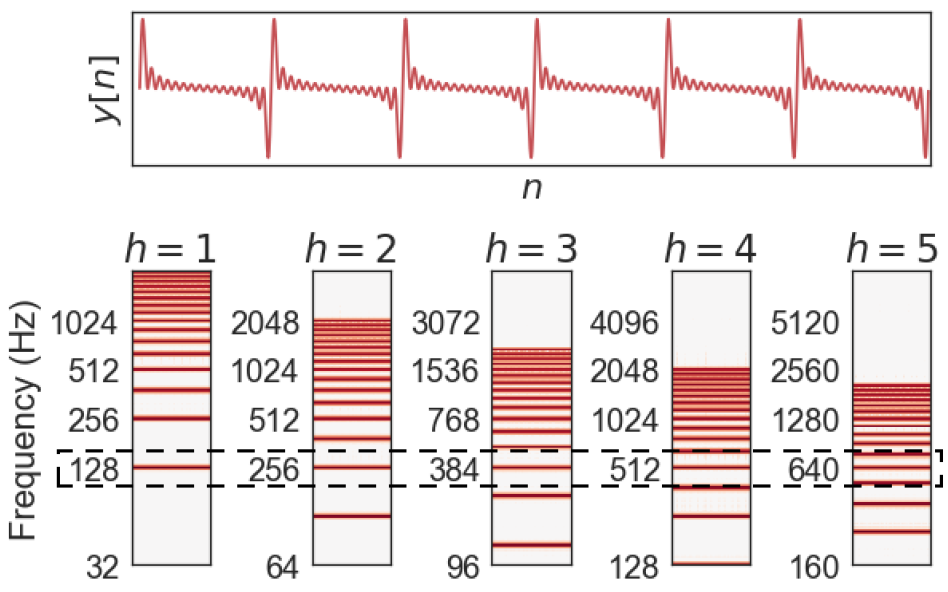}}  
        \caption{Illustration of ideal harmonic signal with F0 of 128 Hz and 16 harmonics (Top). The HCQT of ideal harmonic signal with harmonics $H[1]$ through $H[5]$ (Bottom). We can see the alignment of F0 (128 Hz) and its first four harmonics in the $h$ dimension \cite{bittner2018multitask}.}
        \label{fig:hcqt_rep}
 \end{figure} 

A novel Harmonic Constant-Q Transform (HCQT) is presented in \cite{bittner2017deep}. The HCQT is used as the input representation to capture the harmonic relationship in polyphonic music.  
The HCQT is a three dimensional feature map represented by $H[h, t, f]$. Where $h$, $t$, $f$ are the harmonic, time, and frequency respectively. The $H[h, t, f]$ measures the harmonic $h$ of frequency $f$ at time $t$. For any $h > 0$, the $H[h]$ is computed as the standard Constant-Q Transform (CQT) where the minimum frequency is scaled by the harmonic: $h \times f_{min}$. The HCQT computed for five harmonics for a perfectly harmonic signal with a fundamental frequency of 128 Hz and 15 harmonics above fundamental is shown in Fig.~\ref{fig:hcqt_rep}. We can see that the harmonically related frequencies are aligned in the harmonic or HCQT channel dimension. 

\begin{figure}[!tbp]
        \centering
		\resizebox{4cm}{4cm}{\includegraphics{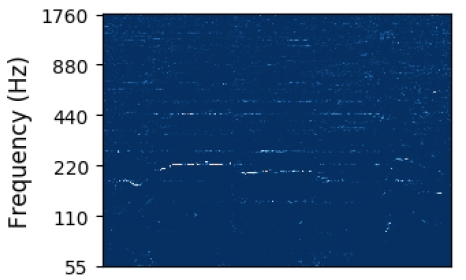}}  
        \caption{Illustration of the source-filter non-negative matrix factorization decomposed salience representation~\cite{basaran2018main} of the source of a music excerpt.}
        \label{fig:sf_nmf_rep}
 \end{figure}

Unlike other approaches, in \cite{basaran2018main}, musically motivated source-filter non-negative matrix factorization (NMF) \cite{durrieu2011musically, durrieu2010source} decomposition source representation is chosen as input for its ability to capture the pitch and timbre content of the lead instrument compared to classical time-frequency representations. Specifically, the F0 saliency representation of the music is obtained by source-filter NMF decomposition. Where the mixture signal is modelled as the source, filter and accompaniment parts. The source part, which consists of the melody's salience representation, is used as input to the model. The salience representation obtained by the source-filter NMF decomposition is shown in Fig.~\ref{fig:sf_nmf_rep}. We can see that the source-filter NMF decomposition gives a good initial F0 salience representation for the melody extraction model.

\begin{figure}[!tbp]
        \centering
		\resizebox{7cm}{6cm}{\includegraphics{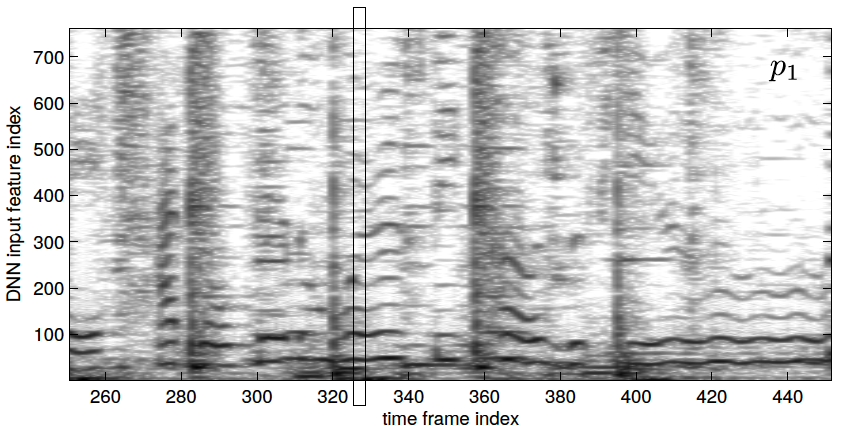}}  
        \caption{HPSS separated signal which contains melody and drum as main sources used as input to melody extraction network~\cite{rigaud2016singing}.}
        \label{fig:p1_hpss_rep}
 \end{figure}

\begin{figure}[!tbp]
        \centering
		\resizebox{7cm}{6cm}{\includegraphics{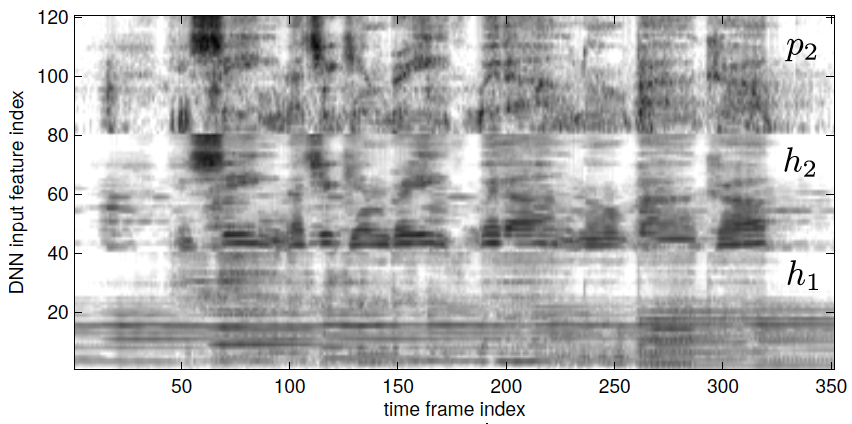}}  
        \caption{The concatenated $p2$, $h1$ and $h2$ feature representations used by vocal classification model~\cite{rigaud2016singing} for voice activity detection.}
        \label{fig:p2_h2_h1_hpss_rep}
 \end{figure}

The double stage harmonic/percussive source separated signals~\cite{fitzgerald2010single, tachibana2010melody} are used as input to the network in~\cite{rigaud2016singing}. In double stage HPSS, the input mixture is first decomposed into a sum of two signals $h1$ and $p1$ with a window size of 300 ms STFT. Where $p1$ will mainly contain drum and melody and $h1$ contains remaining stable instruments. Again, $p1$ is further decomposed into $p2$ and $h2$ with low resolution STFT, typically with 30 ms window size. Where $h2$ mainly contains melody and $p2$ contains drums. For voice activity detection network, the concatenated mel-spectrograms obtained from $h1$, $h2$, and $p2$ is used as input. Whereas the feature map $p1$ is used as input to the singing melody extraction network. The HPSS representations used by~\cite{rigaud2016singing} for voice activity detection and melody extraction are shown in Figs.~\ref{fig:p2_h2_h1_hpss_rep}, and ~\ref{fig:p1_hpss_rep}.

The most common parameters used in computing the input representations for the melody extraction models are shown in Table~\ref{tab:input_feature_parm}. We can observe that the parameters used to compute the input representation varies from one author to other. It would be interesting to see how the different neural network models and architectures perform with the same input parameter settings and the input feature dimensions. Also, it would be interesting to see how the same model performs for different input representations because each representation is designed based on a different domain/prior knowledge. For detailed information on the feature-specific parameters (e.g., number of bins used per octave in computing CQT, HCQT), the readers are requested to consult the corresponding manuscripts.           

\begin{table*}[h]
  \begin{center}
    \caption{The standard parameters used in computing the input representation. MS: Magnitude Spectrogram, LogFS: Log-Frequency Spectrogram, LogS: Log Spectrogram, CFP: Combined Frequency and Periodicity, LogMS: Log Magnitude Spectrogram, HCQT: Harmonic Constant-Q Transform, SF-NMF: Source-Filter Non-Negative Matrix Factorization, HPSS: Harmonic-Percussion Source Separation, IA: Integrated Autocorrelation, CQT: Constant-Q Transform}
    \label{tab:input_feature_parm}
    \resizebox{\textwidth}{!}{\begin{tabular}{l c c c c c c c }
      \toprule 
       {First Author} & \textbf{Feature} & \textbf{Sampling Rate (kHz)} & \textbf{Window Length (samples)} & \textbf{Hop Length/overlap (samples)} &\textbf{Input Feature Length (frames)} \\
      \midrule 
       Xingjian 2021~\cite{du2021singing} & MS & 44.1 & 1024  & 256 & 128 \\
       Hsin 2018~\cite{chou2018hybrid} & LogFS and IA & 16  & 48  & 24  & 1136 dim (IA) \\ 
       Ping	2020~\cite{gao2020multi} & LogS & NA & NA & NA & NA \\ 
       Ping	2019~\cite{gao2019multi} & CFP & 16 & NA & NA & NA \\
       Tsung-Han 2019~\cite{hsieh2019streamlined} & CFP & 44.1 & 2048 & 256 & 256 \\  
       Yongwei	2021~\cite{gao2021hrnet} & MS & 22.05  & 1024 & 256 & 32 \\
       Sangeun	2019~\cite{kum2019joint} & LogMS & 8 & 1024 & 80 & 31 \\
       Sangeun	2017~\cite{kum2017classification} & LogMS & 8 & 1024 & 80 & 11 \\
       Ming-Tso	2019~\cite{chen2019cnn} & Raw audio samples & 16 & 140 ms & 120 ms & NA \\ 
       Rachel	2017~\cite{bittner2017deep} & HCQT & NA & NA & 11 ms & NA \\
       Yongwei	2021~\cite{gao2021vocal} & MS & 16 & 1024 & 512 & 64 \\ 
       Shuai	2021~\cite{yu2021frequency} & CFP & 44.1 & 2048 & 256 & 128 \\
       Shuai Yu	2021~\cite{yu2021hanme} & CFP & 8 & 786 & 80 & NA \\
       Tomoyasu	2019~\cite{nakano2019joint} & HCQT, MS and Harmonic Spectrogram & NA & 4096 & 1024 & 512 \\
       Andreas	2019~\cite{jansson2019joint} & MS & 22.05 & 1024 & 256 & NA \\
       Dogac	2018~\cite{basaran2018main} & SF-NMF salience & NA & NA & NA & 25 or 50 \\ 
       Hyunsin	2017~\cite{park2017melody} & MS & 16 & 48 ms & 10 ms & NA\\
       Sangeun	2016~\cite{kum2016melody} & LogMS & 8 & 1024 & 80 & 11 \\
       Rachel	2018~\cite{bittner2018multitask} & HCQT & NA & NA & NA & NA \\
       Sangeun	2020~\cite{kum2020semi} & LogMS & 8 & 1024 & 80 & 31 \\ 
       Fran{\c{c}}ois	2016~\cite{rigaud2016singing} & HPSS & 16 & 300 ms, 30 ms & NA & NA \\
       Li	2018~\cite{su2018vocal} & CFP & 16  & 2048 & 320 & 25 $\times$ 25 patch \\
       Yongwei	2019~\cite{gao2019vocal} & CQT & 44.1 & NA & 23.2 ms & 41 \\
       Wei Tsung	2018~\cite{lu2018vocal} & CFP & 16 & 2048 & 320 & NA \\
       Zhe-Cheng	2016~\cite{fan2016singing} & MS & NA & 1024 & 512 & NA \\
       \bottomrule 
    \end{tabular}}
  \end{center}
\end{table*}

\section{Target Representation} \label{sec:output_rep}

Based on the target or output representation used for training the melody extraction models, we can categorize the melody extraction neural network models, into two categories. (i) The models that use quantized pitch labels as targets, (ii) The models that use spectrograms as targets.  \newline

\begin{figure}[!tbp]
\centering
\includegraphics[scale=0.5]{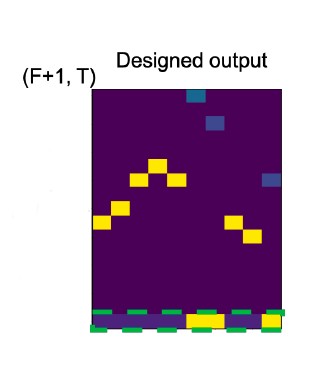}
\caption{Quantized pitch frequency target representation for melody extraction~\cite{hsieh2019streamlined}. T is the frame time, F is the quantized pitch frequency and, dotted green box at the bottom of the figure indicates the voicing labels.}
\label{fig:output_rep}
\end{figure}

\noindent \textbf{(i) Quantized pitch target:} In this type of target representation, the plausible vocal pitch range is quantized into finite number of pitch classes mostly on the cent scale with different resolutions. Further, an additional voicing label is also stacked on the quantized pitch labels in some models to treat melody extraction as a multi-class classification problem that simultaneously performs melody extraction and voicing detection. For example, in~\cite{hsieh2019streamlined}, the vocal pitch frequency range 31 Hz (B0) to 1250 Hz (D\#6) is divided into 320 frequency bins with 60 bins per octave, and for general melody extraction (vocals and lead melody instrument), the frequency range from 20 Hz (E0) to 2048 Hz (C7) is divided into 400 frequency bins with 60 bins per octave. Also, an additional voicing (melody) detection label is added on top of the pitch labels to determine if the frame contains melody. An example quantized pitch frequency target representation is shown for staked time frames in Fig.~\ref{fig:output_rep}. The figure shows that the horizontal axis represents the time frames, and the vertical axis represents the quantized pitch labels. At the bottom of the figure, the values in the dotted green box represent the voicing labels. Here, the voicing label 0 represents the presence of melody, and 1 represents the absence of melody. A simple argmax (index with maximum value) function is applied on the columns to obtain pitch labels and voicing detection during prediction.      

A slightly different approach is adopted in~\cite{bittner2017deep, kum2019joint} to create the pitch labels to minimize the error caused by the loss function by Gaussian blurring the one hot target labels. That is, the target representation is created by assigning a magnitude equal to 1 for the ground truth fundamental frequency quantized to the nearest time-frequency bin. Further, targets are Gaussian blurred in frequency axis such that the energy near the ground truth decays within the quarter tone. Gaussian smoothing is done to soften the penalty for near-correct predictions during training and take into account the human labeling error, which accounts for up to 20 cents.

The number of pitch classes and the pitch resolution considered for target representation varies across the melody extraction models. Please note that increasing the number of pitch levels also increases pitch resolution and results in label imbalance problems if there are not adequate frames (examples) to represent the particular label. In contrast, considering fewer labels may solve the class imbalance problem but results in low pitch resolution contours. Also, we have to note that not all authors stack the voicing decision labels on top of the pitch labels. Some consider it a standalone task and train it as a separate binary classification task. Later, the voicing detection is combined with the melody extraction to get the overall clean melody output. The number of pitch classes, pitch resolution, and other parameters adopted for creating the target pitch classes for training various models is shown in Table~\ref{tab:class_opr}. 

\noindent Note that mostly one-hot vector representation is adopted for target representation where the location of the pitch is assigned with value 1 and the rest of the values will be zeros in the one-hot array of the pitch. \newline   
 
\begin{figure*}[!tbp]
    \centering
    \subfloat[]{{\includegraphics[height= 5cm, width=4.6cm]{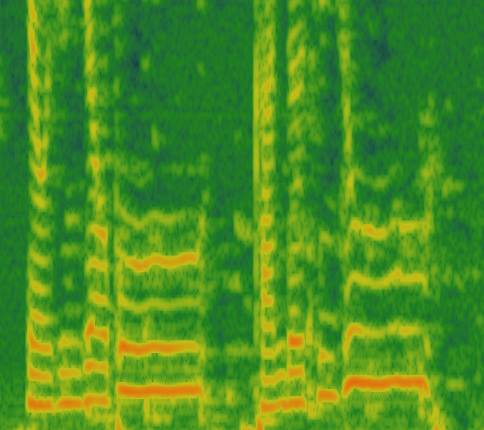}}}%
    \qquad
    \subfloat[]{{\includegraphics[height= 5cm, width=4.6cm]{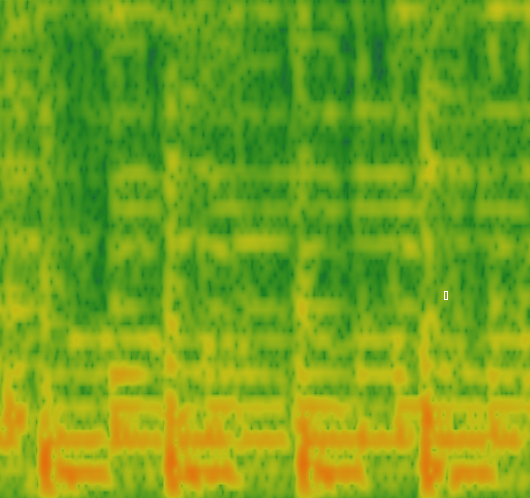}}}%
    \caption{Illustration of real valued magnitude vocal and accompaniment spectrogram used in source separation approaches for melody extraction.}%
    \label{fig:vocal_acc}%
\end{figure*}

\noindent \textbf{(ii) Spectrogram as target:} These are mostly the vocal source separation models that uses real valued vocal magnitude spectrogram as target. Few approaches also try to jointly separate vocal and instrument accompaniment from the input polyphonic mixture by utilizing both vocal and accompaniment spectrograms as targets. An example of vocal and accompaniment sample spectrograms used in source separation is shown in Fig.~\ref{fig:vocal_acc}. The segment size, frequency resolution, and various other parameters considered in creating the spectrogram targets are shown in Table~\ref{tab:sep_opr}.     

\begin{table*}[h]
  \begin{center}
    \caption{The frequency range, resolution, number of classes, and voicing decision labels used by the classification based model. VM and GM stands for vocal melody and general melody extraction. NA stands for either not mentioned or not available. Voicing column indicates is the voicing label included with the pitch labels during training.}
    \label{tab:class_opr}
    \resizebox{\textwidth}{!}{\begin{tabular}{l c c c c c c }
      \toprule 
       {First Author} & \textbf{Frequency range} & \textbf{Resolution in semitone} & \textbf{Pitch Classes} &\textbf{Voicing} \\
      \midrule 
       Xingjian 2021~\cite{du2021singing} & D2 (73.416 Hz) to B5 (987.77 Hz) & 1/2 & 360 & Yes \\
       Hsin 2018~\cite{chou2018hybrid} & D2 (73 Hz) to F\#5 (740 Hz) & 82 & 1/2 & Yes \\ 
       Ping	2020~\cite{gao2020multi} & D2 (73 Hz) to F\#5 (740 Hz) & 82 & 1/2 & Yes \\ 
       Ping	2019~\cite{gao2019multi} & NA & NA & NA & NA \\
       Tsung-Han 2019~\cite{hsieh2019streamlined} & B0 (31 Hz) to D\#6 (1250 Hz) for VM; E0 (20 Hz) to C7 (2048 Hz) for GM & 320 for VM; 400 GM & 60 bins per octave for both & Yes \\  
       Yongwei	2021~\cite{gao2021hrnet} & D2 (73.41 Hz) to B5 (987.77 Hz) & 1/8 & 361 & Yes \\
       Sangeun	2019~\cite{kum2019joint} & D2 (73.41 Hz) to B5 (987.77 Hz) & 1/16 & 721 & Yes \\
       Sangeun	2017~\cite{kum2017classification} & D2 (73.41 Hz) to B5 (987.77 Hz) & 1/32 & NA & Separate voicing classifier \\
       Ming-Tso	2019~\cite{chen2019cnn} & D2 (73 Hz) to F\#5 (740 Hz) & 1/2 & 82 & Yes \\ 
       Rachel	2017~\cite{bittner2017deep} & NA & NA & 360 & NA \\
       Shuai	2021~\cite{yu2021frequency} & 31 Hz to 1250 Hz & 60 bins per octave & 320 & Yes \\
       Shuai Yu	2021~\cite{yu2021hanme} & 31 Hz to 1250 Hz & 60 bins per octave & 320 & Yes \\
       Dogac	2018~\cite{basaran2018main} & A1 to A6 & semitone & 61 & NA \\ 
       Hyunsin	2017~\cite{park2017melody} & 55 Hz to 1760 Hz & NA & 600 & NA \\
       Sangeun	2016~\cite{kum2016melody} & D2 to F\#6, semitone, 1/2, 1/4 semitone & NA & NA & NA \\
       Rachel	2018~\cite{bittner2018multitask} & NA & 360 & NA & NA \\
       Sangeun	2020~\cite{kum2020semi} & 82.4 Hz to 1975.7 Hz & semitone & 441 & Yes \\ 
       Fran{\c{c}}ois	2016~\cite{rigaud2016singing} & 69.29 Hz to 1108.73 Hz & NA & 193 & Yes \\
       Li	2018~\cite{su2018vocal} & NA & NA & Binary & NA \\
       Yongwei	2019~\cite{gao2019vocal} & 55 Hz to 1.76 kHz & NA & 60 & Yes \\
       Wei Tsung	2018~\cite{lu2018vocal} & NA & NA & 128 and 384 bins for symbolic and audio & NA \\
       \bottomrule 
    \end{tabular}}
  \end{center}
\end{table*}

\begin{table*}[h]
  \begin{center}
    \caption{Output representations used by source separation based methods. The Separation column shows the output representations used for separating the vocal source. The Extraction column shows the frequency range, resolution, number of classes used in the model to extract melody from the separated vocal source. Voicing column indicates whether the voicing label is included along with the pitch labels.}
    \label{tab:sep_opr}
    \resizebox{\textwidth}{!}{\begin{tabular}{l c c c c c c }
      \toprule 
       {First Author} & \textbf{Separation} & \textbf{Extraction} &\textbf{Voicing} \\
      \midrule 
       Yongwei	2021~\cite{gao2021vocal} & Vocal magnitude spectrogram & 73.41 Hz to 987.77 Hz, 1/8 semitone, 362 classes & Yes \\ 
       Tomoyasu	2019~\cite{nakano2019joint} & Vocal and Accompaniment magnitude spectrogram & 360 & NA \\  
       Andreas	2019~\cite{jansson2019joint} & Vocal magnitude spectrogram & 360 & NA & \\
       Zhe-Cheng	2016~\cite{fan2016singing} & Vocal magnitude spectrogram & NA & NA \\
       \bottomrule 
    \end{tabular}}
  \end{center}
\end{table*}

\section{Loss Functions} \label{sec:loss_fun}

The loss function, cost function, or error function is the neural network objective function that we want to minimize to get the optimal candidate (neural network parameters) solution for the given task. We can broadly categorize the loss functions used to minimize the model error for melody extraction into four categories such as (i) General loss functions, (ii) Joint loss functions, (iii) Melody specific loss functions, and (iv) Class balancing loss functions. \newline

\noindent \textbf{(i) General loss functions:} These are the commonly used loss functions to train the deep learning models irrespective of the task (image classification or speech classification), and type of data (speech or image or text). The common loss functions are the binary cross-entropy (BCE), categorical cross-entropy (CE), L1 (Mean absolute error (MAE)), and L2 (Mean squared error (MSE)). The binary and categorical cross-entropy loss functions are used in melody extraction neural network models that consider melody extraction as a classification task. In comparison, L1 and L2 loss functions are used by melody (primarily vocals) source separation models to predict the melody source spectrogram from the input polyphonic spectrogram. 

\begin{equation}
BCE = - \frac{1}{N} \sum_{i=1}^{N} y_i \cdot \log(\hat{y_i}) + (1 - y_i) \cdot \log(1 - \hat{y_i})    \end{equation}

\begin{equation}
CE = - \frac{1}{N} \sum_{i=1}^{N} \log \hat{y_i} \left[  y_i \in C_{y_i} \right]
\end{equation}

\begin{equation}
L1 = \frac{1}{N} \sum_{i=1}^{N} \left| y_i - \hat{y_i} \right|
\end{equation}

\begin{equation}
L2 = \frac{1}{N} \sum_{i=1}^{N} \left( y_i - \hat{y_i} \right)^2
\end{equation}

\noindent where $y_i$ is the $i_{th}$ ground truth label, $\hat{y_i}$ is the model prediction, $C_{y_i}$ is the ground truth label belonging to class $C$, $N$ is the number of examples. \newline

In~\cite{chou2018hybrid, kum2017classification, chen2019cnn, bittner2017deep, gao2021vocal, yu2021frequency, yu2021hanme, basaran2018main, kum2020semi, rigaud2016singing, su2018vocal, gao2019vocal}, CE loss is used for error minimization. BCE is used in~\cite{gao2020multi, gao2019multi, hsieh2019streamlined, kum2016melody}. L1 loss between the predicted and the target vocal spectrogram is used in~\cite{gao2021vocal}. \newline

\noindent \textbf{(ii) Joint loss functions:} These are the loss functions which combines primary and auxiliary loss functions to form the total loss. For example, the primary melody extraction loss function (categorical cross-entropy) is clubbed with the voicing detection loss function (binary cross-entropy) to minimize the total loss in joint models, which simultaneously optimize the model for melody extraction and melody (voicing) detection. In~\cite{du2021singing}, the model is trained to jointly minimize the melody pitch loss and voicing detection loss as 
\begin{equation}
L_{total} = CE(y_{pitch}, \hat{y}_{pitch}) + BCE(y_{voice}, \hat{y}_{voice})
\end{equation}

\noindent A slightly modified joint loss function is proposed in ~\cite{kum2019joint} such as
\begin{equation}
L_{total} = L_{pitch} + \alpha L_{voice}
\end{equation}

\noindent where $\alpha$ is the weighting factor, $L_{pitch}$ is the CE loss of melody pitch classification model, $L_{voice}$ is the loss of auxiliary voicing detection model. The $L_{voice}$ is computed as 
\begin{equation}
L_{voice} = CE(softmax(O_{sv}), y_{voice})
\end{equation}

\noindent where 
\begin{equation}
    O_{sv} = O_{mv} + O_v
\end{equation}

\noindent where $O_{mv}$ is obtained by summing all the pitch classes of the main network and $O_v$ is the output of the auxiliary network. 

\noindent A joint loss function which included both vocal and accompaniment separation along with main melody pitch loss function is proposed in~\cite{nakano2019joint} as 
\begin{equation}
    L_{total} = L_{pitch} + L{sep}
\end{equation}

\noindent where
\begin{equation}
    L_{sep} = |Y_{vocal} - \hat{Y}_{vocal}| + |Y_{acc} - \hat{Y}_{acc}|
\end{equation}

\noindent where $Y_{vocal}$ and $\hat{Y}_{vocal}$ are the target and predicted vocal spectrograms. $Y_{acc}$ and $\hat{Y}_{acc}$ are the target and the predicted accompaniment spectrograms. In~\cite{jansson2019joint}, the vocal separation network is optimized with an L1 loss between the predicted and target vocal magnitude spectrogram, and the $f0$ salience network is optimized with L2 loss between the predicted and target saliency maps. In~\cite{bittner2018multitask}, the sum of the CE loss functions are proposed for multitask melody, multi-f0, bass, and vocal extraction. Whereas in~\cite{fan2016singing}, the sum of L2 loss between the predicted vocals and target vocals and the predicted instruments and target instrument spectrograms is proposed for separating vocals and instruments from input polyphonic spectrogram.  \newline

\noindent \textbf{(iii) Class balancing loss functions:} These loss functions are used to address the class imbalance problem which occurs when melody extraction is considered as classification problem by quantizing the pitch into fixed number of pitch classes. The class imbalance is addressed by suitably weighting the imbalanced classes in the loss function. 
Focal loss is proposed in~\cite{lu2018vocal} to counter the class imbalance problem of negative classes for training the encoder-decoder model for melody extraction. The focal loss is given by 
\begin{equation}
    FL(\hat{y}_t) = - \alpha_t(1 - \hat{y}_t)^\gamma \log(\hat{y}_t) 
\end{equation}
where $\hat{y}_t$ is the model predicts probability for class $t$. $\alpha_t$ is the weighting factor value between $[0, 1]$, $(1 - \hat{y}_t)^\gamma$ is the modulating factor and $\gamma$ controls the weights assigned to dominating examples. \newline

\noindent \textbf{(iv) Melody specific loss functions:} These category of loss functions are developed specifically for melody extraction. For example, few methods exploit the harmonic structure of the melody source while minimizing the total loss of the model. These loss functions can also be categorized as joint loss functions; however, since these loss functions are designed keeping melody extraction in mind, we categorize them as melody specific loss functions. \newline

\noindent A new loss function is introduced in~\cite{gao2021hrnet} to overcome the class imbalance problem, which occurs due to an imbalance between voiced and unvoiced frames in the music excerpt. Total loss is proposed as the sum of two losses, i.e., pitch classification loss denoted by $L_{pitch}$ which is the cross-entropy loss between the predicted and target one-hot encoded pitch classes. 
\begin{equation}
L_{total} = L_{pitch} + L_{voicing}
\end{equation}
\begin{equation} \label{eq:pitch_loss}
    L_{pitch} = 
\begin{cases}
    CE(y, \hat{y}),& \text{if the frame is voiced}\\
    0,              & \text{otherwise}
\end{cases}
\end{equation}

\noindent As shown in Eq.~\ref{eq:pitch_loss}, the pitch loss does not take unvoiced frames for calculating the loss function. The voicing decision loss $L_{voicing}$ is the binary cross-entropy loss that considers all frames, including unvoiced frames for loss calculation.    
\begin{equation}
L_{voicing} = BCE(y_{voice}, \hat{y}_{voice})
\end{equation}

\noindent A novel loss term is introduced in~\cite{park2017melody} to reduce the octave errors caused by the melody extraction model along with the main CE loss function. The total loss function is given by 
\begin{equation}
    L_{total} = L_{CE} + \alpha_h L_{h}
\end{equation}
\noindent where $L_{CE}$ is the cross entropy loss between the predicted and the target labels, $\alpha_h$ is the weighting factor and $L_{h}$ is the harmonic sum loss. The harmonic sum loss is given by 
\begin{equation}
    L_{h} = \sum_t p^T \hat{y}_t
\end{equation}

\noindent where $p \in {0, 1}^{|B|}$ is a column vector to sum the all harmonics of the ground truth melody pitch $y_t$. The $j_{th}$ element of the $p$ is given by
\begin{equation}
  p_{j} = 
\begin{cases}
    1,& \text{if $mod(j, w) = mod(y_{t}, w)$  $and$  $j \neq y_{t}$}\\
    0,              & \text{otherwise}
\end{cases}    
\end{equation}
\noindent where $mod(j, w)$ is obtained by the remained of $j$ division by $w$ and $w < |B|$ which is the octave length. 

\section{Optimizer} \label{sec:optimizers}

The optimizer is used to adjust the weights of the model parameters to minimize the loss function. As we can observe from Table~\ref{tab:optimizers}, most of the authors used Adam optimizer to adjust the model weights for its ability to fast convergence, adaptive learning rate, and variance stabilization. We can also see that few authors used other optimization algorithms such as RMSprop and SGD with momentum. The batch size used for the batch gradient descent varies across the proposed models, mainly depending on the available GPU memory. Few authors also reduce the initial learning by some factor if the validation accuracy plateaus to bump the model from local optima. Also, not all models are trained for all the epochs specified in the table. Some authors adopted an early stopping strategy to stop the training if the validation accuracy saturates after a few epochs.          

\begin{table*}[h]
  \begin{center}
    \caption{The optimization algorithms used by various authors to adjust the models parameters.}
    \label{tab:optimizers}
    \resizebox{\textwidth}{!}{\begin{tabular}{l c c c c c c }
      \toprule 
       {First Author} & \textbf{Optimizer} & \textbf{Learning Rate} & \textbf{Batch Size} &\textbf{Epoch} \\
      \midrule 
       Xingjian 2021~\cite{du2021singing} & Adam & 0.0001  & NA & NA \\
       Hsin 2018~\cite{chou2018hybrid} & Adam & NA & NA & 15 \\ 
       Ping	2020~\cite{gao2020multi} & Adam & 0.001 & NA & 15 \\ 
       Ping	2019~\cite{gao2019multi} & Adam & 0.001 & NA & NA \\
       Tsung-Han 2019~\cite{hsieh2019streamlined} & Adam & NA & NA & NA \\  
       Yongwei	2021~\cite{gao2021hrnet} & Adam & 0.0001 & 64 & NA \\
       Sangeun	2019~\cite{kum2019joint} & Adam & 0.002 & NA 45 \\
       Sangeun	2017~\cite{kum2017classification} & SGD with Nesterov momentum & 0.02 & NA & 100 \\
       Ming-Tso	2019~\cite{chen2019cnn} & NA & NA & NA & NA \\ 
       Rachel	2017~\cite{bittner2017deep} & Adam & NA & NA & NA \\
       Yongwei	2021~\cite{gao2021vocal} & Adam & 0.0001 & 5 & NA \\ 
       Shuai	2021~\cite{yu2021frequency} & Adam & 0.0001 & NA & NA \\
       Shuai Yu	2021~\cite{yu2021hanme} & Adam & NA & NA & NA \\
       Tomoyasu	2019~\cite{nakano2019joint} & Adam & NA & NA & NA \\
       Andreas	2019~\cite{jansson2019joint} & NA & NA & NA & NA \\
       Dogac	2018~\cite{basaran2018main} & Adam & NA & 16 & NA \\ 
       Hyunsin	2017~\cite{park2017melody} & SGD with momentum & 0.00001 & NA & 20\\
       Sangeun	2016~\cite{kum2016melody} & RMSprop & NA & NA & NA \\
       Rachel	2018~\cite{bittner2018multitask} & Adam & NA & 4 & 25 \\
       Sangeun	2020~\cite{kum2020semi} & Adam & 0.003 & NA & 70 \\ 
       Fran{\c{c}}ois	2016~\cite{rigaud2016singing} & SGD with momentum & 0.000001 & 30 & 1000 \\
       Li	2018~\cite{su2018vocal} & Adam & NA & NA & NA \\
       Yongwei	2019~\cite{gao2019vocal} & SGD & NA & NA & NA \\
       Wei Tsung	2018~\cite{lu2018vocal} & Adam & NA & NA & NA \\
       Zhe-Cheng	2016~\cite{fan2016singing} & RMSprop & NA & NA & 1000 \\
       \bottomrule 
    \end{tabular}}
  \end{center}
\end{table*}


\begin{figure}[!tbp]
        \centering
		\resizebox{4cm}{5cm}{\includegraphics{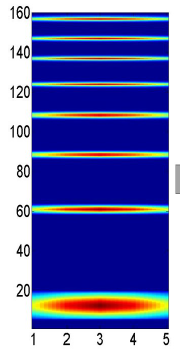}}  
        \caption{CNN kernel is initialized by the log-frequency harmonic pattern~\cite{chou2018hybrid}.}
        \label{fig:init_hsain}
 \end{figure} 

\begin{figure}[!tbp]
        \centering
		\resizebox{6cm}{5cm}{\includegraphics{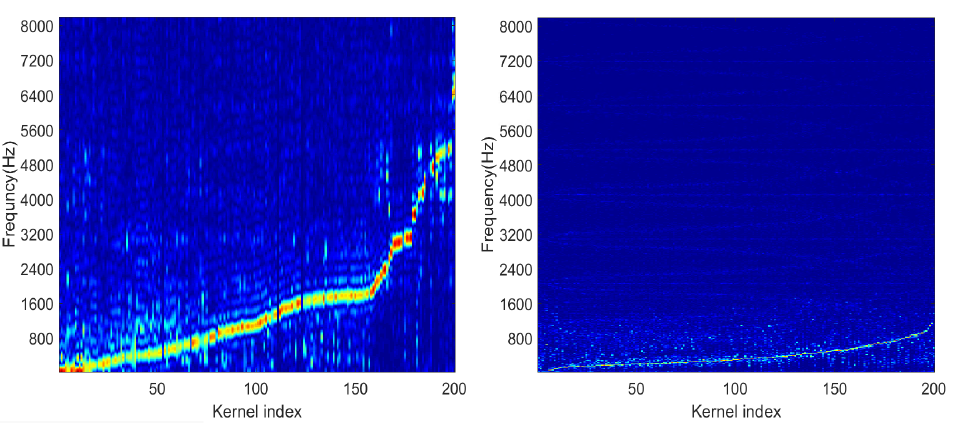}}  
        \caption{Sorted pre-trained kernels~\cite{chen2019cnn} of length 64 and 960 used to initialize the first stage 1D CNN kernels.}
        \label{fig:init_chen2019cnn}
 \end{figure} 

\section{Parameter Initialization} \label{sec:param_init}

The parameter initialization plays a vital role in fast convergence and the likelihood of obtaining lower training and generalization error rates. We can see the various initialization strategies adopted by the authors of the proposed melody extraction models. Most of the authors use the default initialization schemes provided by the deep learning frameworks. In~\cite{chou2018hybrid}, the CNN kernel is initialized by the pattern inspired by the excitation of the auditory spectrum~\cite{shamma2000case} of the first eight resolvable harmonics as shown in Fig.~\ref{fig:init_hsain}. The authors also found that the model with CNN kernel initialized with the custom weights converges faster than the model initialized with random weights. The network parameters are initialized by using He uniform distribution in~\cite{kum2019joint, kum2017classification}. The 1D CNN kernels of the two-stage melody extraction model in~\cite{chen2019cnn} are initialized from the sorted pre-trained 1D CNN model kernels. Specifically, they pre-train 1D CNN kernels with lengths 64 and 960 and sort them to mitigate the kernel permutation problem. And then, they initialize the first stage 1D CNN kernels with sorted pre-trained filters as shown in Fig.~\ref{fig:init_chen2019cnn}. Whereas in~\cite{park2017melody, kum2016melody}, the model parameters are initialized from the uniform distribution. The model weights are initialized from Gaussian distribution with zero mean and 0.1 standard deviation in~\cite{rigaud2016singing}. 


\section{Explainability} \label{sec:explainability}

The deep neural networks are mostly trained as black-box models across various domains, either for regression or classification or for both. In melody extraction, too, most of the authors treat the neural networks as black-box and try to improve the model performance of the networks by mostly proposing new architecture's, input and output representations, and loss functions. In this section, we briefly discuss the melody extraction methods, which include explainability.        

\begin{figure}[!tbp]
        \centering
		\resizebox{9cm}{4.5cm}{\includegraphics{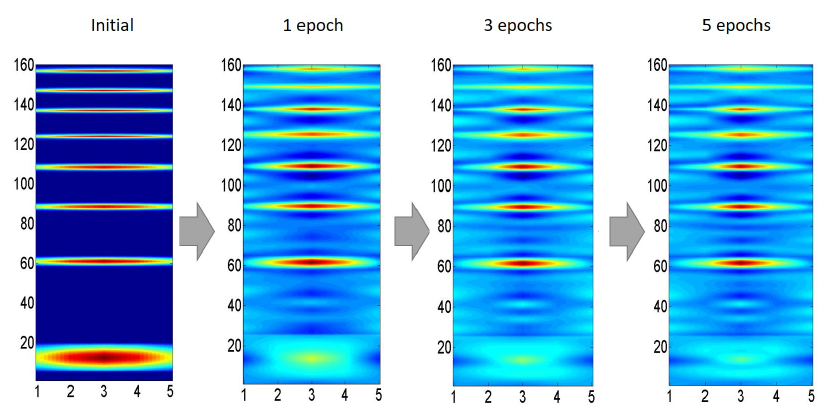}}  
        \caption{The evolution of the harmonic pattern of the CNN kernel during training after every few epoch~\cite{chou2018hybrid}.}
        \label{fig:exp_chou2018hybrid}
 \end{figure}

In~\cite{chou2018hybrid}, it is shown that the CNN kernel will finally converge to the resolvable harmonic pattern even if the CNN kernel is initialized with the random initial values or with the kernel initialized with the resolvable harmonics. The evolution of the CNN kernel after every few epochs is shown in Fig.~\ref{fig:exp_chou2018hybrid}. The figure shows that the kernel converges to the resolvable harmonic pattern.  

An observation is made in~\cite{chen2019cnn} that the 1D CNNs with 64 kernel size filters learn to resolve the high frequencies with low resolution, whereas 1D CNNs with 960 kernel size learn to resolve low frequencies with high resolution as shown in Fig.~\ref{fig:init_chen2019cnn}.      

\begin{figure}[!tbp]
        \centering
		\resizebox{10cm}{6cm}{\includegraphics{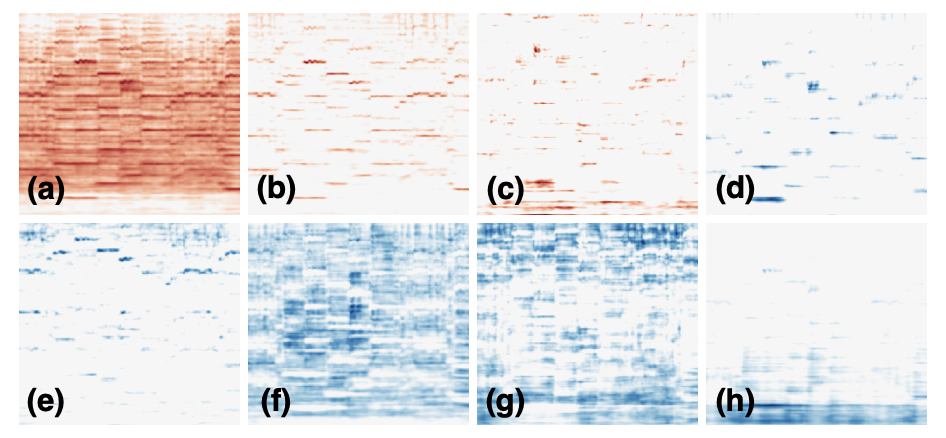}}  
        \caption{The enhancing harmonic content, deemphasizing octave errors and denoising activation properties observed in the final CNN layer activations~\cite{bittner2017deep}.}
        \label{fig:exp_bittner2017deep}
 \end{figure}

An interesting observation is made in~\cite{bittner2017deep} by plotting the eight feature maps of the penultimate CNN layer. It is observed that some activations try to emphasize the harmonic content, while others try to deemphasize the octave error mistakes and while some other activations try to deemphasize the broadband and low-frequency noise as shown in Fig.~\ref{fig:exp_bittner2017deep}.  

\begin{figure}[!tbp]
        \centering
		\resizebox{8cm}{8cm}{\includegraphics{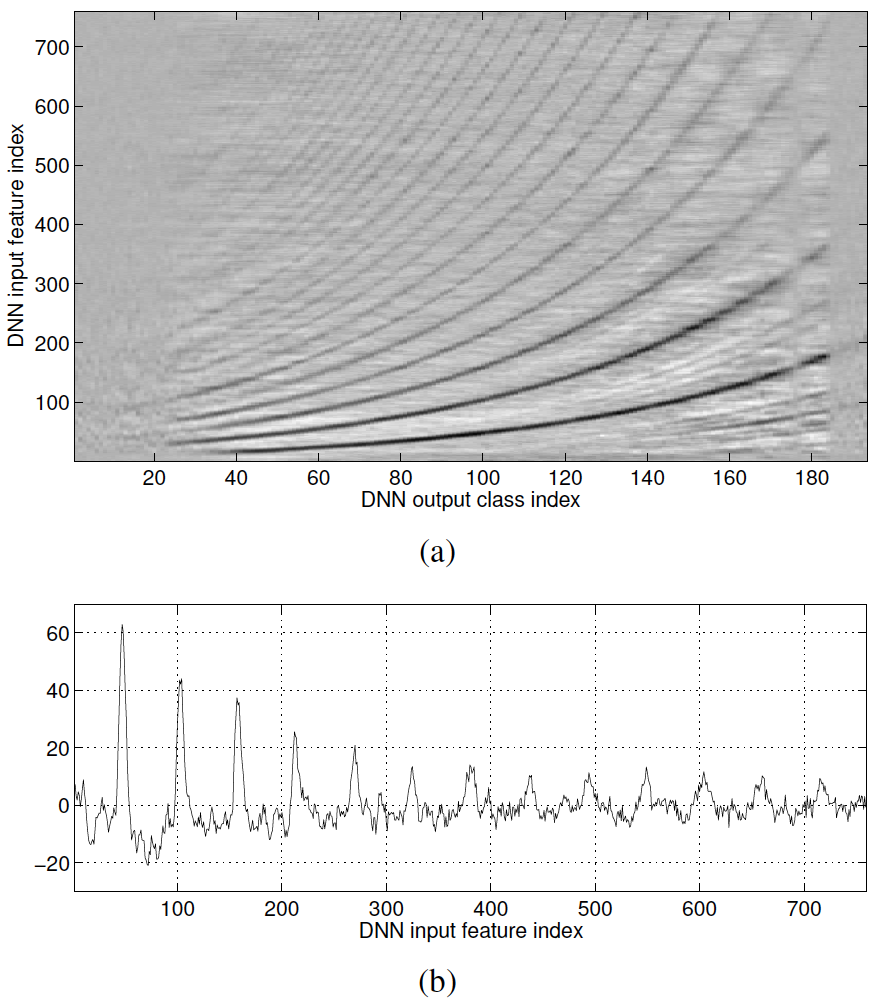}}  
        \caption{Illustration of the transposed weight matrix and the weight vector for the F0 with index 100~\cite{rigaud2016singing}.}
        \label{fig:exp_rigaud2016singing}
 \end{figure} 

A simplified linear version of the non-linear DNN model is analyzed by removing all non-linear operations to understand what the neural network is learning is presented in~\cite{rigaud2016singing}. It is found that the learned weight matrix exhibits the harmonic structure for most of the F0s except for low and high-frequency range, which is because of no or few training examples in that range that can be observed in Fig~\ref{fig:exp_rigaud2016singing}.   

\section{Data Augmentation} \label{sec:data_augmentatio}

Data augmentation is a technique used in deep learning to automatically create augmented versions of the original training data to improve the generalization accuracy of the model by reducing overfitting. However, data augmentation is a widely used technique in the Computer Vision community to improve the model performance, not much popular among the melody extraction community. However, few authors used data augmentation techniques to reduce the model overfitting. The training data is augmented by a phase vocoder~\cite{laroche2002time} based pitch-shifting method, which modifies the pitch of the audio independent of time-stretching to augment the audio data in~\cite{kum2016melody, kum2017classification, kum2019joint, kum2020semi}. The data augmentation for the training set is introduced in~\cite{bittner2018multitask} by adopting various mixing strategies of the multitrack dataset. Inspired by the RandAugment~\cite{cubuk2019randaugment}, RandAudioAugment (RAA) is proposed in~\cite{kum2020semi} that randomly augments the audio with the effects: audio equalizer, phaser, low-pass, high-pass filtering, overdrive, and reverb. The RAA is implemented using $pysndfx$ Python library \url{https://github.com/carlthome/python-audio-effects}. Audio Degradation Toolbox~\cite{mauch2013audio} is used in~\cite{rigaud2016singing} to augment the data by randomly introducing the various degradation types available in the toolbox by preserving the alignment between the audio and the annotations. In order to increase the symbolic training data, the MIDI files are augmented by pitch-shifting up and down and also by shifting half of the training data melody down by one-octave in~\cite{lu2018vocal} to cover a wide pitch range. 

\section{Evaluation Measures} \label{sec:eval_measur}

The five global measures introduced in MIREX2005~\cite{poliner2007melody} are used to evaluate the performance of the melody extraction methods. Representing the estimated melody pitch frequency as a vector $f$, and the ground truth melody sequence as  $f^*$. The vocal indicator vector as $V$, and its $\tau ^{th}$ element $\upsilon = 1$ when the algorithm estimates the $\tau^{th}$ frame as vocal, with the corresponding ground truth $V^*$. Also, defining the non-vocal indicator as $\bar{\upsilon _{\tau}}$ = 1 - ${ \upsilon_{\tau}}$. The five global measures are given as below: \newline

\noindent \textbf{Voicing Recall Rate} (VR): the amount of frames estimated as voiced (melody frames) by the algorithm which are labelled as voiced (melody frames) in the ground truth.

\begin{equation}
VR =  \frac{\sum_{\tau}^{} \upsilon_{\tau} \upsilon_{\tau}^*}{\sum_{\tau}^{} \upsilon_{\tau}^*}   
\end{equation} \\

\noindent where $\upsilon_{\tau}$ and $\upsilon_{\tau}^{*}$ are the $\tau^{th}$ elements of $V$ and $V^*$ respectively. \newline

\noindent \textbf{Voicing False Alarm Rate} (VFA): the amount of frames estimated as voiced by the algorithm that are labelled as unvoiced in the ground truth. 

\begin{equation}
VFA =  \frac{\sum_{\tau}^{} \upsilon_{\tau} \bar{\upsilon_{\tau}^{*}}}{\sum_{\tau}^{} \bar{\upsilon_{\tau}^{*}}}   
\end{equation} \\

\noindent \textbf{Raw Pitch Accuracy} (RP): is the probability of a correct pitch value (to within $ \pm 1/4$ tone) given that the frame is indeed pitched. This will also includes the pitch guesses for frames that were judged unvoiced. \newline

\begin{equation}
RP =  \frac{\sum_{\tau}^{} \bar{\upsilon_{\tau}^{*}}  \mathcal{T} \left[C(f_\tau) - C(f_{\tau}^{*}) \right]}{\sum_{\tau}^{} \bar{\upsilon_{\tau}^{*}}}   
\end{equation}
\noindent where $f_\tau$ and $f_{\tau}^{*}$ are the  $\tau^{th}$ elements of $f$ and $f^*$ respectively. The threshold function $\tau[z]$ is defined by

\[
 \tau[z] =
  \begin{cases} 
      \hfill 1    \hfill & \text{ if $|z|$ $<$ 50} \\
      \hfill 0 \hfill & \text{ if $|z|$ $\geq$ 50} \\
  \end{cases}
\]

\noindent where the function $C$ maps the frequency in Hz to perceptually motivated logarithmic frequency scale. Each semitone is divided into 100 $cents$. The frequency values are represented as real-valued cents above a reference frequency $f_{ref}$:

\begin{equation}
C(f) = 1200 \log_{2} \left(\frac{f}{f_{ref}}\right) 
\end{equation}

\noindent The reference frequency is $f_{ref}$ = 55 Hz (piano note A1). Note that since 100 $cents$ = 1 semitone, the 50 $cent$ threshold is equivalent to half a semitone. \\

\noindent \textbf{Raw Chroma Accuracy} (RCA): Similar to raw pitch accuracy except that the predicted and the ground truth f0 is mapped to single octave. This measures the pitch accuracy which by ignoring octave errors which is a common error made by melody extraction models. \newline

\begin{equation}
RP =  \frac{\sum_{\tau}^{} \bar{\upsilon_{\tau}^{*}}  \mathcal{T} \left[ \langle C(f_\tau) - C(f_{\tau}^{*}) \rangle_{12}] \right]}{\sum_{\tau}^{} \bar{\upsilon_{\tau}^{*}}}   
\end{equation}

\noindent The octave equivalence is obtained by finding the difference between the semitone-scale pitch value modulo 12 where,

\begin{equation}
\langle a \rangle_{12} = a - 12 \lfloor \frac{a}{12} + 0.5 \rfloor   
\end{equation}

\noindent \textbf{Overall Accuracy} (OA): combines both the voicing detection and the pitch estimation to give the proportion of frames that are correctly labeled with both pitch and voicing. \newline

\begin{equation}
OA = \frac{1}{L} \sum_{\tau}^{} \bar{\upsilon_{\tau}^{*}} \mathcal{T} \left[C(f_\tau) - C(f_{\tau}^{*}) \right] + \bar{\upsilon_{\tau}^{*}} \bar{\upsilon_{\tau}} 
\end{equation}

\noindent where $L$ represents the total number of frames. \\

All measures represent the worst case (0) to best case (1), except for (\textbf{VFA}) where 0 represents the best case and 1 the worst case. The performance of the given method is determined as the average of all music excerpts scores for the measure considered on the whole music dataset. The open-source Python library \textit{mir\_eval}~\cite{raffel2014mir_eval}\footnote{https://github.com/craffel/mir\_eval} is mostly used by authors of the latest models to evaluate the performance of melody extraction.

\section{Performance} \label{sec:performance}

In this section, we discuss the performance of the melody extraction models based on the raw pitch accuracy (RPA) and overall (OA) accuracy metrics. The raw pitch accuracy gives the idea of how well the melody extraction algorithm tracks the melodic source's pitch. The overall accuracy also includes the voicing detection and pitch detection accuracy to give the overall system performance. The performance of the deep learning models vastly depends on the data with which the models are trained. That is because of the assumption on the data that is independent and identically distributed (iid) assumption on which the model is being trained. For example, a model trained only on MIR\_1K train split dataset is expected to perform well on similar data, i.e., MIR\_1K test split, and it may perform poorly on the MedleyDB dataset, which consists of songs with a variety of musical genres very different from MIR\_1K. Hence, it will be very difficult to directly compare the models on the reported results by authors of the proposed models because models trained (e.g., ~\cite{gao2019vocal}) on the large amount of in-house dataset gives undue advantage over other models which are trained on the small openly available datasets. Also, few models split the same dataset into train and test splits and evaluate the model on the test split. The split from the same dataset again gives an undue advantage for the model performance. Also, a direct comparison is difficult because the models are trained on different combinations of datasets, e.g., MIR\_1K and MedleyDB, RWC, and MedleyDB. Since no standard protocol is followed for training and testing the models with the common dataset, a fair comparison of the performance of the models is challenging. Hence, here we present the general trend in the performance results achieved by various models without pointing out that the so and so model performs well or is poor. Please note that some authors only reported RPA or OA in their papers. 

\begin{figure}[!tbp]
        \centering
		\resizebox{12cm}{7cm}{\includegraphics{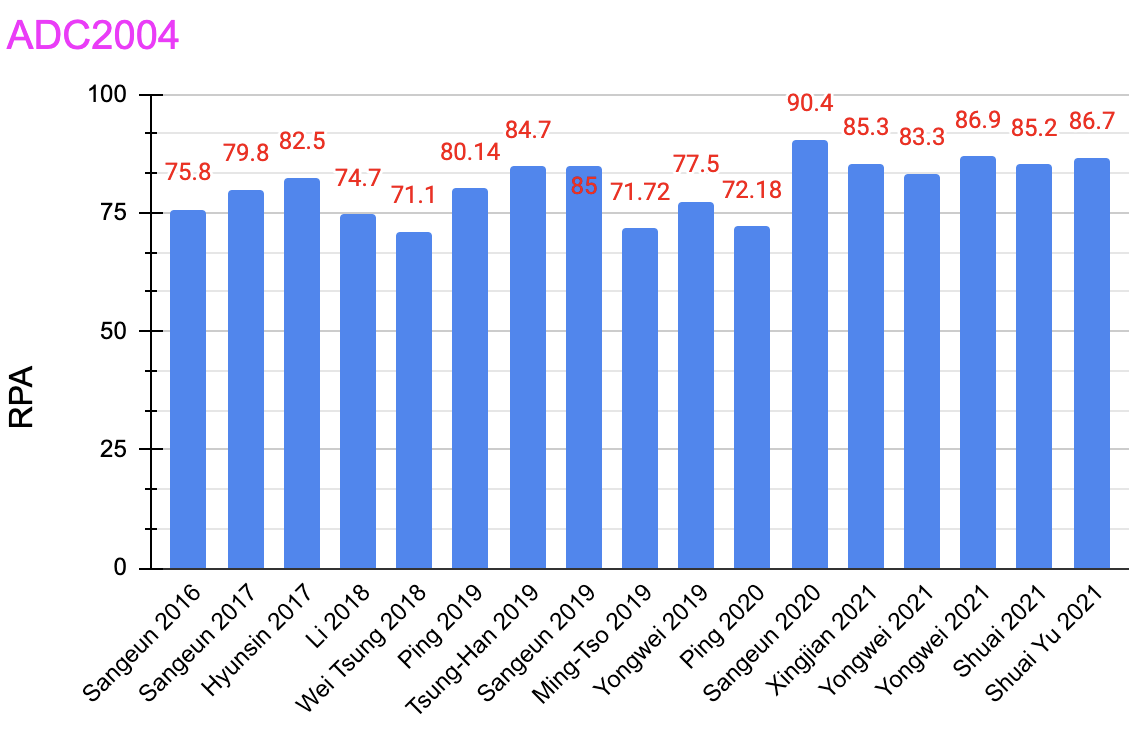}}  
        \caption{Raw pitch accuracy of the melody extraction models on ADC2004 dataset.}
        \label{fig:adc2004_RPA}
\end{figure}

\begin{figure}[!tbp]
        \centering
		\resizebox{12cm}{7cm}{\includegraphics{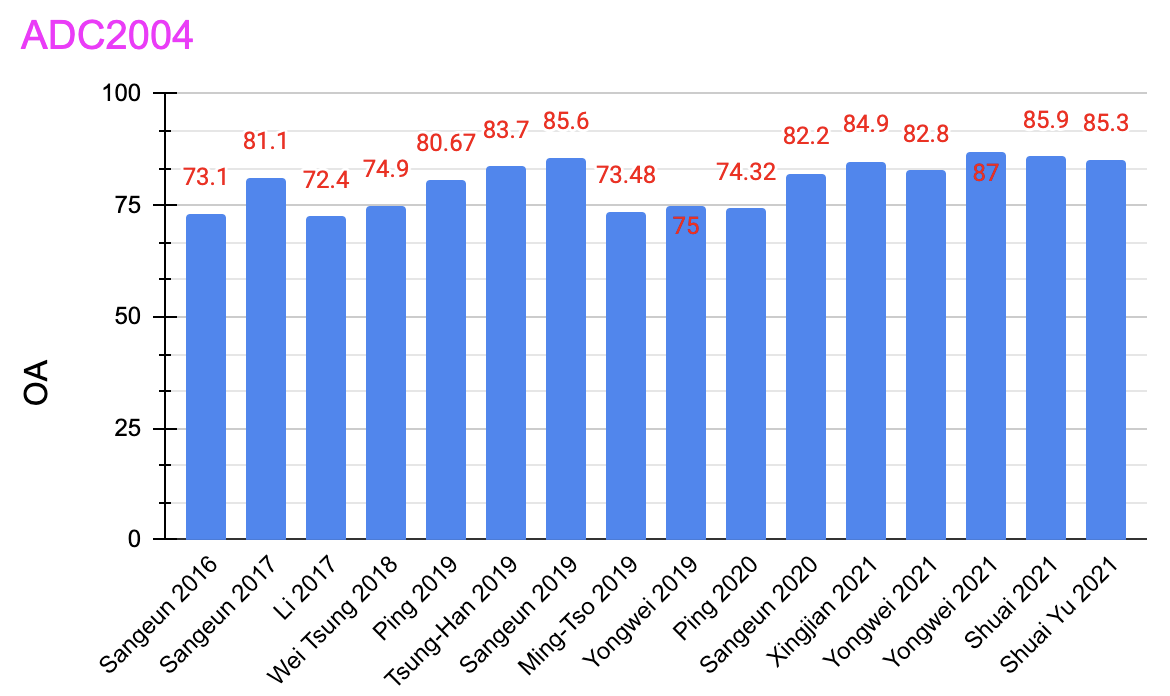}}  
        \caption{Overall accuracy of the melody extraction models on ADC2004 dataset.}
        \label{fig:adc2004_OA}
\end{figure}

\begin{figure}[!tbp]
        \centering
		\resizebox{12cm}{7cm}{\includegraphics{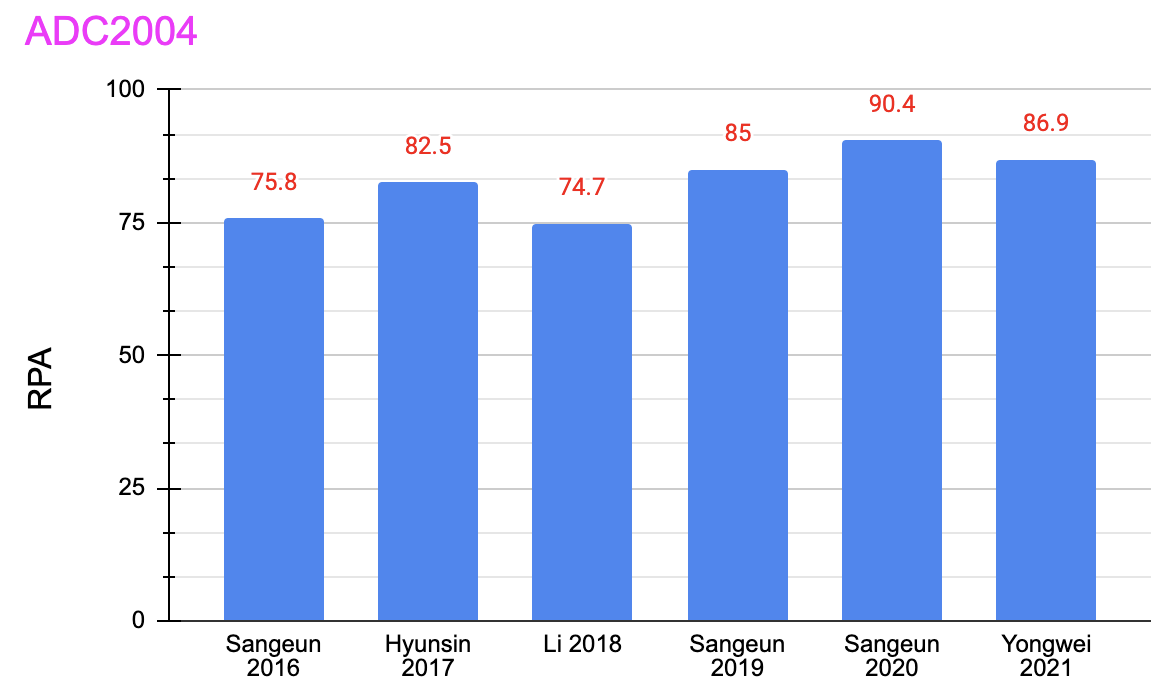}}  
        \caption{Best raw pitch accuracy of the melody extraction models on ADC2004 dataset over the years.}
        \label{fig:adc2004_year_RPA}
\end{figure}

\begin{figure}[!tbp]
        \centering
		\resizebox{12cm}{7cm}{\includegraphics{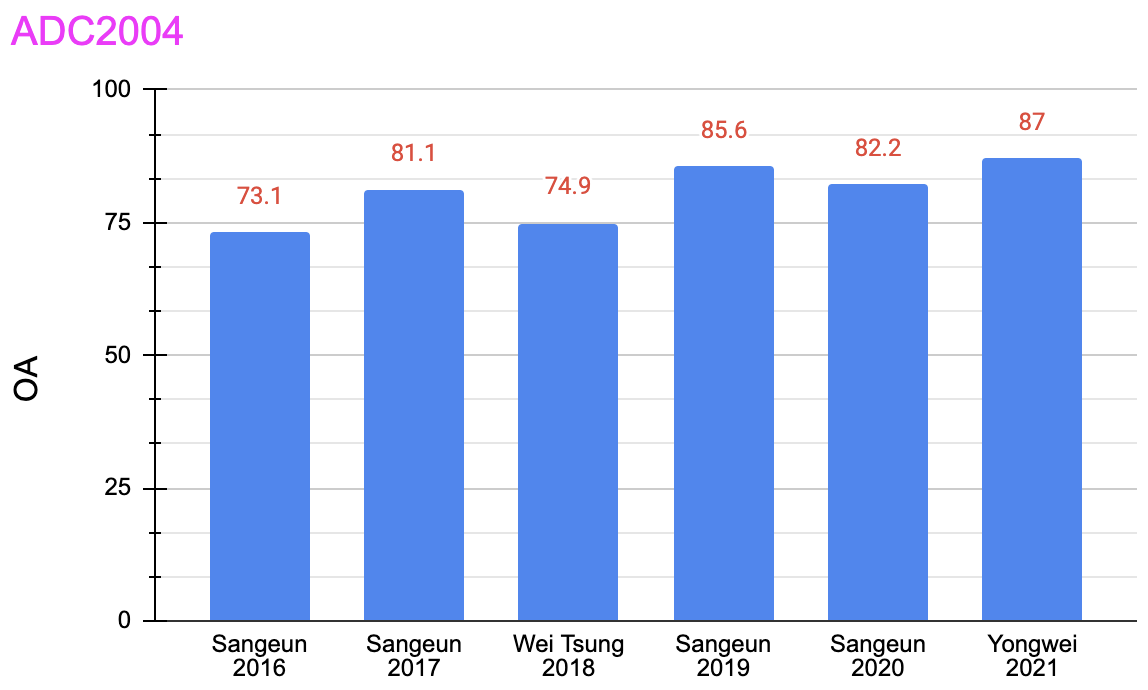}}  
        \caption{Best overall accuracy of the melody extraction models on ADC2004 dataset over the years.}
        \label{fig:adc2004_year_OA}
\end{figure}

The raw pitch accuracy and the overall accuracy for various models on the ADC2004 dataset are plotted in Figs.~\ref{fig:adc2004_RPA} and~\ref{fig:adc2004_OA}. The general trend is that all models have a decent RPA and OA. The minimum RPA is 71\%, and the maximum goes up to 90\% on average. Similarly, the reported OA for ADC2004 dataset goes up to 87\%, with the minimum OA being approximately 72\%. We also plot the best RPA and OA of the models among the models proposed in the same year for ADC2004 dataset in Figs.~\ref{fig:adc2004_year_RPA} and~\ref{fig:adc2004_year_OA}. We see that over the years, the RPA and OA is significantly improved for ADC2004 dataset. However, we can also see that there is still room for improving the results for ADC2004 dataset by proposing new architectures and neural network models. 

\begin{figure}[!tbp]
        \centering
		\resizebox{12cm}{7cm}{\includegraphics{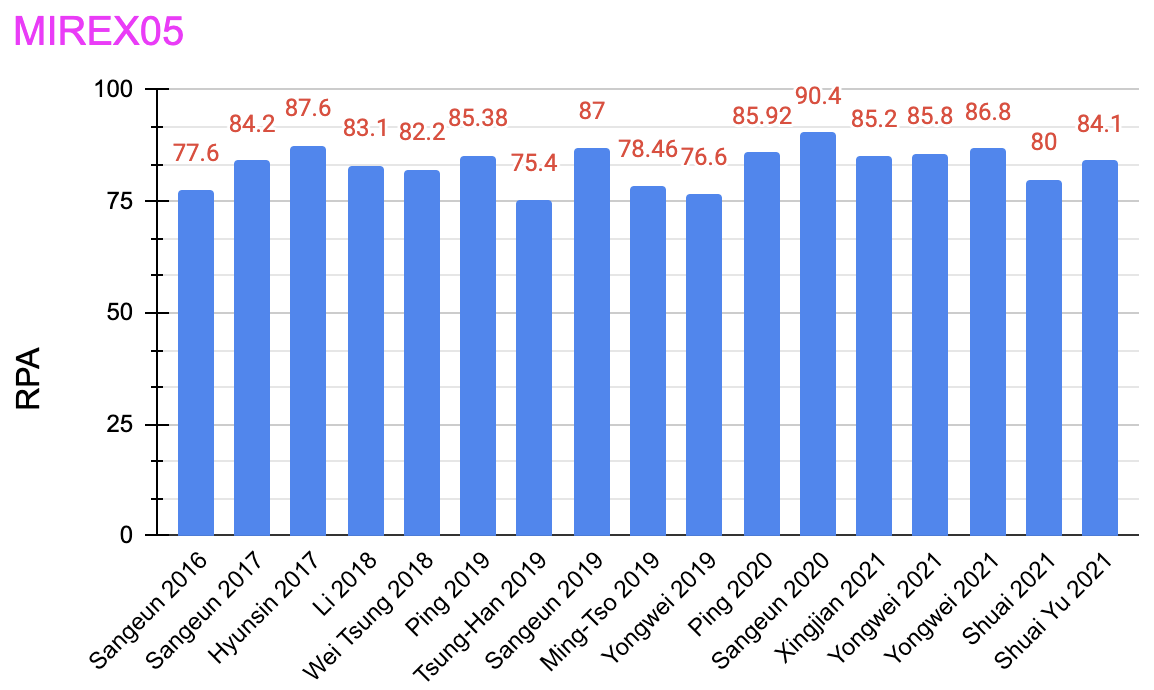}}  
        \caption{Raw pitch accuracy of the melody extraction models on MIREX05 dataset.}
        \label{fig:MIREX05_RPA}
\end{figure}

\begin{figure}[!tbp]
        \centering
		\resizebox{12cm}{7cm}{\includegraphics{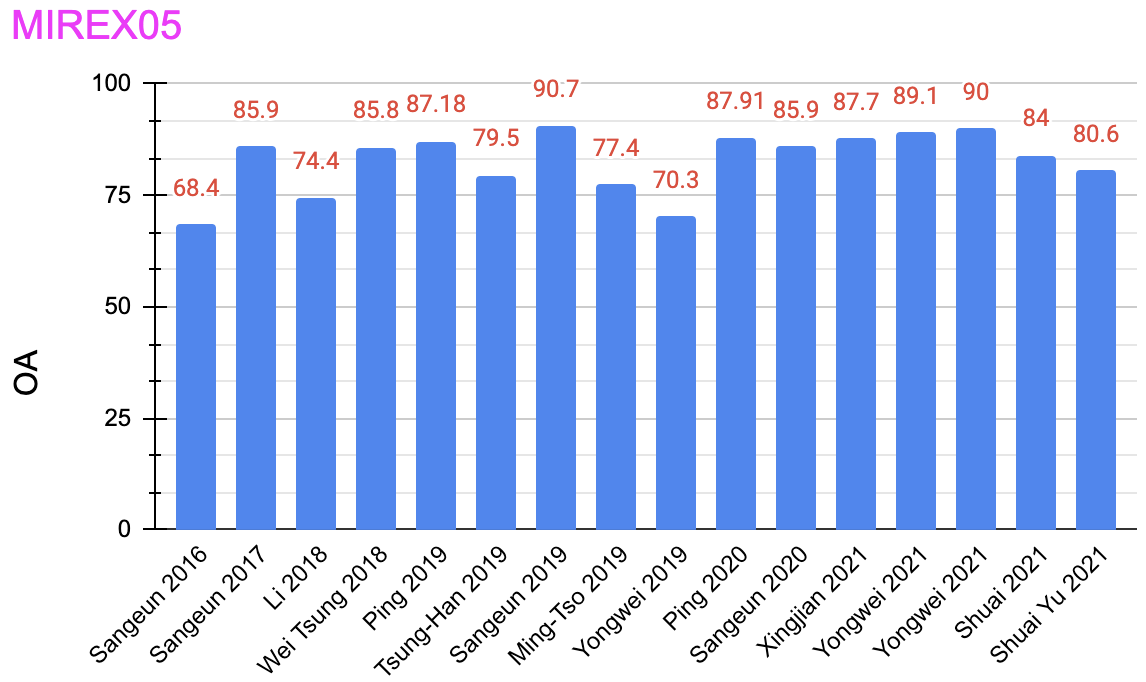}}  
        \caption{Overall accuracy of the melody extraction models on MIREX05 dataset.}
        \label{fig:MIREX05_OA}
\end{figure}

\begin{figure}[!tbp]
        \centering
		\resizebox{12cm}{7cm}{\includegraphics{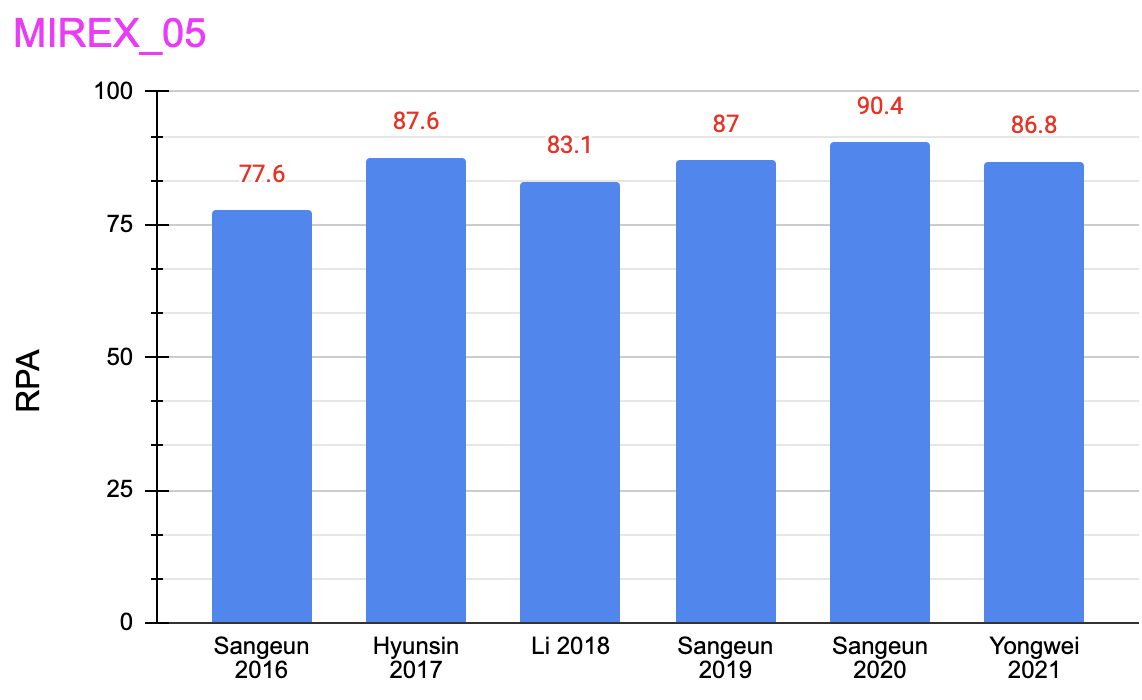}}  
        \caption{Best raw pitch accuracy of the melody extraction models on MIREX05 dataset over the years.}
        \label{fig:MIREX05_year_RPA}
\end{figure}

\begin{figure}[!tbp]
        \centering
		\resizebox{12cm}{7cm}{\includegraphics{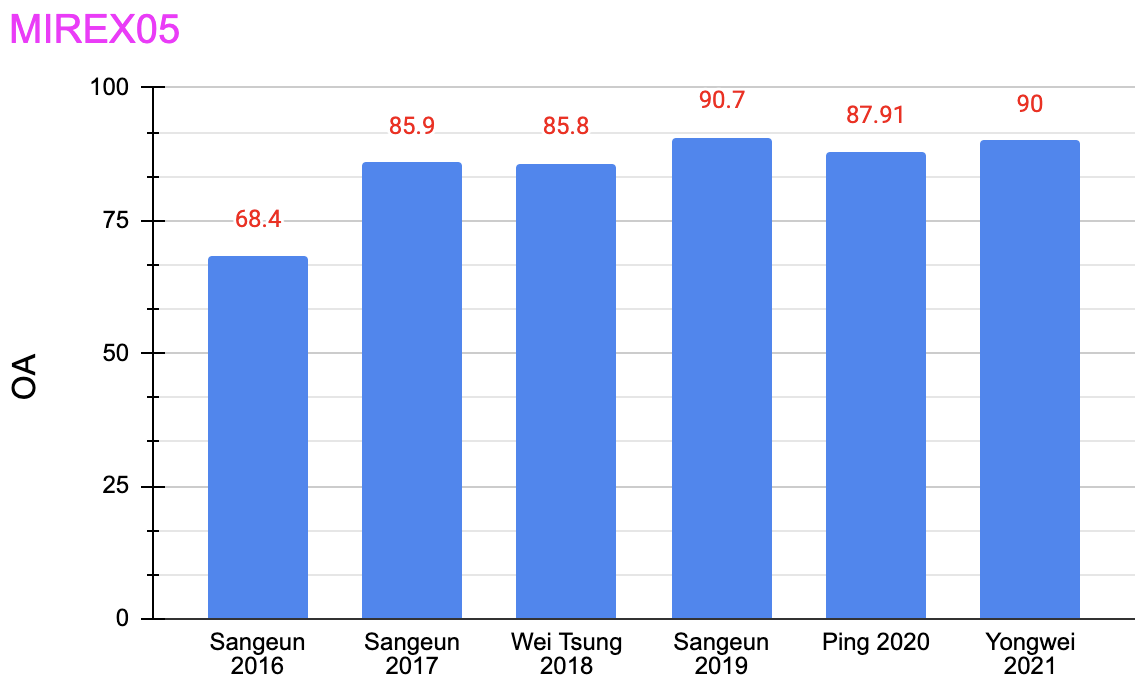}}  
        \caption{Best overall accuracy of the melody extraction models on MIREX05 dataset over the years.}
        \label{fig:MIREX05_year_OA}
\end{figure}

\begin{figure}[!tbp]
        \centering
		\resizebox{12cm}{7cm}{\includegraphics{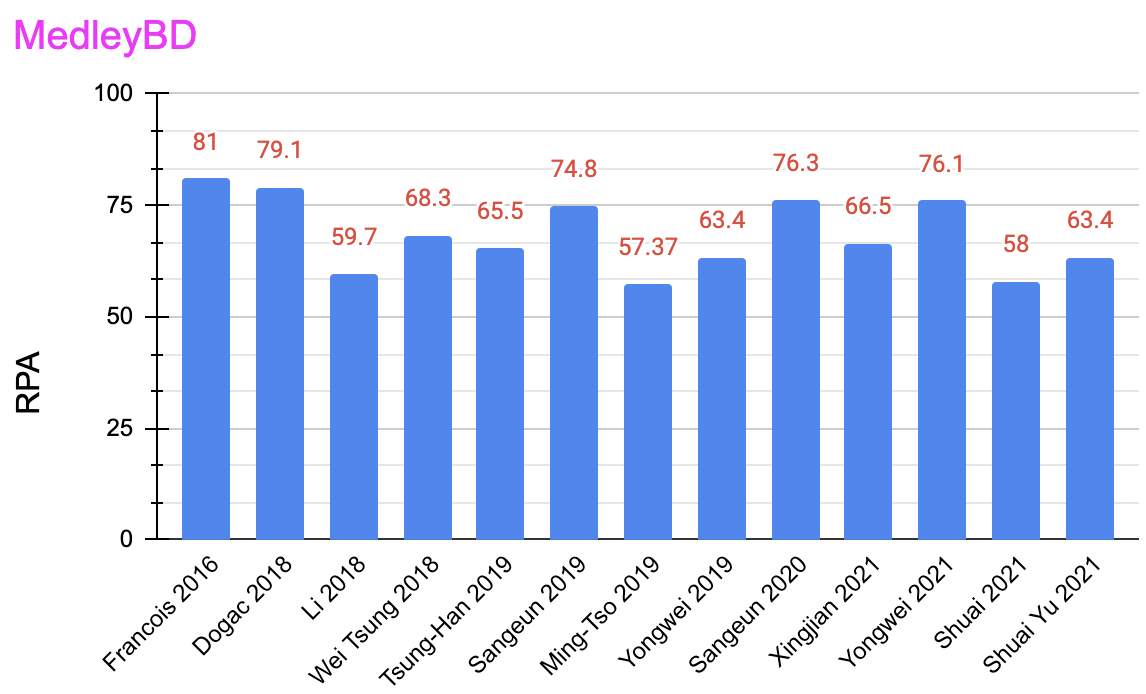}}  
        \caption{Raw pitch accuracy of the melody extraction models on MedleyDB dataset.}
        \label{fig:MedleyDB_RPA}
\end{figure}

\begin{figure}[!tbp]
        \centering
		\resizebox{12cm}{7cm}{\includegraphics{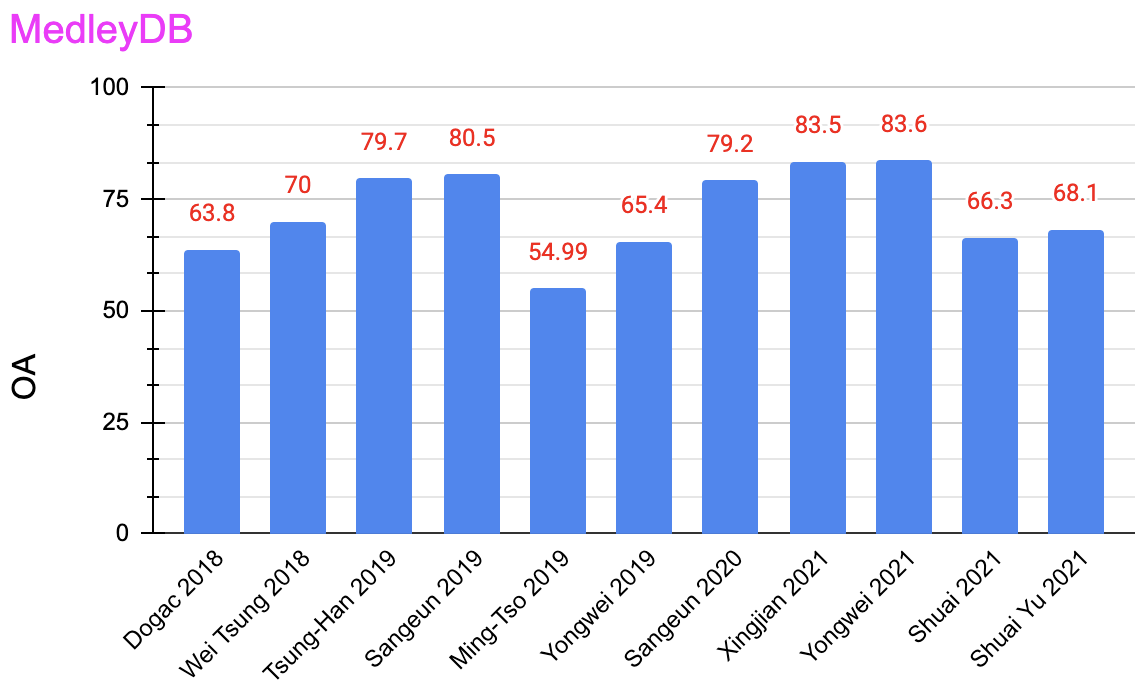}}  
        \caption{Overall accuracy of the melody extraction models on MedleyDB dataset.}
        \label{fig:MedleyDB_OA}
\end{figure}

\begin{figure}[!tbp]
        \centering
		\resizebox{12cm}{7cm}{\includegraphics{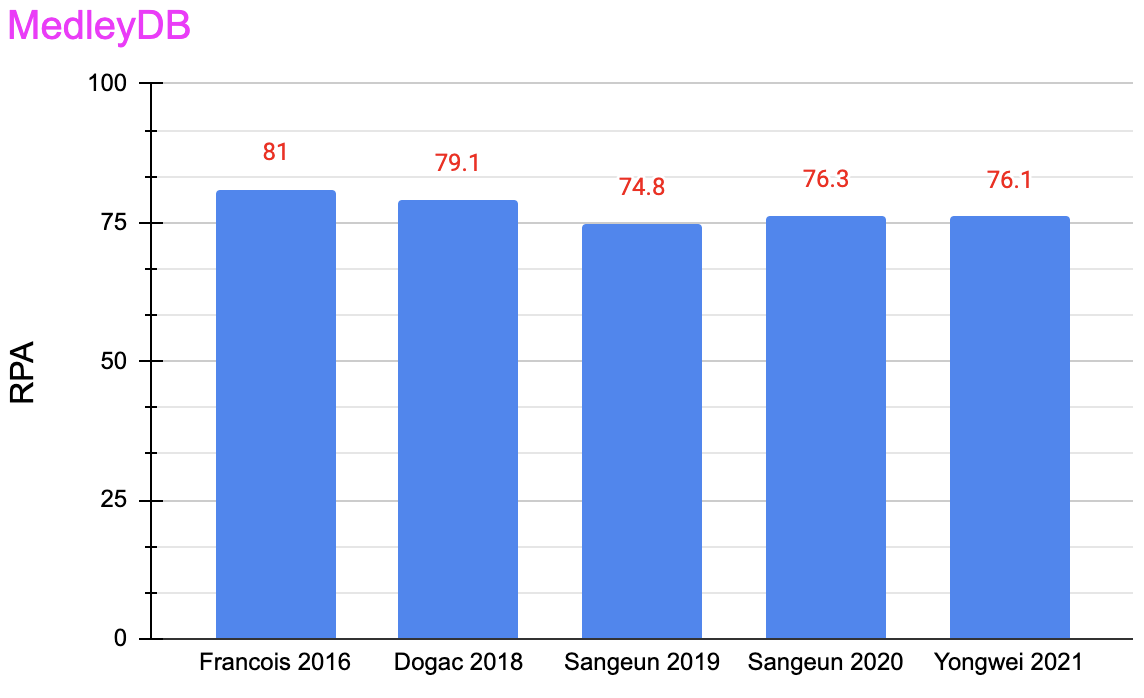}}  
        \caption{Best raw pitch accuracy of the melody extraction models on MedleyDB dataset over the years.}
        \label{fig:MedleyDB_year_RPA}
\end{figure}

\begin{figure}[!tbp]
        \centering
		\resizebox{12cm}{7cm}{\includegraphics{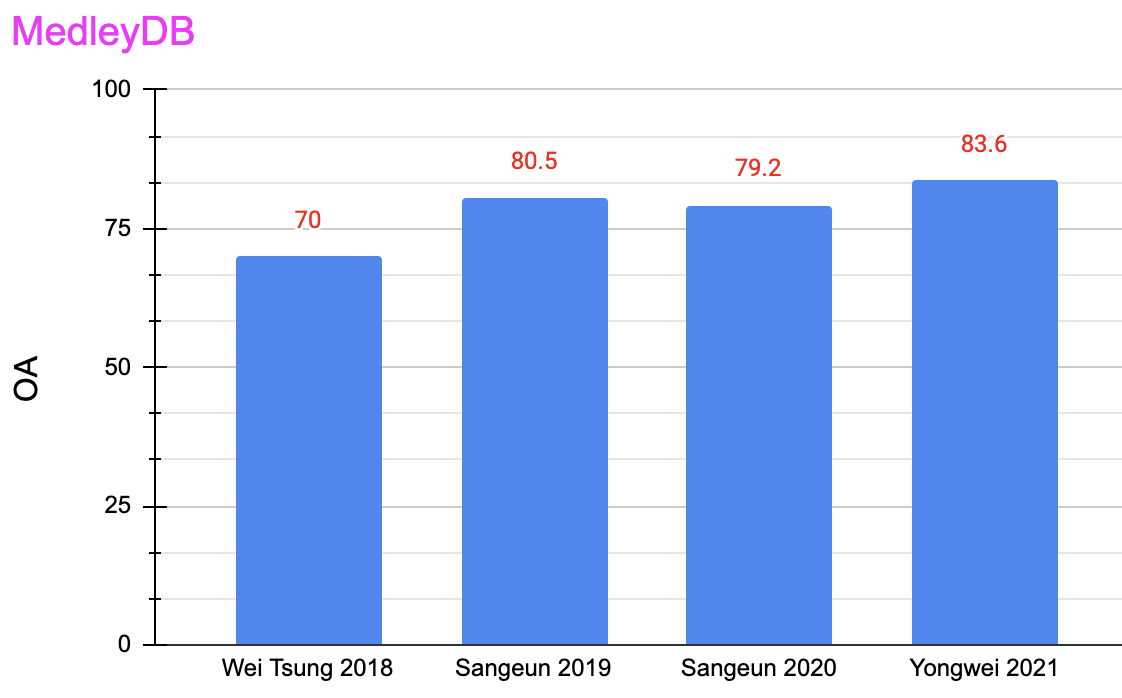}}  
        \caption{Best overall accuracy of the melody extraction models on MedleyDB dataset over the years.}
        \label{fig:MedleyDB_year_OA}
\end{figure}

A similar trend can be observed for MIREX05 dataset as shown in Figs.~\ref{fig:MIREX05_RPA},~\ref{fig:MIREX05_OA},~\ref{fig:MIREX05_year_RPA} and~\ref{fig:MIREX05_year_OA}. However, we can see that the models are performing better on MIREX05 compared to the ADC2004 over the years. The performance of the models for MedleyDB dataset can be seen in Figs.~\ref{fig:MedleyDB_RPA},~\ref{fig:MedleyDB_OA},~\ref{fig:MedleyDB_year_RPA} and~\ref{fig:MedleyDB_year_OA}. MedleyDB being a complex dataset compared to other datasets, we do not see an increasing performance pattern of RPA or OA. Instead, we can see an oscillating pattern in the presented results by various authors. The reason for oscillating behavior can be that few models trained on the part of MedleyDB dataset performed well on the part of data used as test set. Whereas other models that do not use MedleyDB data during training sufferers from the data distribution shift that might have resulted in poor performance. Surprisingly, the best RPA of the year plotted on Fig.~\ref{fig:MedleyDB_year_RPA} shows the decreasing trend. Whereas OA plotted on Fig.~\ref{fig:MedleyDB_year_OA} shows a more or less increasing pattern over the years. The plots in Fig.~\ref{fig:MedleyDB_year_RPA} and Fig.~\ref{fig:MedleyDB_year_OA} shows that there is still lot of space to improve the results. 

\begin{figure}[!tbp]
        \centering
		\resizebox{12cm}{7cm}{\includegraphics{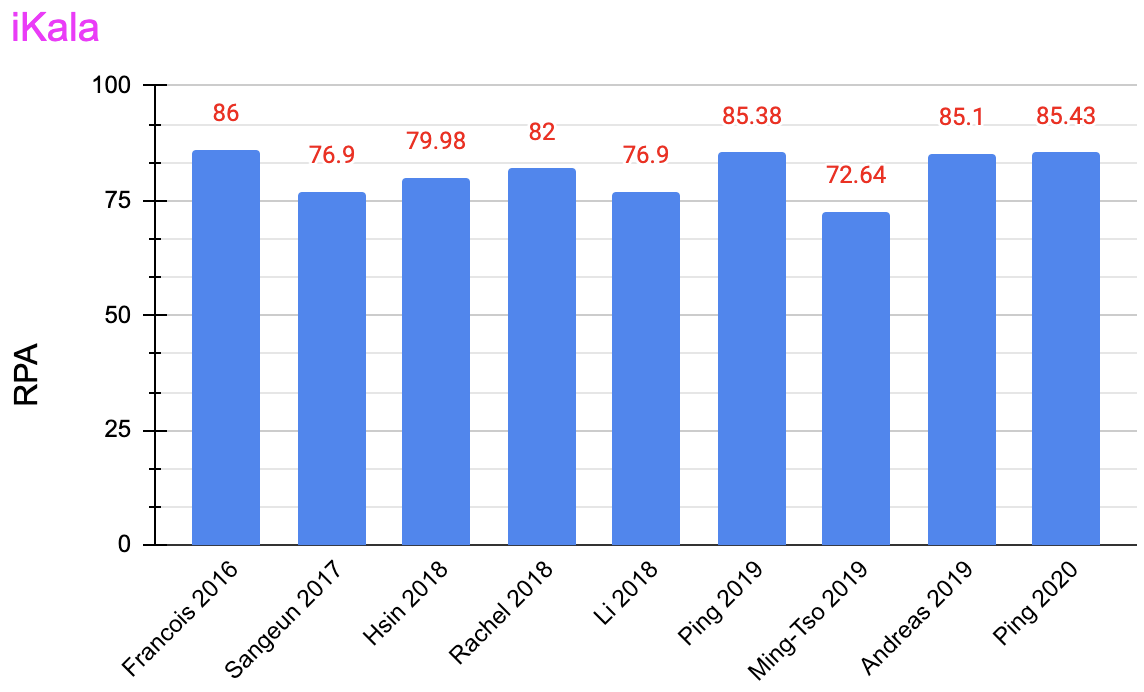}}  
        \caption{Raw pitch accuracy of the melody extraction models on iKala dataset.}
        \label{fig:iKala_RPA}
\end{figure}

\begin{figure}[!tbp]
        \centering
		\resizebox{12cm}{7cm}{\includegraphics{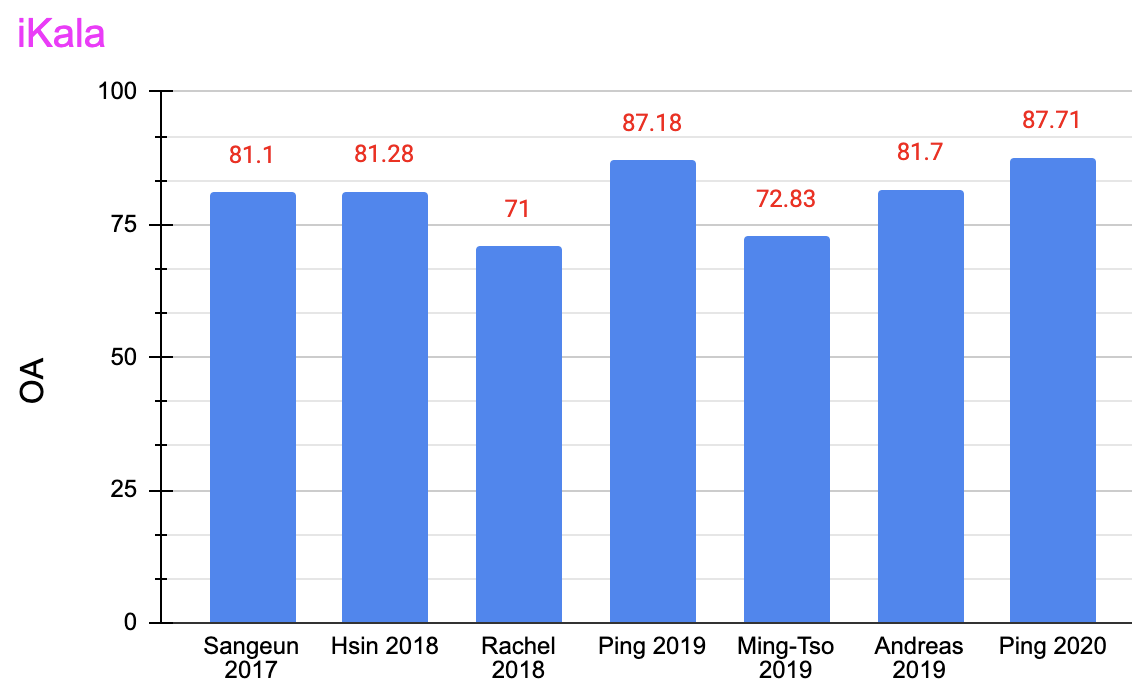}}  
        \caption{Overall accuracy of the melody extraction models on iKala dataset.}
        \label{fig:iKala_OA}
\end{figure}

\begin{figure}[!tbp]
        \centering
		\resizebox{12cm}{7cm}{\includegraphics{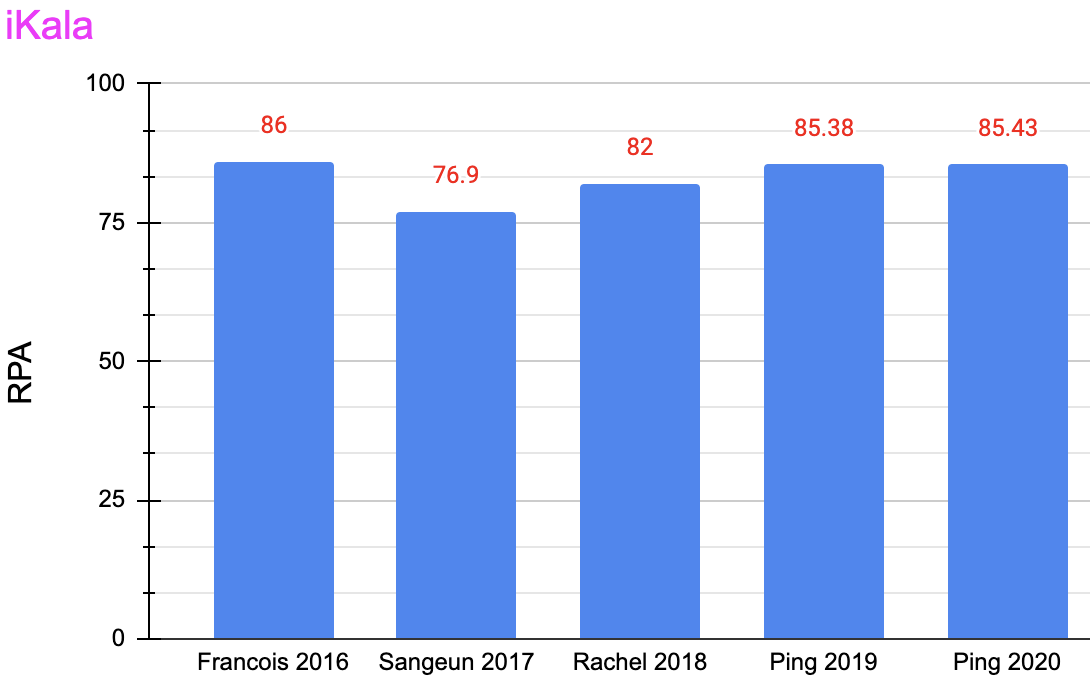}}  
        \caption{Best raw pitch accuracy of the melody extraction models on iKala dataset over the years.}
        \label{fig:iKala_year_RPA}
\end{figure}

\begin{figure}[!tbp]
        \centering
		\resizebox{12cm}{7cm}{\includegraphics{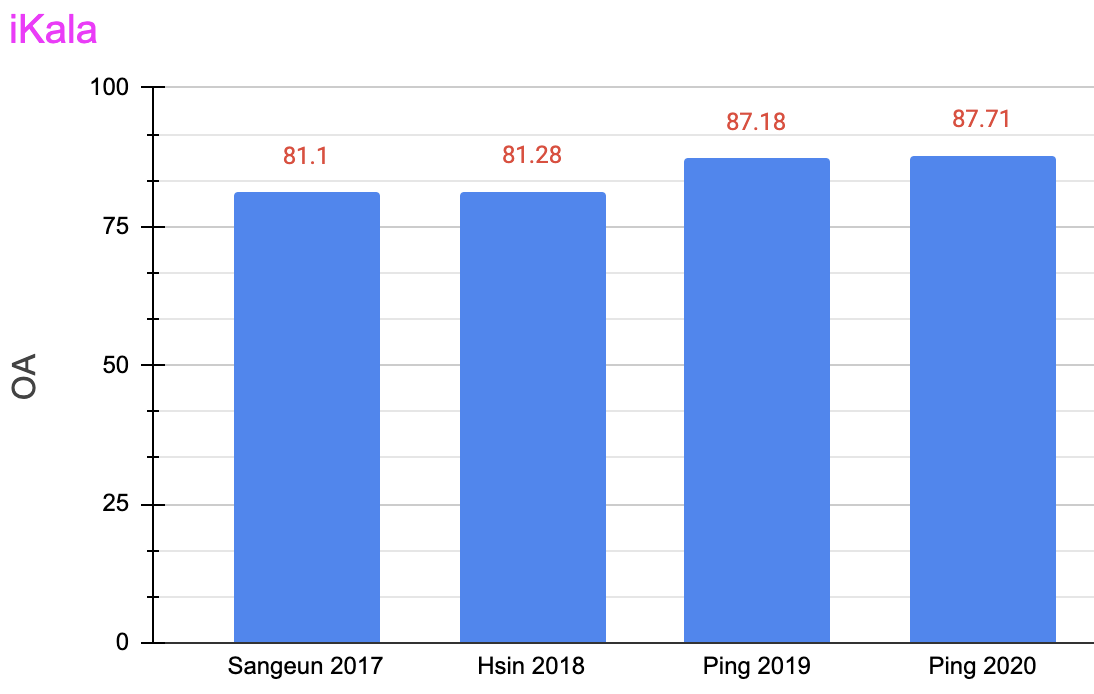}}  
        \caption{Best overall accuracy of the melody extraction models on iKala dataset over the years.}
        \label{fig:iKala_year_OA}
\end{figure}

\begin{figure}[!tbp]
        \centering
		\resizebox{12cm}{7cm}{\includegraphics{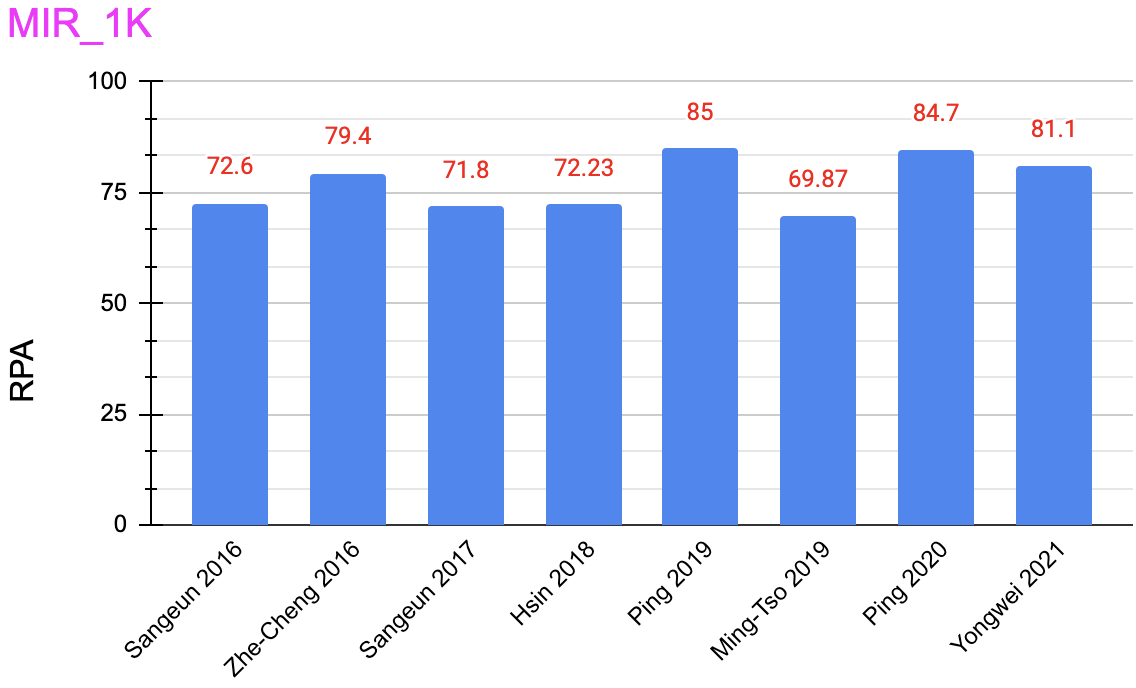}}  
        \caption{Raw pitch accuracy of the melody extraction models on MIR\_1K dataset.}
        \label{fig:MIR_1K_RPA}
\end{figure}

\begin{figure}[!tbp]
        \centering
		\resizebox{12cm}{7cm}{\includegraphics{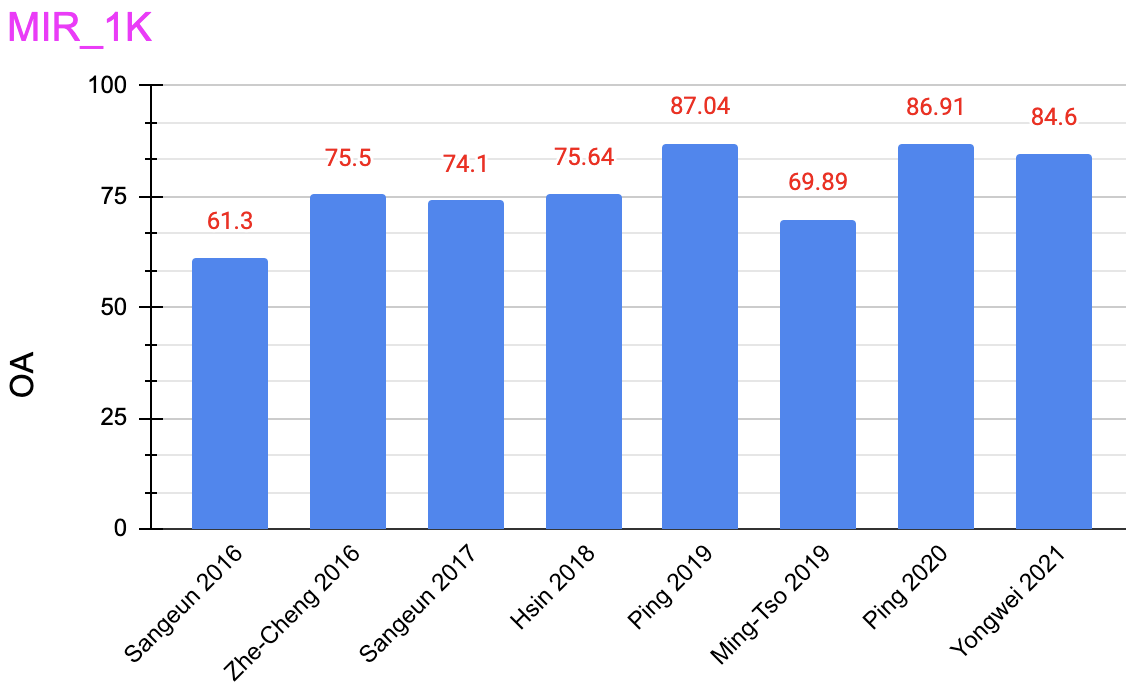}}  
        \caption{Overall accuracy of the melody extraction models on MIR\_1K dataset.}
        \label{fig:MIR_1K_OA}
\end{figure}

\begin{figure}[!tbp]
        \centering
		\resizebox{12cm}{7cm}{\includegraphics{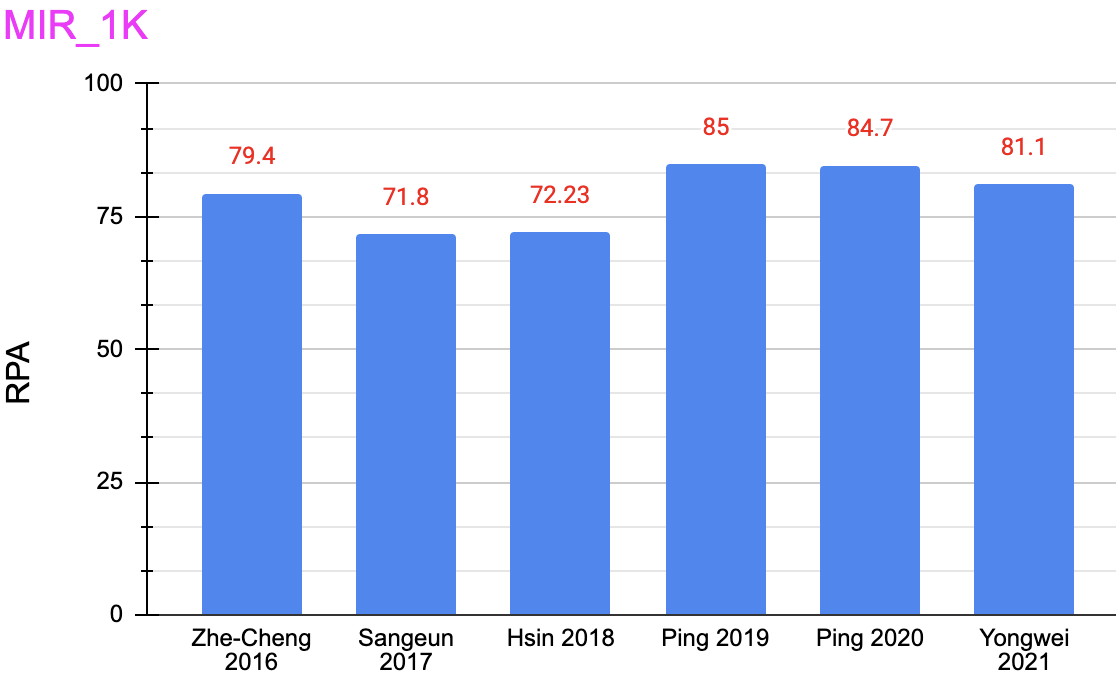}}  
        \caption{Best raw pitch accuracy of the melody extraction models on MIR\_1K dataset over the years.}
        \label{fig:MIR_1K_year_RPA}
\end{figure}

\begin{figure}[!tbp]
        \centering
		\resizebox{12cm}{7cm}{\includegraphics{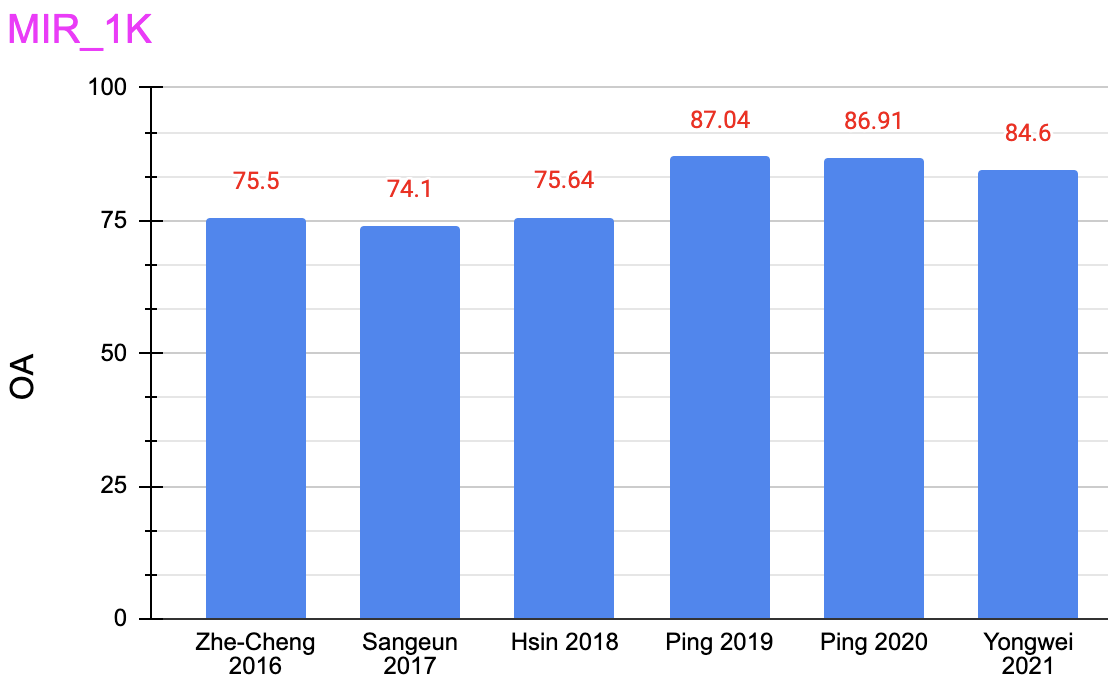}}  
        \caption{Best overall accuracy of the melody extraction models on MIR\_1K dataset over the years.}
        \label{fig:MIR_1K_year_OA}
\end{figure}

Similarly, a slight oscillating results can be seen for the iKala dataset as shown in Fig.~\ref{fig:iKala_RPA}, and Fig.~\ref{fig:iKala_OA}. However, the range of oscillation is not as much as MedledBD. Also, we can see an increasing trend of the OA, and stabilized results for RPA towards the end can be seen on the Figs.~\ref{fig:iKala_year_OA} and~\ref{fig:iKala_year_RPA} respectively. The RAP and OA plots for MIR\_1K dataset for all the reported methods can be seen in Figs.~\ref{fig:MIR_1K_RPA} and~\ref{fig:MIR_1K_OA}. The best RPA and OA of the year can also be seen in Figs.~\ref{fig:MIR_1K_year_RPA} and~\ref{fig:MIR_1K_year_OA}. We can see a common pattern in all the plots that the initially reported results were in the range of 70\% to 80\%. Towards the end of the plots, we can see that the reported results are mostly above 80\%. This indicates that the models are getting better and better over the year on the MIR\_1K dataset. The publicly available open-source codes made available by the authors of the proposed deep learning models to implement/reproduce the results are listed in Table~\ref{tab:source_code}.                
      
\begin{table*}[h]
  \begin{center}
    \caption{Open source codes available to reproduce the deep learning melody extraction models.}
    \label{tab:source_code}
    \resizebox{\textwidth}{!}{\begin{tabular}{l c c }
      \toprule 
       {First Author} & \textbf{URL} \\
      \midrule 
       Sangeun	2016~\cite{kum2016melody} & \url{https://github.com/keums/MelodyExtraction_MCDNN} \\
       Rachel	2017~\cite{bittner2017deep} & \url{https://github.com/rabitt/ismir2017-deepsalience} \\
       Dogac	2018~\cite{basaran2018main} & \url{https://github.com/dogacbasaran/ismir2018_dominant_melody_estimation} \\ 
       Rachel	2018~\cite{bittner2018multitask} & \url{https://github.com/rabitt/multitask-f0} \\
       Wei Tsung	2018~\cite{lu2018vocal} & \url{https://github.com/s603122001/Vocal-Melody-Extraction} \\
       Li	2018~\cite{su2018vocal} & \url{https://github.com/leo-so/VocalMelodyExtPatchCNN} \\
       Ping	2019~\cite{gao2019multi} & \url{https://github.com/eed0650745/singing_melody_extraction} \\
       Tsung-Han 2019~\cite{hsieh2019streamlined} & \url{https://github.com/bill317996/Melody-extraction-with-melodic-segnet} \\  
       Sangeun	2019~\cite{kum2019joint} & \url{https://github.com/keums/melodyExtraction_JDC} \\
       Sangeun	2020~\cite{kum2020semi} & \url{https://github.com/keums/melodyExtraction_SSL} \\ 
       Yongwei	2021~\cite{gao2021vocal} & \url{Vocal Melody Extraction via HRNet-Based Singing Voice Separation and Encoder-Decoder-Based F0 Estimation} \\ 
       \bottomrule 
    \end{tabular}}
  \end{center}
\end{table*}

\section{Possible Future Directions} \label{sec:future directions}

This section briefly discusses the possible future directions to explore and improve the melody extraction performance.
The dataset available to train the melody extraction models is still small compared to the datasets of other tasks in speech (ASR and TTS) and music (source separation) which restricts the researchers to train powerful models such as Transformers. The current dataset does not represent global music. The openly available dataset mainly contains Western and a few Asian music genres. There is a need to add other language music (ex. Bollywood) and more genres to build more inclusive and robust melody extraction models. There is a lot of unlabeled music data on the internet. We can leverage the self-supervised approaches to learn the general representations to improve the performance of the melody extraction models. We can explore GAN’s for both data augmentation and vocal source separation for melody extraction, which is not given much attention. Encoder-decoder based vocal source separation models have the potential to drive the melody extraction performance since we have multiple high performance monophonic pitch detection algorithms which can be directly used on the separated vocal source signal. One can also explore MUSEDB 18 dataset to train melody extraction models by automatically generating melody ground truth from vocal stems using analysis synthesis framework as proposed in CREPE~\cite{kim2018crepe}. We can also experiment if the model architecture performance remains consistent with the change in input representation. There is no study that compares various input representations by keeping the model fixed to learn the significance of input representations for melody extraction. Not much emphasis is given for model parameters initialization, although they play a significant importance in model convergence in the melody extraction literature. The number of pitch class labels used varies with the proposed models. Although the number of pitch class labels ultimately decides the melody pitch resolution, there is a need for standardizing the target pitch classes for universal melody extraction.   

\section{Summary} \label{sec:summary}

In this paper, we have reviewed several data-driven deep learning approaches for melody extraction from polyphonic music. The available deep learning approaches have been broadly categorized into classification-based and source separation-based melody extraction methods. Based on the type of neural networks used, we further categorized the classification based melody extractions methods into feed forward neural network, CNN, RNN, convolutional Recurrent Neural Network, and semantic segmentation (U-net) based melody extraction models. Based on the output representation used for training the melody extraction models, we categorized the melody extraction neural networks into two categories: the models that use quantized pitch labels as targets and the models that use spectrograms as targets and briefly explained the approaches followed to obtains the representations. The architectures of the 25 melody extraction methods are briefly presented, along with the training details. The music datasets used for training and testing the models are presented with great details, along with the differences in the protocols followed for collecting and annotating music data and melody labels. The loss functions used to optimize the model parameters of the melody extraction models are broadly categorized into four categories such as general loss functions, joint loss functions, melody specific loss functions, and class balancing loss functions and briefly described the loss functions used by various melody extraction models. We also presented the various optimization algorithms and the parameter initialization methods adopted for training melody extraction models. We also briefly described the various input representations adopted by the melody extraction models and the parameter settings. We also included the section which describes the explainability of the block-box melody extraction deep neural networks. The melody extraction evaluation measures used to report the performance of the melody extraction models are also discussed in the paper. We have presented the results of 25 melody extraction methods evaluated on the openly available melody extraction music datasets. After comparing and contrasting the results of various melody extraction methods, we found that there are still many scopes to improve the results by proposing advanced architectures, input representations, and training methods. We also briefly discussed the possible future directions to improve the melody extraction results.

\bibliographystyle{elsarticle-num} 

\renewcommand\bibname{References}

\bibliography{Melody_Extraction}

\end{document}